\newif\ifShowKeys
\ShowKeystrue

\ShowKeysfalse

\documentclass[letterpaper,10pt]{article}

\pdfoutput=1
\usepackage[no-natbib-sort]{my-jheppub}

\ifShowKeys \usepackage[notcite]{showkeys} \fi

\usepackage{amsmath, amssymb,amsfonts}

\usepackage{xfrac}   

\usepackage{amsthm}

\usepackage{bm,environ,mathrsfs,array,arydshln,booktabs,float,slashed,hyperref}
\usepackage[mathcal]{euscript}
\usepackage{tensor} 	

\usepackage[nodayofweek]{date time}

\usepackage{graphicx,epsfig}
\usepackage{epic}
\usepackage{tikz} 
\usetikzlibrary{decorations.pathreplacing,decorations.markings,snakes}

\tikzset{middlearrow/.style={
        decoration={markings,
            mark= at position 0.5 with {\arrow{#1}} ,
        },
        postaction={decorate}
    }
}

\usepackage{floatflt}

\usepackage{aurical}
\usepackage[T1]{fontenc}

\usepackage{framed}
\definecolor{shadecolor}{RGB}{255, 230, 204}

\allowdisplaybreaks

\newcommand{\be}{\begin{equation}}
\newcommand{\ee}{\end{equation}}
\newcommand{\mc}{\mathcal }

\newcommand{\la}{\label}
\newcommand{\eps}{\varepsilon}
\def \bz {\mathsf{z}}
\def \bt {\mathsf{t}}

\def \vp {\varphi}
\newcommand{\F}{\mathsf{F}}

\newcommand{\Psib}{{\bar \Psi}}
\newcommand{\Eb}{\overline{\mc E}}
\newcommand{\E}{\mc E}
\newcommand{\zb}{\bar z }

\newcommand{\desl}{\slashed{\partial}}

\def \ov  {\over}

\def \const {{\rm const}}\def \m {\mu}  \def \ed {\end{document}}
 \def \iffa {\iffalse} 
\def \ci {\cite} 
\def\foot{\footnote}
\newcommand{\rf}[1]{(\ref{#1})}
\def\la{\label}
\def\l{\lambda}
 \def \te {\textstyle} 
\def \ha {{\te{1\ov 2}}}\def \del  {\partial } 
\def \adst  {{AdS$_2$ }}

 \def \no {\nonumber}
\def \chix {\zeta}

\newcommand{\EV}[1]{\langle #1 \rangle}
\newcommand{\p}{\partial}

\newcommand{\beqn}{\begin{eqnarray}}
\newcommand{\eeqn}{\end{eqnarray}}
\newcommand{\sft}{{\sf t}}
\newcommand{\sfz}{{\sf z}}
\newcommand{\sfd}{{\sf d}}

\newcommand{\cE}{\mathcal E}

\newcommand{\cD}{\mathcal D}
\newcommand{\cP}{\mathcal P}

\newcommand{\PBK}[1]{\ensuremath{\begin{pmatrix}#1\end{pmatrix}}}

\newcommand{\sfPhi}{{\sf\Phi}}
\newcommand{\sfPsi}{{\sf\Psi}}

\DeclareMathOperator{\tr}{tr}

\newcommand{\sfF}{{\sf F}}
\newcommand{\kzeta}{{\kappa_\zeta}}
 \newcommand{\kpsi}{{\kappa_\psi}}
\newcommand{\cS}{{\sf S}}
\newcommand{\Cppp}{C_{\sfPhi\sfPhi\sfPhi}}

\newcommand{\chads}{AdS$_{2}$/CFT$_{2}^{\sfrac{1}{2}}\,$}

\newcommand{\disk}[1]{\mathsf{d}^{2}#1\,}

\makeatletter
\DeclareFontFamily{OMX}{MnSymbolE}{}
\DeclareSymbolFont{MnLargeSymbols}{OMX}{MnSymbolE}{m}{n}
\SetSymbolFont{MnLargeSymbols}{bold}{OMX}{MnSymbolE}{b}{n}
\DeclareFontShape{OMX}{MnSymbolE}{m}{n}{
    <-6>  MnSymbolE5
   <6-7>  MnSymbolE6
   <7-8>  MnSymbolE7
   <8-9>  MnSymbolE8
   <9-10> MnSymbolE9
  <10-12> MnSymbolE10
  <12->   MnSymbolE12
}{}
\DeclareFontShape{OMX}{MnSymbolE}{b}{n}{
    <-6>  MnSymbolE-Bold5
   <6-7>  MnSymbolE-Bold6
   <7-8>  MnSymbolE-Bold7
   <8-9>  MnSymbolE-Bold8
   <9-10> MnSymbolE-Bold9
  <10-12> MnSymbolE-Bold10
  <12->   MnSymbolE-Bold12
}{}

\let\llangle\@undefined
\let\rrangle\@undefined
\DeclareMathDelimiter{\llangle}{\mathopen}%
                     {MnLargeSymbols}{'164}{MnLargeSymbols}{'164}
\DeclareMathDelimiter{\rrangle}{\mathclose}%
                     {MnLargeSymbols}{'171}{MnLargeSymbols}{'171}
\makeatother

\def \N  {{\cal N}}
\def \RR {{\mathbb R}}

\title{Supersymmetric Liouville theory in \adst and AdS/CFT}
\author[a]{Matteo Beccaria,}
\author[b]{Hongliang Jiang,}
\author[,c]{Arkady A. Tseytlin\footnote{Also at the   Institute  for Theoretical and Mathematical Physics, Moscow State University
and  Lebedev Institute.}}

\affiliation[a]{Dipartimento di Matematica e Fisica Ennio De Giorgi,\\
Universit\`a del Salento \& INFN, Via Arnesano, 73100 Lecce, 
Italy} 
\affiliation[b]{Albert Einstein Center for Fundamental Physics, Institute for Theoretical Physics, 
University of Bern, \\ Sidlerstrasse 5, 3012 Bern, Switzerland}
\affiliation[c]{Blackett Laboratory, Imperial College, London SW7 2AZ, U.K.}
\emailAdd{matteo.beccaria@le.infn.it,   jiang@itp.unibe.ch,  \\
\qquad\quad \   tseytlin@imperial.ac.uk} 

\abstract{\\
In a series of recent papers, a  special kind   of AdS$_2$/CFT$_1$  duality
was observed:  the boundary   correlators of  elementary fields  that
appear in the Lagrangian
of a  2d  conformal theory    in rigid AdS$_2$  background  are  the
same as  the  correlators
of the corresponding   primary operators
in the chiral half of that  2d  CFT in flat   space restricted to the
real line.
The examples considered were:
(i) the Liouville theory where the operator dual to the  Liouville
scalar in AdS$_2$
is  the stress tensor; (ii) the
abelian  Toda theory where the operators  dual to the Toda
scalars are the $\mathcal W$-algebra generators;
(iii) the non-abelian Toda theory
where the Liouville field is  dual to the stress tensor while
the  extra  gauged WZW theory scalars are dual to non-abelian
parafermionic operators.
By   direct Witten diagram computations in AdS$_2$ one can   check that  the
structure of the boundary correlators is indeed   consistent with the 
Virasoro (or higher) symmetry.
Here we  consider  a   supersymmetric generalization:  the
 $\mathcal N=1$ superconformal  Liouville theory in AdS$_2$.
We start  with the  super Liouville  theory  coupled to  2d supergravity
and show that a consistent restriction to  rigid AdS$_2$  background
requires  a non-zero  value
of the supergravity auxiliary field   and thus a  modification of the
Liouville  potential from its  familiar flat-space form. We show  that
the Liouville  scalar and its fermionic partner
 are  dual to the chiral  half of the stress tensor and the
supercurrent of the super Liouville  theory on the plane. We perform
   tests supporting the  duality
 by explicitly computing  AdS$_2$  Witten diagrams  with bosonic  and
fermionic loops.}
\begin{document}


\begin{tabbing}
\hspace*{11.7cm} \=  \kill 
    \> Imperial-TP-AT-2019-07
\end{tabbing}

\maketitle


\def \ed  { \bibliography{BT-Biblio}
\bibliographystyle{JHEP}
\end{document}}

\def \wz {w}
\def \gb {{g_{_\del}} }  \def \sb {{\cS_{_\del}} }
\def \N  {{\cal N}}\def \z  {\zeta} 
\def \DD {{\rm D}}
\def \ssb {{g_{_\del}}}
\def \tC  {\tilde C} 
\def \a   {\alpha}  \def \ep  {\epsilon}  \def \ae {a} 
\def \hW  {\hat{W}}
\def \vPhi  {\varPhi}


\section{Introduction and summary}
\iffa 
\begin{verbatim}
suppose one can prove theorem:
given a CFT in ads2 one may expect  Virasoro acting on boundary correlators 
(realized as reparametrizations of line) 
with 
central charge same as of bulk CFT.
Then matching correlators of m^2=2 fields with 
correlators of T is a consequence-- follows from symmetry 
Bonus is that then boundary correlators of composite operators like exp a phi etc or just 
phii^n should similarly be constrained by Virasoro.
same should be in WZW case 
modulo understanding how KM 
is related to Virasoro. and similar inToda though realization of why W symm should act on 
bndry correlators and how is more subtle probably
\end{verbatim}
\fi

Recent investigations  of  correlators of operators on 1/2 BPS  Wilson loop  
at strong coupling using  AdS$_{5}\times S^5$   superstring action (see \ci{Giombi:2017cqn,
Beccaria:2019dws}  and refs. there) motivate the study of boundary correlators in  conformal quantum  field theories  in 
\adst space. 
 While  for a general QFT in   \adst one expects the boundary 
correlators to be  covariant under the isometry of \adst   or the  1d conformal group $SO(2,1)$, 
in the case  of a  conformal (Weyl-covariant)  theory   one  may expect  this  symmetry to be 
enhanced to the  infinite-dimensional Virasoro symmetry (or reparametrizations of 1d boundary) putting 
strong constraints on  the structure of the boundary correlators. 

This was studied recently on the examples of the Liouville and Toda theories 
 \cite{Ouyang:2019xdd,Beccaria:2019stp,Beccaria:2019ibr,Beccaria:2019mev}.
One  remarkable feature is that  while the S-matrix of elementary   fields of 2d CFT in flat space is 
not well defined, its counterpart  in \adst (i.e. the set of boundary correlators) is. 
In general,  a   scalar   with  $m^{2}=\Delta(\Delta-1)$ 
 in  \adst with metric  $ds^{2}=\frac{1}{\bz^{2}}(d\bt^{2}+d\bz^{2})$
  having   boundary asymptotics $\varphi(\bt, \bz)\big|_{  \bz\to 0  }  = 
\bz^{\Delta}\,\sfPhi(\bt) + \cdots$
should be dual to a 1d conformal field $V_{\Delta}$ with  dimension $\Delta$.
In the case of the Liouville scalar in \adst  expanded  near the  constant  vacuum 
(cf.  \cite{DHoker:1983zwg,DHoker:1983msr,Zamolodchikov:2001ah})  one finds $m^2=2$ and thus $\Delta=2$ 
which is the same as  the dimension of the  stress tensor $T= V_2$  in the  2d CFT in flat space. 
Then the expected  Virasoro symmetry of the boundary scalar correlators implies that they should be 
proportional to the correlators of the  holomorphic stress tensor of the original Liouville theory 
on the complex  $\wz$-plane  formally  restricted to the real line   (boundary of half-plane).\foot{In that sense 
the boundary \adst correlators provide another realization of the 
same Virasoro   symmetry (with the same central charge) as the 
flat-space stress tensor correlators.}
This  correspondence (dubbed ``\chads duality'')   extends also to the  case of higher spin $\mc W$-symmetry generators in Toda  theories. 
  In   general, in the  case of a theory with  several fields  
\begin{align}
\la{1.1}
\langle\sfPhi_{1}(\bt_{1})  \cdots \sfPhi_{N}(\bt_{N})\rangle
  &\equiv 
\lim_{\bz_{i}\to 0}\prod_{i=1}^{N}\bz_{i}^{-\Delta_{i}}\,\langle \varphi_{1}(\bt_{1}, \bz_{1})\cdots \varphi_{N}(\bt_{N}, \bz_{N})
\rangle_{_{\text{AdS}_{2}}}\notag \\
& =  \prod_{i=1}^{N}\kappa_{\Delta_{i}} \langle V_{\Delta_{1}}(\wz_{1})\cdots V_{\Delta_{N}}(\wz_{N})\rangle 
\Big|_{\wz_{i}\to \bt_{i}} \ . 
\end{align}
The proportionality coefficients $\kappa_{\Delta_{i}}$   are   functions
 of the central charge $c$ (or couplings) of the underlying CFT and are  a non-trivial part of this 
 correspondence  (see  \cite{Beccaria:2019stp,Beccaria:2019ibr,Beccaria:2019mev} for details). 

\iffa The present studies of \chads duality have fully elucidated its details in the Liouville theory 
at tree \cite{Ouyang:2019xdd} and quantum level \cite{Beccaria:2019stp}, in the 
conformal abelian Toda  theories  associated with a finite Lie algebra $\mathfrak{g}$ 
($\mathfrak{g}=A_{2}$ at tree level in \cite{Ouyang:2019xdd} and at quantum level in \cite{Beccaria:2019mev},
$\mathfrak{g}=A_{n>2}$ at tree level in \cite{Beccaria:2019ibr}, non simply-laced $\mathfrak{g}=B_{2}$
at tree level in \cite{Ouyang:2019xdd}). To discuss a case closer to the string theory related \adst  model
 in \cite{Giombi:2017cqn} containing  derivative interactions, the \chads duality has been investigated 
 in the non-abelian conformal  Toda  theory of \cite{Gervais:1992bs} in \cite{Beccaria:2019mev}. In that case, 
 the 2d CFT has a derivative part that is not free and is instead 
described  by an effective  $\sigma$-model  coming from the $SL(2,\RR)/U(1)$ gauged 
WZW model \ci{Bardacki:1990wj,Witten:1991yr}. 
\fi

The aim of the present paper 
is  to extend the analysis  of the  bosonic Liouville theory in   \cite{Beccaria:2019stp} to 
  the case of $\mc N=1$ super Liouville theory  containing 2d fermions. 
  One   motivation  is  to learn  how to systematically compute fermionic loops  in \adst in a way consistent with 
  all relevant symmetries  which should be important  also in the  AdS$_{5}\times S^5$   superstring  
  context \ci{Giombi:2017cqn,Beccaria:2019dws}. 
  Below   we shall summarize   our main results.

  \iffa 
The main motivation for this investigation is to make a further step toward the \adst theory associated with the 
Wilson loop defect theory. A missing ingredient in the analysis of  \cite{Giombi:2017cqn,Beccaria:2019dws}
are loop corrections with fermion exchanges as well as correlators of fermionic operators. 
The extension of \chads duality to a $\mc N=1$ superconformal theory is useful in order to see how much
supersymmetry is able to simplify the treatment of fermionic corrections by relating them as far as possible to 
their bosonic counterparts. Beside, it  seems interesting to address and clarify various important technical 
aspects. Examples are how supersymmetry is inherited from the bulk to the boundary and 
what kind of regularization is compatible with supersymmetry in (curved) \adst.
\fi

\subsection{Bosonic  Liouville theory}

Let us first recall the  main duality  relations     in the  bosonic Liouville theory.
The quantum Weyl-covariant Liouville action  in  a   general curved  2d space is 
 \cite{Polyakov:1981rd,Nakayama:2004vk}
\be
\la{1.2}
\mc S = \frac{1}{4\pi}\int d^{2}x\,\sqrt{g}\, \big[\partial^\l\phi  \del_\l  \phi 
+\mu^2\,e^{2\,b\,\phi}+Q\,R\,\phi\big]\ ,\qquad \qquad \qquad  Q={ 1 \ov b} +b\ ,
\ee
with the central charge   being 
\be
\la{1.3}
c=1+6\,Q^{2}\  .
\ee 
Let  us specify the background to be the  Euclidean AdS$_{2}$   space  
with coordinates $x=(\bt, \bz)$  and the 
 Poincar\'e  metric $ds^{2} = \frac{1}{\bz^{2}}(d\bt^{2}+d\bz^{2})$  
(with curvature $R=-2$). 
In this case  a non-trivial constant vacuum  value of the 
 the scalar field is  $\langle\phi\rangle=  \phi_{0} = { 1 \ov 2 b} \log { Q\ov b\mu^2}$. Expanding near it, 
 $\phi = \phi_{0}+\chix$,  one finds that the fluctuation  $\chix$ has   classical mass 
 given by $m^2= 2$. 
Assuming  Dirichlet boundary
conditions  for $\zeta$, we have   
  \be
\la{1.4}
\chix(\bt, \bz)\big|_{\bz \to 0}  =\bz^2 \sfPhi (\bt) + \mc O(\bz^{3})\ . 
\ee 
Identifying the corresponding $\Delta=2$ boundary conformal field with the
real-line restriction of the 2d  holomorphic  stress tensor $T(\wz)$ on the complex $w$-plane 
one can then verify that the key \chads \,  relation (\ref{1.1})
is satisfied at the quantum level (i.e. for any  coupling $b$) 
\begin{align}
\la{1.5}
\langle\prod_{i=1}^{N}\sfPhi(\bt_{i}) & \cdots \sfPhi_{N}(\bt_{N})\rangle= 
\kappa^{N}\ \langle \prod_{i=1}^{N} T(\wz_{i})\rangle 
\Big|_{\wz_{i}\to \bt_{i}}\ .  
\end{align}
The all-order expression for $\kappa=\kappa(b)$ was  determined in \cite{Beccaria:2019stp} 
\be
\la{1.6}
\kappa  = -4\frac{Q}{c} = - \frac{ 4\,b  (1 + b^2)}{(3+2\,b^{2})(2+3\,b^{2})} = 
-\frac{2}{3}\,b+\frac{7}{9}\,b^{3}+\cdots\ .
\ee
The structure of the correlators of $T(w)$  is completely determined 
by the Virasoro algebra and then the same    expressions should be found 
for the boundary correlators; this can be checked directly   using  small $b$ (or large $c$)   perturbation theory. 
 For  example, the well-known expression for the 4-point function 
of the stress-tensor ($\wz_{ij}=\wz_{i}-\wz_{j}$)
 \begin{align}
 \la{1.7}
 \langle T(\wz_{1}) \cdots  T(\wz_{4})\rangle = &\frac{c^{2}}{4}\,\Big(\frac{1}{\wz_{12}^{4}\,\wz_{34}^{4}}
+\frac{1}{\wz_{13}^{4}\,\wz_{24}^{4}}+\frac{1}{\wz_{14}^{4}\,\wz_{23}^{4}}\Big)\notag \\
&+c\,\Big(
\frac{1}{\wz_{12}^{2}\,\wz_{23}^{2}\,\wz_{34}^{2}\,\wz_{14}^{2}}
+\frac{1}{\wz_{13}^{2}\,\wz_{24}^{2}\,\wz_{14}^{2}\,\wz_{23}^{2}}
+\frac{1}{\wz_{12}^{2}\,\wz_{24}^{2}\,\wz_{34}^{2}\,\wz_{13}^{2}}
\Big),
\end{align}
corresponds to the decomposition of $\langle \sfPhi(\bt_{1})\cdots\sfPhi(\bt_{4})\rangle$ into 
the  disconnected and connected parts 
\begin{align}\la{1.8}
\langle \sfPhi(\bt_{1})\,\cdots\,\sfPhi(\bt_{4})\rangle &  = 
\langle \sfPhi(\bt_{1})\,\cdots\,\sfPhi(\bt_{4})\rangle_{\rm disc}  +
\langle \sfPhi(\bt_{1})\,\cdots\,\sfPhi(\bt_{4})\rangle_{\rm conn},
\\  
\la{1.9}
\langle \sfPhi(\bt_{1})\cdots \sfPhi(\bt_{4})\rangle_{\rm disc} &= (C_{\sfPhi\sfPhi})^2\,\Big(
\frac{1}{\bt_{12}^{4}\,\bt_{34}^{4}}
+\frac{1}{\bt_{13}^{4}\,\bt_{24}^{4}}
+\frac{1}{\bt_{14}^{4}\,\bt_{23}^{4}}\Big) \ ,  \notag \\
\langle \sfPhi(\bt_{1})\cdots \sfPhi(\bt_{4})\rangle_{\rm conn} &= C_{\sfPhi\sfPhi\sfPhi\sfPhi}\,\Big(
\frac{1}{\bt_{12}^{2}\,\bt_{23}^{2}\,\bt_{34}^{2}\,\bt_{14}^{2}}
+\frac{1}{\bt_{13}^{2}\,\bt_{24}^{2}\,\bt_{14}^{2}\,\bt_{23}^{2}}
+\frac{1}{\bt_{12}^{2}\,\bt_{24}^{2}\,\bt_{34}^{2}\,\bt_{13}^{2}}
\Big)\ , 
\end{align}
where
\begin{align}
\la{110}
C_{\sfPhi\sfPhi} &= \kappa^{2}\,\frac{c}{2} = \frac{8\,(1+b^{2})^{2}}{(3+2b^{2})(2+4b^{2})} = 
\frac{4}{3}-\frac{2}{9}\,b^{2}+\frac{13}{27}\,b^{4}+\cdots\ ,  \\
C_{\sfPhi\sfPhi\sfPhi\sfPhi} &= \kappa^{4}\,c = \frac{256\,b^{2}\,(1+b^{2})^{4}}{(3+2b^{2})^{3}(2+3b^{2})^{3}} = 
\frac{32}{27}\,b^{2}-\frac{80}{27}\,b^{4}+\cdots\ . \la{1.10}
\end{align}
These relations were  checked explicitly in \cite{Beccaria:2019stp} at the one-loop level. 
Similar relations  are found   for the 2-, 3-point and higher point boundary correlators.

\subsection{$\mc N=1$ super Liouville theory}

The  action of  $\N=1$   supersymmetric  Liouville theory on a curved background  can be  found  by constructing a    locally supersymmetric   generalization of  \rf{1.2}, i.e. 
by coupling the Liouville matter multiplet to $\N=1$   2d   supergravity
   \ci{Polyakov:1981re,Distler:1989nt}.
   The  condition of Weyl  invariance on a general curved background   fixes  the value of 
   the analog of $Q$  in \rf{1.2}
   and thus   the central charge to be 
\be
\la{1.12}
Q= \frac{1}{b}+\frac{b}{2}
\ , \qquad \qquad  \qquad 
c = \frac{3}{2}+6Q^{2}\ .
\ee
Restricting to the  \adst   background  in a way preserving global AdS supersymmetry 
requires  setting  gravitino to zero and fixing  the supergravity auxiliary  field $A$
to a non-zero constant value in terms of the \adst curvature  ($A=2=-R$). 
This leads to the following  action in \adst background  (see section \ref{sec3})
\be
\la{1.11}
S={1 \ov 4 \pi}  \int d^2 x \sqrt{g}\; \Big( \p^\l \phi \p_\l \phi
+ \bar\Psi\slashed D \Psi 
+\mu^2 b^2 e^{2b\phi} - 2  Q  \phi -    \mu  (b Q -1) e^{b\phi}
-  \mu  b^2 e^{b\phi}  \bar\Psi \Psi    \Big) \ , 
 \ee
 where $\Psi=\PBK{\psi \\ -i\bar \psi}$ is the 2-component Majorana fermion. 
 Here the $- 2  Q  \phi$  term  is  the same as  $Q R \phi$  in \rf{1.2}  restricted to \adst 
 while  the (possibly  unfamiliar)  extra 
  $\mu  (b Q -1) e^{b\phi}$ potential   term   originates from the  non-zero value of the supergravity auxiliary  field   and is thus  necessary for the  global  AdS  supersymmetry of  the action \rf{1.11}. 
 The final action (\ref{1.11}) is a special case of the general (rigidly) supersymmetric
 Wess-Zumino actions on \adst derived in \ci{Higashijima:1983wy,Bardeen:1984hm}.
 The  special choice of the superpotential is dictated by the supergravity construction and
 assures superconformal invariance.
 
As in the bosonic case  \rf{1.2},  there is again a constant  minimum of the potential 
  $\langle\phi\rangle=\phi_{0}$, $\langle\Psi\rangle=0$, and expanding near it 
gives a massive bosonic fluctuation with $m_\zeta^{2}=2$  
and a massive fermion with $m_\psi=1$. 
The asymptotic behaviour of the bosonic fluctuation 
 field $\z$ is  again as in   \eqref{1.4}, while for the (upper component of) the  fermion field  we  have 
\be\la{1.13}
\psi(\bt, \bz)\big|_{\bz \to 0}   = \bz^{3/2}\,\sfPsi(\bt)+\mc O(\bz^{2})\ . 
\ee
The associated conformal fields   (with  $\Delta_\zeta=2$  and    $\Delta_\psi=\frac{3}{2}$)  
 appearing in (\ref{1.1})
are to be identified with the stress tensor $T(w)$ and its partner 
 the supercurrent $G(w)$ realizing the $\mc N=1$ superconformal algebra
with the central charge $c$ in  (\ref{1.12}). 

The boundary correlators  of $\z$  and $\psi$  are defined as  
\begin{align}\la{1.14}
\langle\prod_{i=1}^{N}\sfPhi(\bt_{i})\ \prod_{j=1}^{M}\sfPsi(\bt_{j}')\rangle
 \equiv 
\lim_{\bz_{i}, \bz_{j}'\to 0}\prod_{i=1}^{N}\bz_{i}^{-2}\prod_{j=1}^{M}\bz_{j}'^{-3/2}\,\langle \zeta(\bt_{i}, \bz_{i})
\,\psi(\bt_{j}', \bz'_{j})\rangle_{_{\text{AdS}_{2}}}.
\end{align}
The goal of this  paper is to show that these  correlators
are   expressed in terms of the  correlators of the  generators $T,G$ of the  
  $\mc N=1$ super-Virasoro  algebra   as  in \rf{1.1} 
\begin{align}
\la{1.16}
\langle\prod_{i=1}^{N}\sfPhi(\bt_{i})\ \prod_{j=1}^{M}\sfPsi(\bt_{j}')\rangle= 
(\kzeta)^{N}\, (\kpsi)^{M}\ 
\langle\prod_{i=1}^{N}\langle T(\wz_{i})\ \prod_{j=1}^{M}G(\wz_{j}')\rangle 
\Big|_{\wz_{i}\to \bt_{i}, \wz'_{j}\to \bt_{j}'}.
\end{align}
As we  will find below,  the expression for $\kzeta(b)$    has the same form  in terms of $Q$   and $c$ 
in \rf{1.12} 
as in \rf{1.6}   while $\kpsi(b)$ is  proportional  to it as a consequence of 
supersymmetry 
\be
\la{1.17}
\kzeta= -4 \frac{Q}{c},\qquad \qquad\qquad  \kpsi =  {3\ov2 \sqrt2}\,  \kzeta  =    -3\,\sqrt{2}\,\frac{Q}{c}\ .
\ee
Below  we shall demonstrate the validity of  (\ref{1.16}),(\ref{1.17})
 for various correlators at the tree and one-loop   level.

The  content of the  rest of this paper is as   follows. 
In section \ref{sec:sLflat} we  briefly review the superconformal Liouville theory in flat 2d space
and the underlying $\mc N=1$ superconformal algebra.

 In section \ref{sec3} we discuss how to formulate 
the super Liouville theory on \adst space, preserving both conformal invariance and 
global   supersymmetry. 
We start in section \ref{sec3.1} with a locally supersymmetric theory of a scalar multiplet coupled to the $\N=1$  2d supergravity including both general superpotential and curvature coupling. 
We  show how to  restrict this theory  to  a  rigid \adst    background  getting a particular 
supersymmetric Wess-Zumino  theory. 
 In section \ref{sec3.2} we  find  the  special conformal  superpotential  that 
 solves the condition of (super) Weyl invariance in a  general 2d supergravity background. 
 

In section \ref{sec4} we present the main features of the 
\chads correspondence for the super Liouville theory. In section \ref{sec:osp12} 
we show how the  global \adst supersymmetry is holographically projected to the $\mathfrak{osp}(1|2)$
superconformal symmetry on the boundary of AdS$_2$. 
In section \ref{sec4.2} we determine the exact 
expressions for the two proportionality coefficients $\kappa_{\zeta}(b)$ and $\kappa_{\psi}(b)$ appearing in 
the key relation (\ref{1.16}). 
Finally, in section \ref{sec4.3} we summarize the precise predictions that follow
from  (\ref{1.16}) for the 2-, 3- and 4-point  boundary correlation functions
of the super Liouville   fields. 

In section \ref{sec5} we set up  perturbation theory in \adst and 
discuss in particular
the explicit form of the  fermionic propagator. 
Then in  section \ref{sec6} we compute the  2-point functions  for the bosonic and 
fermionic fields in   the one-loop approximation. 
This  computation provides a   non-trivial check 
 of  the  duality predictions  \rf{4.40},\rf{441}. 

Section \ref{sec7} is devoted to a similar calculation for the  3-point  scalar correlation function. 
Again, we confirm  the predictions  \rf{444},\rf{462}  at the one-loop level. 
In section \ref{sec8} we consider 
the non-vanishing 4-point  boundary correlators. 
Starting from 4-point level  the  (super)conformal invariance 
alone  does not a priori fix the  structure of the correlators so their  direct computation 
provides a check of the Virasoro symmetry.
We  check agreement with the duality 
(\ref{1.16}) explicitly at the  tree level, i.e. the leading   order in the weak-coupling expansion.


There are several  technical Appendices. 
We recall our conventions in Appendix  \ref{app:conventions}. In  Appendix  \ref{secB}
we summarize  information about different forms of the  fermionic 
propagator in AdS$_{d+1}$   and specifically in \adst. In Appendix  \ref{4pt2B2F}
 we present a lengthy calculation of the mixed 
4-point function of two scalars and two fermions discussed in section \ref{sec8}.

\section{Superconformal  Liouville theory in flat space}
\la{sec:sLflat}

 Let us first  briefly review 
the super Liouville theory  in   flat  2d  space   \cite{Polyakov:1981re,Martinec:1983um,Nakayama:2004vk}. Its  action is given by\footnote{Our conventions are discussed  in  Appendix~\ref{app:conventions}.
 Note that compared, e.g., to  \ci{Nakayama:2004vk}   
 we rescale $\phi$ and $Q$ by a factor of $\sqrt2$ for convenience. }
  \be
 \la{2.1}
\mc S=  \frac{1}{4\pi} \int d^2 x  \; \Big( \p^\l \phi \p_\l \phi+ \bar\Psi\slashed \p \Psi  
+\mu^2 b^2 e^{2b\phi}
-  \mu  b^2 e^{b\phi}  \bar\Psi \Psi    \Big) .
 \ee
This is  a special case  of the 2d Wess-Zumino theory for $\N=1$    scalar multiplet 
$(\phi, \Psi, F)$ 
 \be
\la{2.2}
\mc S = \frac{1}{4\pi}\int d^2 x \, \Big[\p^{\l}\phi\p_{\l}\phi + \Psib
\desl\Psi -F^{2}+2F W'(\phi)
-W''(\phi)\Psib\Psi
\Big],
\ee
with  the exponential  superpotential $W$ 
\be \la{2.3}
W(\phi) = \frac{\mu}{b}\,e^{b\phi}\ .
\ee
The action (\ref{2.2}) is thus invariant under the  standard global $\N=1$  supersymmetry transformation
\be
\delta\phi = \Eb\,\Psi, \qquad\ \  \delta\Psi = \desl \phi\,\E+F\,\E,\qquad\ \ 
\delta F = \Eb\,\desl \Psi\ ,
\ee
where $\E$ is a constant Majorana spinor.  
Eliminating $F$ in \rf{2.2} gives \rf{2.1}. 
%

The  holomorphic components of the conserved  conformal stress tensor and the 
 supercurrent admit the following free-field representation\footnote{The free   fields  may be  related  to the fields in \rf{2.1} by a 
 Backlund transformation, cf. \ci{DHoker:1983xdu}.
 The improvement $Q$-terms may be  either postulated to satisfy the superconformal algebra  or 
 derived from coupling  the  scalar multiplet to the supergravity background.}
\be
\la{2.5}
T = -(\p\phi)^{2}+Q\p^{2}\phi-\frac{1}{2}\psi\p\psi,\qquad \qquad G = i\,\sqrt{2}\,(\psi\p\phi-Q\,\p\psi)\ . 
\ee
 $Q$ and $c$  are related to  the coupling $b$  as in \rf{1.12}.
The singular part of the  OPE  of $T$  and $G$   corresponds to  the $\mc N=1$ superconformal algebra
\begin{align}
T(w')T(w) &\sim \frac{c}{2}\,\frac{1}{(w'-w )^{4}}+\frac{2\,T(w )}{(w'-w )^{2}}+\frac{\p T(w )}{w'-w },\label{2.6}\\
T(w')G(w) &\sim \frac{3}{2}\frac{G(w )}{(w-w')^{2}}+\frac{\p G(w )}{w'-w },\label{2.7} \\
G(w') G(w) &\sim \frac{2c}{3}\,\frac{1}{(w'-w )^{3}}+  \frac{2T(w )}{w'-w }. \label{2.8}
\end{align} 
Note that  \rf{2.7}  can be rewritten as 
\be\label{2.9}
G(w') T(w)  \sim   \frac32 \frac{ G(w)}{ (w'-w)^2}+\frac12 \frac{\p G(w)}{ w'-w}.
\ee
The  exponential  NS primary field and its superpartner in super Liouville theory  are ($\alpha$ is a constant  parameter)\foot{Here we consider a complex plane with independent  holomorphic and antiholomorphic sectors. 
For a discussion of the super Liouville  theory on a plane  with  boundary 
 see \cite{Fukuda:2002bv}.} 
\be
\la{2.10} 
V_\alpha =e^{ 2 \alpha \phi   },\qquad  \qquad \Lambda_\alpha =-i \sqrt2 \alpha\,  e^{ 2  \alpha \phi  }\, \psi\ ,
\ee
and their OPEs with $T $  and $G$ are 
\beqn  \label{2.11}
T(w')V_\alpha(w) &\sim& \frac{h_\alpha V_\alpha(w) }{(w'-w)^2}+\frac{\p V_\alpha(w)}{w'-w}
\ , 
\qquad T(w')\Lambda_\alpha(w) \sim \frac{(h_\alpha+\frac12) \Lambda_\alpha(w) }{(w'-w)^2}+\frac{\p \Lambda_\alpha(w)}{w'-w}\ , 
\\    \label{2.13}
G(w') V_\alpha(w) &\sim& \frac{\Lambda_\alpha(w)}{ w'-w }\ , \qquad \qquad \qquad \qquad
G(w') \Lambda_\alpha(w) \sim \frac{2h_\alpha V_\alpha}{(w'-w)^2}
+\frac{\p V_\alpha(w)}{w'-w}\ , 
\eeqn
where the chiral dimension is $h_\alpha =\alpha (  Q-\alpha )$.

\section{$\N=1$ super  Liouville theory in AdS$_2$  background}\label{sec3}

While   putting bosonic Liouville theory in  \adst   background is straightforward
by just  specifying   the metric, to do the same for  the  super Liouville theory 
in a way  consistent with global $\N=1$  supersymmetry in AdS  space is more  non-trivial.

One is to start with the general locally supersymmetric coupling of the  WZ model 
for the scalar multiplet \rf{2.2} to the  $\N=1$   2d supergravity multiplet $(e^a_\mu, \chi_\mu, A)$ 
\ci{Deser:1976rb,Brink:1976sc,Howe:1978ia}, set the gravitino $\chi_\m$ to zero and 
 the metric to be the  \adst  one,  and  finally  determine the value of the supergravity auxiliary field $A$ from the integrability condition  for existence of Killing spinors (i.e. from  the  vanishing of the gravitino 
 supersymmetry variation).\foot{A similar approach  is used in higher dimensions 
 when putting a supersymmetric  field  theory on a specific  curved  background  in way 
 preserving some global supersymmetres  (see  \ci{Buchbinder:1998qv,Festuccia:2011ws,Kuzenko:2015lca}   
 and refs. there).}
 
 The resulting  action for a  WZ  theory  in \adst  
was found in  \ci{Higashijima:1983wy,Bardeen:1984hm,Uematsu:1984zy,Uematsu:1986de} (and used also in 
 \ci{Sakai:1984fg,Inami:1985di}). However, these papers did not include the  possibility of 
 the curvature  coupling (or ``dilaton term''  in string context) 
  which is crucial for the conformal invariance of the Liouville theory. 
  The manifestly supersymmetric   form of the $R \phi$ coupling in \rf{1.2}  on a  general  supergravity  background    was  given in   \ci{Distler:1989nt} (see also \ci{DHoker:1988pdl,Abdalla:1991hp})
  but  the restriction of the resulting theory  to the \adst background  was not  explicitly 
  constructed   in the past.\foot{However, some  related  discussion  appeared in \ci{DHoker:1983xdu}.}
  
    Below   we will 
    close  this gap, 
     explaining the derivation of the action \rf{1.11}  of the  $\N=1$ super Liouville   action  in  \adst\!\!.  
We shall  start  with  the  manifestly  locally supersymmetric   form of the  2d scalar  multiplet 
 theory  with generic superpotential  $W$ (``tachyon'')   and  curvature coupling function  $U$ (``dilaton'') 
  on a general supergravity   background and   consider its  supersymmetric reduction to \adst background. 
   We shall  also  determine the  conditions on the functions  $W$ and $U$  required for the 
     (super) Weyl  covariance of  such  model on a general background
     implying   its   superconformal invariance  and then consider the resulting super Liouville theory in \adst.

\subsection{Scalar  multiplet  theory 
in     2d supergravity   background  and  restriction  to \adst}
\la{sec3.1}

 The  general  action for the  scalar superfield 
  \be\la{3.2}
 \Phi = \phi + i \theta \psi + i\bar\theta \bar \psi +i \theta\bar \theta F \ , 
 \ee
 coupled   to 2d $\N=1$   supergravity  may be written as \cite{Howe:1978ia,Uematsu:1984zy,DHoker:1988pdl} 
 \be
 \la{3.1}
\mc S= - {  i\ov 4 \pi}  \int d^2 {Z}\,  E\,  \Big[i \cD^\a \Phi \cD_\a \Phi 
+ 2 W(\Phi)+ \mathcal R\,  U(\Phi)     \Big]\  ,
 \ee
 where   $  d^2 {Z}\,   \equiv  d^2 x d\theta d\bar \theta$   and   $E=\text{sdet}\, E_M^A$.
   $E^A_M$ is the supervielbein (containing the supergravity multiplet  $(e^a_\mu, \chi_\mu, A)$)
 and $ \mathcal R$  is the supercurvature\foot{$\mc R = R_{+-}$ in the notation of \cite{DHoker:1988pdl}.
 We assume $\int d\theta d\bar \theta \,  \theta\bar \theta =1$.}
 \begin{align}\la{333}
 \mc R= A + i \theta \bar \theta (R+\tfrac12 A^2)+...\ ,\qquad\qquad 
 E 
  = e\big(1 -\tfrac{i}{2} \theta \bar \theta A+ ... \big)\ ,
 \end{align}
  where $R$ is the scalar curvature of the metric  and dots  stand  for the terms depending on  the gravitino.  
  The bosonic part of \rf{3.1} thus contains the terms (cf. \rf{1.2},\rf{2.2}) 
  $\mc S=  {  1\ov 4 \pi}  \int d^2 x \big[ - F^2 + 
  2 F W'(\phi)  +   R U(\phi) + ...\big].$

 To restrict this theory  to \adst preserving rigid  supersymmetry we first  set 
 the gravitino to zero $\chi_\mu=0$. Then   under the supersymmetry variation  $\delta e_\mu{}^a=\delta A=0$  while the condition of the 
 vanishing of the local   supersymmetry variation of  gravitino becomes 
  \be\la{3.3} 
 \delta \chi_\mu  =2 (D_\mu -\tfrac14 \gamma_\mu A)\, \E=0 \ . 
 \ee
 Specifying the  metric to be the \adst one (with unit radius)
 \be
 \la{3.5}
 ds^2 =
 \frac{  d\sfz^2 +d\sft^2}{\sfz^2} \  , \qquad   \qquad   R=-2  \ , 
 \ee
  to preserve the  global $SO(2,1)$  symmetry 
    we  also choose the background  value 
 of the auxiliary field $A$  to be constant. Then  the integrability condition of \rf{3.3} 
 gives 
 \be  \la{33}  R+ \ha A^2 =0 \ , \ \ \ \ \ \  \ \ \ \ A= 2 \ae \ , \ \ \ \  \ \ \  \ae =\pm 1 \ , \ee
 and \rf{3.3}   becomes equivalent to  the \adst Killing spinor equation 
  \be 
 \la{3.4}
 D_\mu \E= \tfrac12\, \ae\,\gamma_\mu \E\ .
 \ee
To find the  form of the action \rf{3.1} on this \adst background  we observe that (cf. \rf{333},\rf{33})
 \be
\int d\theta d\bar \theta   E\Big[2 W(\Phi)+ \mathcal R\, U(\Phi)  \Big]  
  =ie\Big[ 2F\, W' (\phi) -  2a  \, W(\phi)  -  2  U(\phi) +   2 a F\, U'(\phi)  \Big]\ .
 \ee
Then the  bosonic potential  part of \rf{3.1}  
  may     be written as 
\begin{align}\la{3.110}
V(\phi,F)&=  -  F^2  +  2 F\hW'   -  2\ae  \, \hW(\phi) \ ,  \\
\hW&\equiv W(\phi)  +  \ae\, U(\phi) \ . \la{3.10}
\end{align}
where $- F^2 $ comes from the kinetic part of the action \rf{3.1}. 

Including also the fermionic terms we finally 
 get the following  supersymmetric  action on AdS$_2$   
\be
\la{3.11}
\mc S= {  1\ov 4 \pi} \int d^2x \sqrt g \,  
\Big( \p^\lambda \phi \p_\lambda\phi+ 
\bar\Psi\slashed D \Psi - F^2+2 F\hW' -2\ae\hW-\hW'' \bar\Psi \Psi  \Big).
\ee
where $ \int d^2x \sqrt g  = \int \frac{d\bt\,d\bz}{\bz^{2}}$ (cf. \rf{3.5}). 
This  action is invariant under the following supersymmetry transformations
\be\la{311} 
\delta\phi = \Eb\,\Psi, \qquad\ \ \  \delta\Psi = \desl \phi\,\E+F\,\E,\qquad\ \ \ 
\delta F = \Eb\,\slashed D \Psi\ ,
\ee
where  $\E $ is the solution to the Killing spinor equation~\eqref{3.4}.  
Note that if  we set  formally the  background value $a$ of the auxiliary field $A$ to zero 
and take the metric flat  then \rf{3.11}  reduces to the flat-space WZ  action \rf{2.2}. 

Solving for the auxiliary field ($F= \hW'$) we get from \rf{3.11} 
\be
\la{310}
\mc S= {  1\ov 4 \pi} \int d^2x \sqrt g \,  
\Big( \p^\lambda \phi \p_\lambda\phi+ \bar\Psi\slashed D \Psi +  \hW'^2 -2\ae\hW-\hW'' \bar\Psi \Psi  \Big).
\ee
The action \rf{310}  has  the same  form as the general supersymmetric WZ action in \adst
found  in  \ci{Higashijima:1983wy,Bardeen:1984hm}.
The fact that the curvature coupling function $U(\phi)$ in \rf{3.1} 
appears in \rf{311},\rf{310}  only  through the generalized superpotential function $\hW$  is a check of consistency of  this coupling with the \adst  supersymmetry.


\subsection{Super Weyl invariance conditions}
\la{sec3.2}

The action \eqref{3.11}  has rigid \adst supersymmetry
 but  does not   correspond to a  superconformal theory  for a generic $W(\phi)$. 
 To determine the  required  conditions  on $W$ and $U$  we  need  to go back to 
  the  action \rf{3.1}  defined on a general supergravity background and impose 
  the condition of decoupling of the superconformal factor at the quantum level.


Let us first recall  the analogous argument in the bosonic case  starting with the 
single scalar action 
\be\la{313}
\mc S= \frac{1}{4\,\pi}\int d^{2}x \sqrt{g} \Big[
\del^\lambda \phi \del_\lambda \phi 
 +R\,U(\phi)+T(\phi)\Big] \  ,  
\ee
with generic  coupling functions $U$ and $T$.  Interpreting  them 
as the ``dilaton'' and the  ``tachyon''  in 1d target space 
we may readily write down the corresponding expressions  for the Weyl anomaly coefficients 
(see, e.g., \cite{Callan:1985ia,Tseytlin:1988rr,Tseytlin:1990mz}; here $\a'=1$)
\be \la{3.13} 
\bar \beta^U = - \ha U''  + U'^2 \ , \qquad \qquad 
\bar \beta^T =  -2 T    - \ha  T''  + U'  T' \ . 
\ee
Here the non-derivative term  originates from the classical  scaling dimension of the corresponding coupling,  the second-derivative term is the quantum anomalous dimension 
(one-loop Weyl anomaly assuming reparametrization covariant regularization of the propagator) 
  and the first-derivative terms  originate  (upon use of the equations of motion) 
  from the classically  non-invariant dilaton coupling.
The solution  of $\bar \beta^U =\const= {1\ov 6} (c-1)$   required to decouple the conformal factor is the linear dilaton $U=  Q \phi $ and then the  solution of $\bar \beta^T = 0$ is 
the exponential  Liouville potential in \rf{1.2}  with $-2 -  2b^2   +  2Q b  =0$, i.e. (cf. \rf{1.2})
\be  \la{314}
T= \mu^2 e^{2b \phi} \ , \qquad \qquad   Q=\frac{1}{b}+{b} \ .
\ee
A similar argument applies in the supersymmetric case of the action \rf{3.1}.\foot{One may 
repeat the  bosonic derivation using covariantly regularized superpropagator, cf. 
\ci{DHoker:1987rxo}.}
The equation for $U$ is the same as in \rf{3.13}  fixing it   again to be $U=  Q \phi $.
The  equation  for the ``super-tachyon''   or the superpotential  $W$  
differs from the one  for $T$ in \rf{3.13} only in first classical dimension term: 
$W$ term in \rf{3.1}  has  half dimension  of the bosonic tachyon coupling, i.e. 
\be \la{3.14} 
\bar \beta^W =  - W    - \ha  W''  + U'  W' \ . 
\ee
The condition of the  vanishing of $\bar \beta^W $  gives again the exponential function 
$W= \mu e^{b \phi} $ with $b$   now  constrained by $-1 - \ha  b^2   +  Q b  =0$. 
Hence, we finish with 
\be \la{3.16} 
W =\mu\,e^{b \phi}\ , \qquad   \qquad  U = Q\,\phi
\ , \qquad \qquad  Q=\frac{1}{b}+\frac{b}{2} \ , 
\ee
thus determining the  structure of  the  
super Liouville theory on a general curved background
(cf. \rf{1.12}).

\subsection{Action for  super Liouville theory in \adst}
 
 Let us now  combine  the discussion of  the previous two sections 
 and  write down the explicit  form of the action of the $\N=1$  supersymmetric Liouville theory in \adst.\foot{One may   wonder  why one  needed  separate   conditions on $W$   and $U$ 
 if the action \rf{3.11}  depends only on their sum  $\hW$ in \rf{3.10}.
 The answer is analogous to that in the  flat space case: having fixed a background  one is still 
  to specify the stress tensor and  its superpartner  and their definition depends 
  implicitly on the  right choice of $U= Q\phi $  (cf. \rf{2.5}).} 
 
 Using \rf{3.16} the generalized   superpotential in \rf{3.10} takes the form
 \be\la{3.18} 
\hW(\phi)=  \mu\, e^{b\phi} +  \ae\, Q\,\phi \ . 
\ee
Then solving for the auxiliary field $F$ in \rf{3.11} gives 
 the following expression for the super Liouville theory on AdS$_2$
 \be
\la{3.20}
\mc S=   \frac{1}{4\,\pi}  \int d^2 x \sqrt{g}\; \Big[ \p^\l \phi \p_\l \phi+ \bar\Psi\slashed D \Psi - 2Q \phi + 2 \ae \mu (bQ-1) e^{b\phi}
+\mu^2 b^2 e^{2b\phi}
-  \mu  b^2 e^{b\phi}  \bar\Psi \Psi    \Big] \ . 
 \ee
 We used that $a^2=1$ in \rf{33}   and  dropped  a constant $Q^2$ term.
For  $\ae=-1$ this  is the action quoted in \rf{1.11}.

The condition for the extremum of the  scalar potential in \rf{310}  $V= \hW'^2 -  2 a \hW$ is 
$(\hW'' - a )\hW'=0$. Since $F= \hW'$,  the  constant vacuum  $\phi=\phi_0$  preserving supersymmetry (cf. \rf{311}) is 
the solution of $\hW'=0$, i.e. 
\be  \la{320}
\mu\,b\,  e^{b\phi_0} +  \ae\, Q=0 \ . \ee
 Expanding near this   vacuum  $\phi=\phi_0+\zeta$ 
 we get 
  \be
 \la{3.22}
\mc S=\frac{1}{4\,\pi} \int d^2 x \sqrt{g}\; \Big[ \p^\l \zeta \p_\l \zeta+ \bar\Psi\slashed D \Psi  - 2Q \zeta
-   2Q(Q- b^{-1})   e^{b\zeta} 
+ Q^2 e^{2b\zeta}
+  \ae\,  b\,  Q  e^{b\zeta}   \bar\Psi \Psi    \Big]\ .  
 \ee
Thus $a=\pm 1$ determines the sign of the fermion mass term, the two options being equivalent
and related by a chiral transformation; in what  follows we shall choose 
\be\la{323} 
  a=-1  \ . \ee
  So far  we did not use the explicit  value  of $Q$ in \rf{314} required for conformal invariance. 
  Expanding \rf{3.22}  to  quadratic order in the fields we find 
   \begin{align}
   \la{3.24}
\mc S^{(2)} &=\frac{1}{4\pi}\, \int d^2 x \sqrt{g}\; \Big[
\p_\mu \zeta \p^\mu\zeta+  ( 2 + \tfrac{1}{2} b^2) (1 + \ha b^2)  \,\zeta^{2}
+\bar\Psi\,\slashed D\Psi  -(1 + \ha b^2) \bar \Psi \Psi     \Big]\ .
\end{align}
Thus the classical ($b \to 0$)   values of the  bosonic and fermionic masses are 
 $m_\zeta^{2}=2$ and  $m_{\psi}=1$. 

 
\section{Super Liouville  theory: \chads  set up}
\la{sec4}

Starting with the \adst action \rf{3.22},\rf{323}   let us now discuss  the  AdS/CFT  duality.
According to the standard AdS$_{d+1} $/CFT$_d$  rules (see, e.g., \ci{Metsaev:2003cu}), 
the scalar and the spin $1\ov 2$   fermion    masses in AdS$_2$  are related to the  global AdS energies or 
 dimensions   of the dual conformal operators as, respectively,  
$m^2=\Delta(\Delta-1)$ and  $m=\Delta -\frac12$.  Using the   values of  the masses  $m^2_\zeta$ and $m_\psi$  in \rf{3.24}   this gives 
$\Delta_\zeta= 2 +  \ha b^2, \ \   \Delta_\zeta= {3\ov 2}  +  \ha b^2$. 
The fact that  $\Delta_\zeta - \Delta_\psi= {1\ov 2} $ for   any $b$ is a  consequence of  AdS supersymmetry. 

To leading order in $b \to 0$ we thus have 
\be\la{4.1}
\Delta_\zeta = 2\ ,\qquad\qquad  \Delta_\psi = \tfrac{3}{2},
\ee
with the   expansions near the \adst boundary thus  given by
(recall that in \rf{3.22} we have  $\Psi=\PBK{\psi \\ -i\bar \psi}$)
\begin{align}
\la{4.2}
&\qquad \qquad \qquad \zeta(\bt, \bz) = \bz^{2}\,\sfPhi(\bt)+\mc O(\bz^{3})\ , \\
  \la{4.11}
& \psi(\sft,\sfz)= \sfz^{3/2} \sfPsi(\sft) +\sfz^{5/2} \tilde \sfPsi(\sft) +  \mc O(\bz^{7/2})
\ , \qquad 
\bar \psi(\sft,\sfz) =  \sfz^{3/2} \bar \sfPsi(\sft) +\sfz^{5/2}\bar{ \tilde \sfPsi} (\sft)+
+  \mc O(\bz^{7/2})\ . 
\end{align}
To preserve   supersymmetry we will later need to impose the  
 boundary condition  on   $\Psi$ at $\bz=0$:
\be\la{4.356}
\sfPsi = - \bar\sfPsi   \ . 
\ee
As we will see, this boundary condition is also compatible with the standard bulk-to-boundary propagators of the fermion \eqref{5.12}. 

  The conformal dimensions \rf{4.1}  are the same as 
   the  dimensions of the   generators $T$ and $G$  of  the $\N=1$ super-Virasoro algebra. 
 As in the Liouville  theory \ci{Beccaria:2019dws}, 
we    thus expect the   \chads cor\-respondence
(in the sense of the matching of the correlation functions in \rf{1.16})\foot{In view  of \rf{4.3} 
the boundary correlators  involving $\bar \psi$   are determined by those  with  $\psi$.}
\be
\la{4.3}
\sfPhi(\bt)\  \to \  \kzeta\, T(\bt)\ ,\qquad\qquad 
\sfPsi(\bt)\  \to  \kpsi \, G(\bt) \ , 
\ee
where the $b$-dependent coefficients $\kzeta$ and $\kpsi$ are to be determined  below.


\subsection{From bulk supersymmetry  to   $\mathfrak{osp}(1|2) $   superconformal symmetry at the boundary}
\la{sec:osp12}

 To establish the  duality, it is necessary to show that the symmetries of two sides are the same. 
 Let us first  show that the boundary global superconformal symmetry 
 follows  indeed  from the supersymmetry of the   bulk super Liouville  theory. 
 This is an example    of the familiar relation between  the supersymmetry of  a theory in AdS$_{d+1}$  and global superconformal symmetry of the CFT$_d$. 
 In the  special  case of the locally conformal (e.g. Liouville)   theory  in \adst this  global 
 symmetry  is expected to be further enhanced to the super-Virasoro symmetry.

 The supersymmetry parameter $\cE $ in \rf{311}   is the  solution of the  \adst 
 Killing spinor equation \rf{3.4}   which for  $\ae=-1$  is given by  
 (see Appendix~\ref{app:conventions} for spinor conventions):  
 \be\label{4.6}
 \cE(\sft,\sfz)= \PBK{   \epsilon  \\ - i\bar   \epsilon}
 =Z(\sft,\sfz) \Lambda \ , \qquad  \Lambda =\PBK{   \lambda  \\  - i\bar\lambda }\ , \qquad 
 Z=\frac{1}{2\sqrt \sfz}  \PBK{ 1+i\sft+\sfz  &\  - i- \sft+i \sfz   \\ 
  i-\sft-i\sfz     & \ 1-i\sft+\sfz }   \ .   \ee 
  Here  $\Lambda$ 
 is a constant Majorana spinor  
  and $Z$   satisfies   $ \bar Z Z=1 $ with $\bar Z=\mathcal C  Z^T \mathcal C$.\footnote{
 This guarantees that $ \bar\cE \cE=\bar\Lambda\Lambda=\const$, i.e.  $  \p_\mu (\bar \cE \cE)=0$,  which  
 follows from \rf{3.4}.}
Then the   supersymmetry transformation that  leaves invariant the super Liouville  action  \rf{3.22} 
expanded near the supersymmetric vacuum \rf{320}  follows from \rf{320},\rf{311} 
     \be 
     \la{4.9}
  \delta \zeta = \bar  \Psi \cE\ , \qquad\ \ \ \ 
  \delta \Psi = \slashed \p \zeta  \cE + F \cE\ , \qquad \ \ \ \  F=Q \big( e^{b\zeta}-1\big)\ , 
  \ee 
  where $F=\hW'$ is the ``on-shell''  value of the auxiliary field in \rf{3.11} 
  (that vanishes in the vacuum $\z=0$). 
 Note that in  the classical limit ($b \to 1, \  Q \to b^{-1}$)    one  has 
 $F=\zeta+\mathcal O(b^2, \zeta^2)$.
 
 Using  the asymptotic expansion  \rf{4.2},\rf{4.11}   in the r.h.s. of \rf{4.9} 
  we learn  that  near the  boundary 
   the supersymmetry transformations  \rf{4.9}   reduce  to 
   \begin{align}
 \delta {\zeta}(\sft,\sfz) &=  -\ha i  \sfz  \xi (\sfPsi+\bar\sfPsi)+\ha 
\sfz^2  \big[    \p_\sft \xi (\sfPsi-\bar\sfPsi)-  i\xi   (\tilde\sfPsi+\bar {\tilde\sfPsi}) \big] +\mathcal O(\sfz^3)
\la{4.14}
\\
 \delta\psi(\sft,\sfz) &= \tfrac{3}{2}   \sfz^{ 3/2}    \, \xi \sfPhi  
 + \ha  i \sfz^{ 5/2}   \p_\sft (\xi \sfPhi    ) +\mathcal O(\sfz^{7/2})\ ,  \quad 
 \delta\bar\psi(\sft,\sfz) = -   \tfrac{3}{2}  \sfz^{ 3/2}  \, \xi \sfPhi  
 + \ha i  \sfz^{ 5/2}   \p_\sft (\xi \sfPhi    ) +\mathcal O(\sfz^{7/2})\no\ , \\
  \xi (\sft) &=  \l - \bar \l  + i  \sft (\l + \bar \l) \equiv \epsilon +\sft\, \eta    \ . \la{4.15} 
\end{align}
 Comparing this with the    variation of  \rf{4.2},\rf{4.11} 
  implies that we should  set $\sfPsi+\bar\sfPsi=0$, i.e. assume the boundary condition \rf{4.3}. 
  Then we  also find that   $\delta \sfPsi=- \delta \bar\sfPsi$, \  $\delta \bar\sfPsi= \delta \bar{\tilde\sfPsi}$  and 
  $\delta \tilde \sfPsi=\tfrac{i}{3}\delta \p_\sft \sfPsi$.
 As a result, we arrive at the  following  consistency  conditions  required for preservation of the supersymmetry near the boundary\footnote{
 Note  that  the fermion equation of motion  $ \slashed D \Psi =\hW''(\phi) \Psi $ following from 
 \rf{310}  reduces to the last relation in \rf{4.18}   near the boundary (in   the classical limit $b\rightarrow 0$). }
\be
\la{4.18}
 \sfPsi(\sft) = -\bar\sfPsi(\sft) , \qquad  \qquad   \tilde\sfPsi(\sft) =\bar{\tilde \sfPsi}(\sft), \qquad    \qquad 
 \tilde\sfPsi(\sft)   =  \tfrac{i} {3}\p_\sft\sfPsi(\sft) 
\ee 
The resulting supersymmetry  transformation rules for the  boundary  fields are 
 \beqn 
 \la{4.20}
  \delta  \sfPhi(\sft) =  \p_\sft \xi    \sfPsi(\sft)  +\tfrac13 \xi \p_\sft \sfPsi (\sft)   \ , \qquad \qquad \qquad 
   \delta  \sfPsi(\sft)  = \tfrac{3}{2}   \xi \sfPhi(\sft)    \ .
 \eeqn
  Since here $\xi=\epsilon +\sft \,\eta$ is  not constant (cf. \rf{4.15}) 
 we thus   get  two independent  global supersymmetry transformations  with constant parameters 
 \begin{align}
 \label{4.22}
   \delta_\epsilon  \sfPhi(\sft)  &=  \tfrac13 \epsilon \p_\sft \sfPsi (\sft), \qquad\qquad \qquad  \qquad
   \delta_\epsilon  \sfPsi(\sft)  = \tfrac{3}{2}   \epsilon \sfPhi(\sft) \ , 
\\  \label{4.23}
   \delta_\eta  \sfPhi(\sft)  &= \eta\sfPsi(\sft)+ \tfrac13 \eta\sft  \p_\sft \sfPsi (\sft), \qquad\qquad 
   \delta_\eta  \sfPsi(\sft)  = \tfrac{3}{2} \eta \sft    \sfPhi(\sft) \ . 
 \end{align} 
 The transformation in \rf{4.22}  is   the same as   the standard  Q-supersymmetry
  in flat 2d space 
 (relating  also $T$  and $G$ in \rf{2.5}, cf. \rf{4.3}) restricted to the real line, i.e. with $w \to \sft$.
 The transformation in  \rf{4.23}  is  the conformal S-supersymmetry  for fields with conformal 
 dimensions as in \rf{4.1}. The commutators  of the two $\eta$-transformations  are
\be 
 [\delta_{\eta_2},\delta_{\eta_1}] \sfPhi =\eta_1\eta_2 \big( 4\sft+ \sft^2 \p_\sft\big) \sfPhi , \qquad \qquad
 [\delta_{\eta_2},\delta_{\eta_1}] \sfPsi  =  \eta_1\eta_2 \big( 3\sft+ \sft^2 \p_\sft\big) \sfPsi \ , 
 \la{4.24}
\ee
where the  right-hand sides are  recognized  to be  the  special conformal  K-transformations:
$  \{S,S\}= K$.\footnote{Here we use    the standard notation
 $$
 [\delta_{\eta_2},\delta_{\eta_1}]\equiv [ {\eta_2} S,  {\eta_1}S ] =   \eta_1\eta_2  \{S,S\}, \quad
  [\delta_{\eta },\delta_{\epsilon}]\equiv [ {\eta } S,  {\epsilon}Q ] =   \epsilon \eta    \{S,Q\},  \quad
   [\delta_{\epsilon_2},\delta_{\epsilon_1}]\equiv [ {\epsilon_2} Q,  {\epsilon_1}Q ] =   \epsilon_1\epsilon_2  \{Q,Q\}.
 $$
 }
The generator $K$  acting  on a  field $\varphi$  with dimension $\Delta_\varphi $   gives 
\be
 K \varphi =(2\Delta_\varphi  \sft +\sft^2 \p_\sft) \varphi \ , \qquad 
\ee 
so here 
 $
\Delta_\sfPhi=2, \  \Delta_\sfPsi={3\ov 2} $  (cf. \rf{4.1}). 
 Computing the commutators of Q- and S-supersymmetries 
\be
\la{4.29}
[\delta_{\eta},\delta_{\epsilon }] \sfPhi =\epsilon\eta \big( 2+ \sft\,  \p_\sft\big) \sfPhi, \qquad\qquad 
[\delta_{\eta},\delta_{\epsilon }] \sfPsi =\epsilon\eta \big( \tfrac32  + \sft\,  \p_\sft\big) \sfPsi ,
\ee
one checks also that $ \{S,Q\}=D$ where $D$ is the generator of dilations
\be\la{430}
 D \varphi =( \Delta_\varphi   +\sft  \p_\sft) \varphi \ . 
\ee 
Finally,  the commutator of two Q-supersymmetries gives 
  $
 {} [\delta_{\epsilon_2},\delta_{\epsilon_1}] \sfPhi= {\epsilon_1} {\epsilon_2} \p_\sft \sfPhi, \ \ 
  {} [\delta_{\epsilon_2},\delta_{\epsilon_1}] \sfPsi= {\epsilon_1} {\epsilon_2} \p_\sft \sfPsi
 $
 i.e. 
 \be
 \{Q,Q\}=H,\qquad   \qquad H=\p_\sft\ . 
 \ee
 Thus the  generators  $(Q, S, H, D, K)$    form the $\mathfrak{osp}(1|2) $ super Lie algebra~(see,  e.g.,  \cite{Frappat:1996pb,Cunha:2016fnr}), i.e. the super-extension of the 
 conformal algebra $so(2,1)$ in  one dimension. 
 
 This  is the finite-dimensional  sub-algebra of the infinite dimensional super-Virasoro algebra  (the same as  generated by $T$ and $G$ in 
\rf{2.5}). 
It is   natural to expect that given  a super Weyl   invariant theory in the bulk   the  boundary   global superconformal symmetry 
 should get extended to the  full super-Virasoro  symmetry. To  demonstrate  this directly 
  beyond the classical level will require a choice of a symmetry-preserving regularization.

 
 

\subsection{Determination of the proportionality coefficients  $\kappa_{\zeta}(b)$ and $\kappa_ \psi(b)$} 
\la{sec4.2}

Since the supersymmetry relating  the  boundary fields $\sfPhi$ and $\sfPsi$ is the same as the one  relating the  operators 
$T$  and $G$ one  may expect that  the ratio of the  coefficients  $\kzeta$ and $\kpsi$    in \rf{4.3}  should
be universal, i.e. should not depend on $b$.  To  fix these coefficients  we may follow the arguments in 
\ci{Beccaria:2019stp,Beccaria:2019mev}. 

Let us start with a  heuristic  semiclassical argument generalizing the one  for  the Liouville theory in \ci{Beccaria:2019stp}.
Assuming $b\to 0$ (i.e. $Q \to b^{-1}$) the \adst  
 super Liouville action \rf{3.22}   may be related to the   flat space action \rf{2.1} 
by the special Weyl transformation 
 \be\label{4.34}
 g_{\mu\nu} \rightarrow  e^{2\sigma} g_{\mu\nu} , \qquad
 \phi\rightarrow \phi- b^{-1}   \sigma, \qquad   \Psi \rightarrow  e^{-\frac12 \sigma}  \Psi \  ,  \qquad \sigma=\log \sfz
 \ee
 Then using \rf{4.2},\rf{4.11}  the    boundary   asymptotics  of the corresponding   fields 
 on the  flat half-plane  is found to be 
 \beqn\la{420}
\phi (\sft,\sfz)\big|_{\sfz\rightarrow 0} =\sfz^2 \sfPhi(t) -  b^{-1} \log \sfz+... \ , \qquad 
\psi (\sft,\sfz)\big|_{\sfz\rightarrow 0} = \sfz^{3/2}  e^{-\frac12 \log \sfz} \sfPsi(\sft)+... =\sfz  \Psi(\sft)  +...\ .   
\eeqn
Evaluating $T$  and $G$ in \rf{2.5}  on these fields  
and taking the boundary limit   we get 
\beqn\la{421} 
T(\sft,\sfz)\big|_{\sfz\rightarrow 0}  = -\tfrac{3}{2b} \sfPhi(\sft)+... \ , \qquad \qquad 
G(\sft,\sfz)\big|_{\sfz\rightarrow 0} =   -\tfrac{\sqrt2}{ b} \Psi(\sft)+... \ . 
\eeqn
Comparing this with  \eqref{4.3}, we conclude that in the semiclassical limit 
\be\label{4.39}
\kzeta  =-\tfrac23 b + \mc O(b^2)\ , \qquad \ \ \ \ \qquad \kpsi=-\tfrac{1}{\sqrt2} b  + \mc O(b^2) \ ,
\ee
in agreement with the leading asymptotics of the general expressions quoted in \rf{1.17}.


Let us now  give another argument  based on  the OPE in the  boundary CFT  
 which  leads  to the exact expressions 
for $\kzeta$ and $\kpsi$ in \rf{1.17}. 
Let us   first determine $\kzeta$  starting with the OPE  of $T$  with the primary field $V_\a$ in \rf{2.10}
following  the discussion in \cite{Beccaria:2019stp} for the Liouville theory.
 Expanding  the expression for $T V_\a$ in \rf{2.11}   to the leading order in  small $\alpha$  gives 
\be
T(w') \, \phi(w)\    \sim\   \frac{Q }{ 2(w'-w)^2 }+ \frac{\p\phi(w)}{ w'-w }\ . 
\ee
 Setting $w'=\sft', w=\sft+i\sfz$ and using the asymptotic form 
  of $\phi$ near the boundary  (now we  use $Q$ in place of its  classical  value $b^{-1}$  in \rf{420})
 \be
 \la{4.41}
 \phi(\sft,\sfz)\big|_{\sfz\rightarrow 0}  = \sfz^2 \sfPhi(\sft) -Q \log \sfz+\cdots\, ,
 \ee
one finds (note that $\p \equiv \p_w  =\frac12 (\p_\sft -i \sfz)$)
   \be\la{442} 
T(\sft') \phi(\sfz,\sft)   \sim   \frac{iQ}{2z (\sft'-\sft)}  +\sfz \bigg[  
\frac{iQ}{2 (\sft'-\sft)^3}  -\frac{i \sfPhi(\sft)}{ \sft'-\sft}  \bigg]  +\sfz^2 
\bigg[ - \frac{ Q }{  (\sft'-\sft)^4 }+ \frac{  \sfPhi(\sft) }{  (\sft'-\sft)^2 }+ \frac{2\p_\sft \sfPhi(\sft) }{  (\sft'-\sft)^2 } \bigg]\ . 
\ee
Including  the conjugate part to account for the real-line boundary (i.e.  adding the expression with 
$\sfz \rightarrow -\sfz$), gives the  OPE  for  the boundary fields 
  \be
     \la{4.43}
T(\sft') \sfPhi(\sft)   \sim - \frac{2Q }{  (\sft'-\sft)^4 }+ \frac{2 \sfPhi(\sft) }{  (\sft'-\sft)^2 }+ \frac{\p_\sft \sfPhi(\sft) }{  (\sft'-\sft)^2 }\ . 
\ee
 Comparing  this with the  boundary limit of the  OPE  for  $T(w')$ and $T(w)$    in \eqref{2.6}
 we conclude that  the two  relations   have the same form  provided  we identify $\sfPhi(\sft) \sim \kzeta T(\sft)$  as in \rf{4.3}    with  $\kzeta $ given by 
 \be
\la{4.44}
\kzeta = -\frac{4Q}{c}= -\frac{4\,b\,(2+b^{2})}{3\,(1+b^{2})(4+b^{2})} = 
-\tfrac{2}{3}\,b+\tfrac{1}{2}\,b^{3}+\cdots \ , 
\ee
where we used the expressions  \rf{1.12}   for $Q$ and $c$ in the super Liouville theory.

To determine   $\kpsi$  let us start with  the  OPE of $G$ and $V_\a$  in \eqref{2.13} and again expand  in $\alpha$, getting\footnote{This OPE  can  be found  also using the  free field realization  of the stress tensor  $T$ and supercurrent $G$ in \eqref{2.5}.}
\be\la{445} 
G(w') \, \phi(w)   \sim  -  \frac{i}{\sqrt2}\, \frac{\psi(w)}{ w'-w } \  . 
\ee
  Setting $w'=\sft', w=\sft+i\sfz$, and  using   the asymptotic  near-boundary 
  form of $\phi$  \rf{4.41}    and $\psi$  (cf. \rf{4.11},\rf{4.34}) 
 \beqn\label{4.46}
\psi(\sft,\sfz)\big|_{\sfz\rightarrow 0} =  e^{-\frac12 \log \sfz}  \big[ \sfz^{3/2} \sfPsi(\sft) +\sfz^{5/2} \tilde\sfPsi(\sft)+\cdots   \big]
=\sfz  \Psi(\sft)  +\sfz^{2} \tilde\sfPsi(\sft)+\cdots \, ,
\eeqn
we arrive at 
\be
G(\sft') \sfPhi(\sft) \sim \frac{i \sfPsi(\sft)  }{\sqrt2 \sfz (\sft'-\sft)} +\frac{\sfPsi(\sft) }{\sqrt2(\sft'-\sft)^2}- i \frac{\tilde\sfPsi(\sft) }{\sqrt2(\sft'-\sft )}\ . 
\ee
Adding the $\sfz \rightarrow -\sfz$    conjugate expression to the r.h.s.  gives the  boundary OPE
\be
\la{4.49}
G(\sft') \, \sfPhi(\sft) \sim  \frac{\sqrt 2\, \sfPsi(\sft) }{(\sft'-\sft)^2}- i \frac{\sqrt 2\, \tilde\sfPsi(\sft) }{\sft'-\sft }\ . 
\ee
On the other hand, we may compare this with the boundary  limit of the the  OPE  of $G$ and $T$ in \rf{2.9};   assuming 
the identification $ \sfPhi(\bt) \sim   \kzeta\, T(\bt), \ 
\sfPsi(\bt)\sim   \kpsi \, G(\bt)$   we get from \rf{2.9}
\be\la{435}
G(\sft')\,  \sfPhi(\sft) \sim \frac{\kzeta }{ \kpsi} \bigg[   \frac{3 }{\sqrt 2}     \frac{\sfPsi(\sft) }{(\sft'-\sft)^2}+
 \frac{1}{2} \frac{ \p_\sft \sfPsi (\sft) }{\sft'-\sft } \bigg] \ . 
\ee
The expressions in \rf{4.49} and \rf{435} match provided we identify 
\be
\la{4.51}
 \frac{\kzeta }{\kpsi}   =\frac{2\sqrt2}{3}\ , \qquad\qquad 
 \tilde\sfPsi =\frac{i}{3} \p_\sft \sfPsi \ . 
\ee
Hence  we get,  in agreement with \rf{1.17},\rf{4.39}, 
\be
\la{4.52}
\kpsi =\frac{3}{2\sqrt2}\, \kzeta=  - 3\sqrt{2}\,  \frac{ Q}{c} \ , 
\ee
   The  relation for $\tilde\sfPsi$    in \rf{4.51} is the same as the one  in \rf{4.18} 
   found from the condition of preservation  of the  supersymmetry near the boundary; this  supports  the consistency of 
   the above argument.


\subsection{
Duality predictions for the  boundary correlation functions}
\la{sec4.3}

The \chads correspondence  implies the relations \eqref{1.16} 
between the  \adst  boundary  correlators of the elementary super Liouville fields  $\z$,$\psi$  and
the flat-space CFT  correlators  of the super-Virasoro generators $T$ and $G$ restricted to the real line. 
While the matching of the coordinate dependence  on the two sides of the correspondence should follow 
from the super-Virasoro symmetry, direct computation of the boundary correlators in small $b$ (or large $c$) perturbation theory allows one
to check the expressions  \eqref{4.44} and \eqref{4.52} for the  coefficients $\kzeta$ and $\kpsi$.

The correlation functions of  the chiral $T$ and $G$  operators on a plane   are completely  fixed  by the  super-Virasoro symmetry
or the OPE relations \rf{2.6}--\rf{2.8}. Explicitly, one finds (see, e.g., ~\cite{Zamolodchikov:1985wn})
\begin{align}\la{4.35}
\langle T(w_{1})T(w_{2}) \rangle &= \frac{c}{2}\ \frac{1}{w_{12}^{4}},\qquad\qquad 
 \langle G(w_{1})G(w_{2}) \rangle   = \frac{2c}{3}\frac{1}{w_{12}^{3}},
\\ \la{4.36}
\langle T(w_{1})T(w_{2})T(w_{3})\rangle &= c \frac{1}{w_{12}^{2}\,w_{13}^{2}\,w_{23}^{2}},\qquad 
 \langle T(w_{1})G(w_{2})G(w_{3})\rangle  =c \frac{1}{w_{12}^{2}\,w_{13}^{2}\,w_{23}},
\\  \la{4.37}
\EV{T(w_1)T(w_2)T(w_3) T(w_4 )} &=  
 \frac{1}{w_{13}^{4}w_{24}^{4}}  
 \Big[
 \frac{c^2}{4}   \Big(\frac{1}{  \chi ^4}+\frac{1}{(\chi -1)^4}+1\Big) +2 c \frac{ \chi ^2-\chi +1}{(\chi -1)^2 \chi ^2}
 \Big]\ , 
 \\  \la{4.38}
\EV{G(w_1)T(w_2)G(w_3) T(w_4 )}&=   \frac{1}{w_{13}^{3}w_{24}^{4}} 
\Big[ \frac{c^2}{3}+
   \frac{c}{ 2 }  \frac{ 4\chi^2-4\chi+3}{ \chi^2(\chi-1)^2}\Big]\ ,  \qquad \qquad \qquad 
    \chi \equiv  \frac{w_{12}w_{34}}{w_{13}w_{24}} \ , 
\\ \la{4.399}
\EV{G(w_1)G(w_2)G(w_3) G(w_4 )} &=  \frac{1}{w_{13}^{3}w_{24}^{3}}\Big[\frac{4c^2}{9}
\Big(\frac{1}{ \chi^{3}}-1+\frac{1}{(1-\chi)^{3}}\Big)+    2\,c\,\frac{1}{\chi(1-\chi)}\Big] \ , 
\end{align}
where $\chi$ 
is  the  conformally invariant cross ratio. 
As in \rf{1.7}--\rf{1.9}  the order $c^2$ parts of  the 4-point functions  correspond to  disconnected contributions, 
 while the  order $c$ parts correspond to the   connected  contributions to the boundary correlators. 
 
Then  using the duality relation  \eqref{1.16}, we get  the following  predictions for 2-point    boundary 
correlation functions:
\begin{align}\la{4.40}
&\qquad \langle \sfPhi(\bt_{1})\sfPhi(\bt_{2}) \rangle = \frac{C_{\sfPhi\sfPhi}}{\bt_{12}^{4}},\qquad\qquad \qquad \ \ \
\langle \sfPsi(\bt_{1})\sfPsi(\bt_{2}) \rangle = \frac{C_{\sfPsi\sfPsi}}{\bt_{12}^{3}}, \\
&C_{\sfPhi\sfPhi}= \ha {\kappa_\zeta^2\,c} = \tfrac{4\,(2+b^{2})^{2}}{3\,(1+b^{2})(4+b^{2})}=
\tfrac{4}{3}-\tfrac{1}{3}\,b^{2}+\cdots, 
\quad  C_{\sfPsi\sfPsi} = \tfrac{2}{3}\,\kappa_\psi^2\,c = \tfrac{3}{2}\,C_{\sfPhi\sfPhi} = 2-\tfrac{1}{2}\,b^{2}+\cdots. \la{441}
\end{align}
The  non-vanishing 3-point  boundary correlators   should be  given by 
\begin{align}\la{444}
&\langle \sfPhi(\bt_{1})\sfPhi(\bt_{2})\sfPhi(\bt_{3})\rangle =\frac{C_{\sfPhi\sfPhi\sfPhi}}
{\bt_{12}^{2}\,\bt_{13}^{2}\,\bt_{23}^{2}},\qquad \qquad 
\langle \sfPhi(\bt_{1})\sfPsi(\bt_{2})\sfPsi(\bt_{3})\rangle =\frac{C_{\sfPhi\sfPsi\sfPsi}}{\bt_{12}^{2}\,\bt_{13}^{2}\,\bt_{23}}\ , 
\\
&C_{\sfPhi\sfPhi\sfPhi} = \kappa_\zeta^{3}\,c = -\tfrac{32 b (2+b^{2})^{3}}{9 (1+b^{2})^{2}(4+b^{2})^{2}} = 
-\tfrac{16}{9}\,b+\tfrac{16}{9}\,b^{3}+\cdots\ , \la{462} \\
&C_{\sfPhi\sfPsi\sfPsi}= \kzeta\,\kappa_\psi^2\,c = \tfrac{9}{8}\,C_{\sfPhi\sfPhi\sfPhi} = 
-2\,b+2\,b^{3}+\cdots\ .  \la{4.62}
\end{align}
 As the  disconnected  part of the 4-point  correlators  is determined by the 2-point functions,
 the non-trivial prediction follows only from the  connected (order $c$) part of    \rf{4.37}--\rf{4.39}: 
 \begin{align}\la{4143}
& \langle \sfPhi(\bt_{1})\sfPhi(\bt_{2})\sfPhi(\bt_{3}) \sfPhi(\bt_{4})\rangle_{\rm conn} =
 \frac{ C_{\sfPhi\sfPhi\sfPhi\sfPhi} }{\sft_{13}^4\sft_{24}^4}
  \frac{2( 1-\chi+ \chi^2)}{ \chi^2(1-\chi)^2}
 \ ,  \ \ \ \ \ \ \qquad  \chi \equiv  \frac{\sft_{12}\sft_{34}}{\sft_{13}\sft_{24}} \ , \\ 
&\EV{\sfPsi(\bt_1)\sfPhi(\bt_2)\sfPsi(\bt_3) \sfPhi(\bt_4 )}_{\text{conn}} = 
\frac{C_{\sfPsi\sfPhi\sfPsi\sfPhi}}{\bt_{13}^{3}\,\bt_{24}^{4}}\,
\frac{3 - 4 \chi  + 4\chi^{2}}{2\,\chi^2(1-\chi)^{2}},
\la{443} \\
&\EV{\sfPsi(\bt_1)\sfPsi(\bt_2)\sfPsi(\bt_3) \sfPsi(\bt_4 )}_{\text{conn}}  =  \frac{C_{\sfPsi\sfPsi\sfPsi\sfPsi}}
{\sft_{13}^3\sft_{24}^3} \frac{1}{\chi(1-\chi)}  \ , \la{4.45}
\\ 
\la{4.66}
&C_{\sfPhi\sfPhi\sfPhi\sfPhi} = \kappa^{4}_{\zeta}\,c = \tfrac{128\,b^{2}\,(2+b^{2})^{4}}{27(1+b^{2})^{3}(4+b^{2})^{3}} = 
\tfrac{32}{27}\,b^{2}-\tfrac{56}{27}\,b^{4}+\cdots.
\\
&C_{\sfPsi\sfPhi\sfPsi\sfPhi} = \kappa_\zeta^2\kappa_\psi^2\,c = \tfrac{9}{8} C_{\sfPhi\sfPhi\sfPhi\sfPhi} =
\tfrac{4}{3}\,b^{2}-\tfrac{7}{3}\,b^{4}+\cdots, \la{447} \\
&C_{\sfPsi\sfPsi\sfPsi\sfPsi} = 2\,\kappa_\psi^{4}\,c = \tfrac{81}{32}\,C_{\sfPhi\sfPhi\sfPhi\sfPhi}
=3\,b^{2}-\tfrac{21}{4}\,b^{4}+\cdots\ , \la{4.47}
\end{align}
where, as in \rf{4.62},   we used the supersymmetry implied relation  \rf{4.52}.

On the \adst   side the  relation between the correlators with only bosons and  correlators where  some of the bosons are replaced by fermions should
also  follow  from the supersymmetry  \rf{4.22}.  
In more detail, the   kinematical structure (i.e. $\sft$-dependence) of the  2- and 3- point correlators  in \rf{4.40},\rf{444} is fixed by  the global conformal symmetry
while 
for the 4-point correlators in 
\rf{4143}--\rf{4.45}  it is  fixed  by the expected Virasoro symmetry. 
Then  the relations between   their  coefficients, i.e. 
\be \la{466}
C_{\sfPsi\sfPsi} =  \tfrac{3}{2}\,C_{\sfPhi\sfPhi} \ , \ \ \ \ \ \ 
C_{\sfPhi\sfPsi\sfPsi}= \tfrac{9}{8}\,C_{\sfPhi\sfPhi\sfPhi} \ , \ \ \ \ \ 
C_{\sfPsi\sfPhi\sfPsi\sfPhi} = \tfrac{9}{8} C_{\sfPhi\sfPhi\sfPhi\sfPhi} \ , \ \ \ \ \  C_{\sfPsi\sfPsi\sfPsi\sfPsi} =  \tfrac{9}{4} C_{\sfPsi\sfPhi\sfPsi\sfPhi} 
    =\tfrac{81}{32}\,C_{\sfPhi\sfPhi\sfPhi\sfPhi}
\ , \ee
 follow from the  Ward identities corresponding to the  boundary supersymmetry  \rf{4.22}.
For example, 
 considering  the supersymmetry variation   \rf{4.22}  of   the vanishing correlator 
  $ \langle\sfPhi(\bt_{1})
\sfPhi(\bt_{2})\sfPhi(\bt_{3})\sfPsi(\bt_{4})\rangle = 0$, one obtains
\begin{align}
 \tfrac{3}{2}\langle\sfPhi(\bt_{1})\sfPhi(\bt_{2})\sfPhi(\bt_{3})\sfPhi(\bt_{4})\rangle &+ \tfrac{1}{3}\,\Big[
\partial_{\bt_{1}}\langle\sfPsi(\bt_{1})\sfPhi(\bt_{2})\sfPhi(\bt_{3})\sfPsi(\bt_{4})\rangle+\partial_{\bt_{2}}\langle\sfPhi(\bt_{1})\sfPsi(\bt_{2})\sfPhi(\bt_{3})\sfPsi(\bt_{4})\rangle\notag\\
&\qquad +\partial_{\bt_{3}}\langle\sfPhi(\bt_{1})\sfPhi(\bt_{2})\sfPsi(\bt_{3})\sfPsi(\bt_{4})\rangle\Big] = 0\ .\la{400}
\end{align}
As a result, using \rf{4143},\rf{443} we  find that $C_{\sfPsi\sfPhi\sfPsi\sfPhi} = \tfrac{9}{8} C_{\sfPhi\sfPhi\sfPhi\sfPhi}$. 
One can similarly   show that    $ C_{\sfPsi\sfPsi\sfPsi\sfPsi} = \tfrac{9}{4} C_{\sfPsi\sfPhi\sfPsi\sfPhi}     $ by  starting with 
$\delta\langle\sfPhi\sfPsi\sfPsi\sfPsi\rangle=0$.

Our aim   below will  be to check  explicitly  such  symmetry-implied relations and also the expression  for $\kappa_\zeta(b)$  in \rf{4.44} 
(which does not follow just  from symmetries). In particular,   checking the supersymmetry-implied  relations  beyond tree level 
(where  the  derivation of the boundary supersymmetry in section \ref{sec:osp12}
 directly   applies) 
will rest on a choice  of a  regularization prescription consistent with the underlying symmetry.

\section{Perturbation theory in AdS$_{2}$ }
\la{sec5}

The goal  in the rest of the paper is to check the predictions  summarized in the previous section  starting from the \adst  
super Liouville  action  \rf{3.22}  and  using small $b$ perturbation  theory. 
Expanding \rf{3.22}  in powers of $b$ using the value of $Q$ in \rf{3.16}   and separating out 
the free part from interacting vertices (containing powers of $b$) we get 
  \begin{align}
   \la{xxx}
\mc S  =\frac{1}{4\pi}\, \int d^2 x \sqrt{g}\; \Big\{  &
\p_\mu \zeta \p^\mu\zeta  +  2 \zeta^2  + \bar\Psi\slashed D \Psi  -  \bar \Psi \Psi 
\notag 
\\& 
 +\big(3b^{2}+\tfrac{1}{2}\,b^{4}\big)\, \frac{\zeta^2}{2!}\,
 +\big(8b+7b^{3}+\tfrac{3}{2}\,b^{5}\big)\, \frac{\zeta^3}{3!}\,
 +\big(16 b^{2}+15 b^{4}+\tfrac{7}{2}\,b^{6}\big)\, \frac{\zeta^4}{4!}\,
+ \cdots \notag 
\\& 
-   \frac{1}{2 }  \Big[
b^{2}+\big(2\,b+b^{3}\big)\,\zeta+\big(2\,b^{2}+b^{4}\big)\,\frac{\zeta^{2}}{2!}
+\big(2\,b^{3}+b^{5}\big)\,\frac{\zeta^{3}}{3!}\
+\cdots
\Big]  \,\bar\Psi \Psi  
 \Big\} \ . 
\end{align}
We will treat  the $b$-dependent quadratic $\sim b^2\, \zeta^2$ and $\sim b^2 \, \bar \Psi \Psi $   terms  as interaction terms
leading to self-energy insertions into diagrams.

To develop perturbation theory in  powers of $b$ we will need to know the explicit form of the 
free  propagators for the massive fields 
$\zeta$ and $\Psi$. 
Various representations for  the scalar and fermion propagator in AdS$_{d+1}$  are reviewed  in Appendix~\ref{app:propagators}.
Here we will specify them to the present  \adst  case of the scalar $\zeta$ with mass $m_\zeta^2=2$ and a Majorana  fermion $\Psi$ with mass $m_\psi=1$.

We  will use  the following   sets of  coordinates
in Euclidean \adst   
\be
\la{A17}
x^{a}=x_{a}= (x_0,x_1)= (\bz, \bt)\ ;  \qquad\qquad  w  =  \bt+i\,\bz = -i\,\frac{z+1}{z-1}\ ;  \qquad \qquad z=r e^{i\theta} \ . 
\ee
Here the  complex variable 
$w$ parametrizes  the upper half  of complex plane (with the \adst  metric in \rf{3.5}), while  complex  $z $ is a    coordinate  on a unit disc. 
From (\ref{A17}), the   coordinate $\theta$  of the  boundary of the disk ($r=|z|=1$)   and the coordinate $\bt$ of the 
  boundary     ($\sfz=0$) of the upper half plane  are related  by 
\be\la{53}
\sft(\theta) = -\cot \frac{\theta}{2} \ . 
\ee
The \adst measure is then  given by 
\be\la{54} {\sf d}^2 x\equiv  { d^2 x \ov 4 \pi  \sfz^2} = {\sf d}^2 w \equiv  \frac{  d^2 w }{ 4\pi\sfz^{2}} = 
\mathsf{d}^{2}z\equiv \frac{  d^2 z }{ \pi (1-|z|^{2})^{2}} 
\ , \qquad  d^2 w =d\sft \,d\sfz\ , \qquad  d^2 z  = r dr d\theta\ . \ee

The free  propagator of  the scalar $\zeta$  with  $m^2=2$ (with Dirichlet boundary conditions, i.e. $\Delta_\zeta=2$)  is given by\foot{The propagator 
$G$ 
corresponds to canonical normalization of the action  $\frac{1}{2}   \int d^{2}x\,\sqrt{g}\, ( \p_\mu \zeta \p^\mu \zeta - 2\zeta^2)$.}
\be
\la{5.3}
\langle\zeta(x)\,\zeta(x')\rangle_{0} \equiv 
g(x,x') 
=\begin{tikzpicture}[line width=1 pt, scale=0.8, baseline=-0.1cm,decoration={
    markings,
    mark=at position 0.53 with {\arrow{>}}}]
    \node[above] at (-1,0) {$x$};
    \node[above] at (1,0) {$x'$};
    \draw (-1,0)--(1,0);
\end{tikzpicture} 
=  2\pi G(x,x')   = -\frac{1}{2}\,\Big(\frac{1+\eta}{1-\eta}\log\eta+2\Big), \qquad 
\ee
where
\be\la{56}
\eta= \left|\frac{z-z'}{1-z\zb'}\right|^{2} = \frac{u}{u+2}\ ,\qquad\qquad 
 u \equiv  \frac{(\sfz-\sfz')^{2}+ (\sft-\sft')^{2} }{2\,\bz\,\bz'}.
\ee
The bulk-to-boundary propagator is then  ($g(x,x') \equiv g(\sft,\sfz; \sft',\sfz') =  g(w,w')$) 
\be
\la{57}
\gb (\sft;  x') = \gb (\sft;  \sft',\sfz')\equiv \lim_{\sfz \to 0 } \sfz^{-2} g(\sft,\sfz; \sft',\sfz')=\frac43 \Big[ \frac{\sfz'}{\sfz'^2 +(\sft-\sft')^2} \Big]^2\ . 
\ee
For the fermion with mass $m_\psi=1$   ( and thus $\Delta_\psi= {3\ov 2}$)  
one finds (see, e.g.,  \ci{Kawano:1999au}    and   Ap\-pen\-dix~\ref{app:propagators})
\begin{align}
\no
 \cS(x,x')  & 
 =\begin{tikzpicture}[line width=1 pt, scale=0.8, baseline=-0.1cm,decoration={
    markings,
    mark=at position 0.53 with {\arrow{>}}}]
    \node[above] at (-1,0) {$x$};
    \node[above] at (1,0) {$x'$};
    \draw[line width=0.25mm, postaction={decorate}] (-1,0)--(1,0);
\end{tikzpicture}=  2\pi S(x,x') 
\\&= -\frac{1}{2}\frac{1}{(u+2)^{2}}\frac{1}{\sqrt{\bz\,\bz'}}
\,\Big[
( x^{a}\, \Gamma_{a} \Gamma_{\bz}+\Gamma_{\bz}\,\Gamma_{a}\,x'^{a})\,{\sf F}_{1}(u )
-\Gamma^{a}(x-x')_{a}\,{\sf F}_{2}(u )
\Big],
\la{5.9}
\end{align}
where $\Gamma^a=\Gamma_a = (\Gamma_\sft, \Gamma_\sfz)$ are the flat-space 2d Dirac matrices (see Appendix \ref{app:conventions})  and 
\be
\la{5.10}
{\sf F}_{1}= -\frac{u+2}{2}\,\Big[2+(u+2)\,\log\frac{u}{u+2}\Big],
\qquad \qquad 
{\sf F}_{2} = \frac{(u+2)^{2}}{2\,u}\,\Big[2+u\,\log\frac{u}{u+2}\Big] \ .
\ee
Here  the propagator  $S(x,x') $  corresponds to the canonically normalized   kinetic term for a Majorana  fermion:   $\frac{1}{2}\int d^{2}x\,\sqrt{g}\, \bar\Psi  (\slashed{D}-m)\bar\Psi$. It satisfies  
\be
\la{5.5}
(\slashed{D}_{x}-1)\,\cS(x,x' ) = \cS(x,x' )(-\overleftarrow{\slashed{D}}_{x'}-1) 
= \frac{2\pi}{\sqrt g}\,\delta^{(2)}(x-x').
\ee
The component form of \rf{5.9}  is 
 \begin{align}\notag
  \cS(x,x') =\EV{\Psi (x) \bar \Psi (x')}_0 &=\PBK{ i\EV{ \psi(x) \bar\psi(x')}_0  & \EV{\psi(x)  \psi(x')}_0   
  \\
   \EV{\bar\psi(x) \bar\psi(x')}_0  & - i \EV{\bar \psi(x)  \psi(x')}_0   }
   \equiv \PBK{ i g_{\psi\bar \psi}(x,x')   & g_{\psi  \psi}(x,x')   
  \\
  g_{\bar\psi\bar\psi}(x,x')   & - i g_{\bar\psi\psi}(x,x')    }
  \\&=
  \frac{1}{2} \frac{1}{(u+2)^2} \frac{1}{\sqrt{\sfz\sfz'}} 
  \PBK{ -i (\bar w- w') \sfF_1  &    (\bar w-\bar w')\sfF_2
   \\
    (  w-  w') \sfF_2    &  i ( w-\bar w') \sfF_1
  }\ .\la{555}
 \end{align}
 These component propagators are also independently computed in Appendix~\ref{app:componentPropgators}. 
Taking one leg to the boundary   and adding a $\sfz^{-3/2}$  factor we get  the   bulk-to-boundary propagators
($x'=(\sft', \sfz'), \ w'= \sft' + i \sfz'$)
\begin{align}
\sb(\sft; x') &=\lim_{\sfz \to 0 } \sfz^{-3/2} \cS(x,x')
\equiv \PBK{ i \ssb _{\psi\bar \psi}(\sft;x')   & \ssb _{\psi  \psi}(\sft;x')   
  \\
  \ssb _{\bar\psi\bar\psi}(\sft;x')   & - i \ssb _{\bar\psi\psi}(\sft;x')    }
  = 
   \frac{1}{2 \sfz'^{3/2}}  
  \PBK{  \frac{-i}{( \sft-w')(\sft- \bar w')^2} &   \frac{1}{( \sft-w')^2(\sft- \bar w') } 
   \\
  \frac{ 1}{( \sft-w')(\sft- \bar w')^2}   &   \frac{i}{( \sft-w')^2(\sft- \bar w') } 
  }\ . \notag  \\
  \end{align}
  Especially, they satisfy 
  \be
 \ssb _{\psi  \psi}(\sft;x') =-\ssb _{ \bar \psi \psi}(\sft;x') , \qquad \qquad 
\ssb _{\psi\bar \psi}(\sft;x') =-\ssb _{ \bar\psi\bar \psi}(\sft;x') \ . \la{5.12}
\ee
This is   consistent with the supersymmetric boundary condition  $\sfPsi =-\bar\sfPsi$  in \rf{4.356},\rf{4.18} (cf. \rf{4.11})
implying that in boundary correlators 
  we can freely replace  $\bar \psi$ with $- \psi $, i.e. it is enough to consider boundary correlators of $\psi$ only. 
  
The bulk-to-boundary  fermion propagators may be expressed   in terms of the  bosonic  propagator in \rf{57} as follows 
\begin{align}
\la{Sb1}
\sb (x;\sft') &=\lim_{\sfz' \to 0 } \sfz'^{-3/2} \cS(x,x')=3\,  U(x;\sft')\,  \cP_- \,\gb (\sft';x)\ , 
\\
\sb (\sft;x') &=\lim_{\sfz \to 0 } \sfz^{-3/2} \cS(x,x')=-3\, \cP_+\,  U(x';\sft)\, \gb (\sft ; x')
\la{Sb2}\ , \\
U(x,\sft')& = {1 \ov \sqrt{\sfz}}  \Big[\sfz \Gamma^\sfz+(\sft-\sft')\Gamma^\sft  \Big]\ , \qquad \qquad 
\cP_\pm = \ha (1 \pm \Gamma^\sfz) \ . \la{514}
\end{align}
One  can then prove  the following   useful relation: 
 \be
 \la{SbSbuseful}
\lim_{\sfz_1 , \sfz_2\to 0} \sfz_1^{-3/2}\sfz_2^{-3/2} \cS(x_1 , x)\,\cS(x, x_2) \equiv \sb (\bt_{1}; x)\,\sb (x; \bt_{2}) = -9\,\bt_{12}\,\Gamma^{\bt}\,\cP_{-}\,\gb (\bt_{1}; x)\,\gb (\bt_{2};x)\ .
\ee 
In \rf{5.3}, \rf{5.9}   and below   we will denote  the scalar   propagator by  a line  and the fermion propagator   by a line with an arrow. The  dashed line
will stand  for the  boundary of \adst  (for  convenience represented as a circle, i.e. a boundary in  Poincare disc representation). 
For example, the relation   \rf{SbSbuseful} implies the following  graphical   equality of
the fermion and scalar  Witten  diagrams (with  one point   in the bulk) 
\be
\begin{tikzpicture}[line width=1 pt, scale=0.4, rotate=0,baseline=-0.1cm,decoration={
    markings,
    mark=at position \midpoint with {\arrow{>}}}]
\coordinate (A1) at (90:1);  \coordinate (A2) at (210:2);  \coordinate (A3) at (-30:2);
\draw[thin, densely dashed] (0,0) circle (2);
\def\midpoint{0.53} \draw[postaction={decorate}] (A2)--(0,0);
\def\midpoint{0.63}\draw[postaction={decorate}] (0,0)--(A3);
\draw (0,0)--(A1);
\draw[fill=black] (A2) circle (0.1); \draw[fill=black] (A3) circle (0.1); 
\node[right] at (A3) {$\bt_{2}$};
\node[left] at (A2) {$\bt_{1}$};
\end{tikzpicture} 
= - \frac92   \bt_{12} 
\begin{tikzpicture}[line width=1 pt, scale=0.4, rotate=0,baseline=-0.1cm,decoration={
    markings,
    mark=at position \midpoint with {\arrow{>}}}]
\coordinate (A1) at (90:1);  \coordinate (A2) at (210:2);  \coordinate (A3) at (-30:2);
\draw[thin, densely dashed] (0,0) circle (2);
\draw (0,0)--(A1);\draw (0,0)--(A2);\draw (0,0)--(A3);
\draw[fill=black] (A2) circle (0.1); \draw[fill=black] (A3) circle (0.1); 
\node[right] at (A3) {$\bt_{2}$};
\node[left] at (A2) {$\bt_{1}$};
\end{tikzpicture} 
\la{PsiPhiDiagram}\ . 
\ee
Here  
we assumed  the projection ($\Gamma^{\bt}\,\cP_{-}\to \frac12$)  to the upper  $\psi$ component  of $\Psi$  assuming 
that the boundary field $\sfPsi$ (cf. \rf{4.11}) is inserted at $\sft_1,\sft_2$. 



There is another useful  representation for  the free fermion propagator in terms of the bosonic one 
following from  
 the   Ward identity for  \adst  supersymmetry ~\cite{Bardeen:1984hm,Dusedau:1985uf}. 
Consider the  WZ model \rf{3.11} (with $a=-1$)  with the  superpotential  $\hat W(\phi)=\frac12\phi^2$
which  describes the \adst  supersymmetric  system of 
the   free scalar field $\phi$ with mass $m^2=2$ and the free fermion $\Psi$   with  $m=1$. 
 The  supersymmetric Ward identity   similar to the one in \rf{400} implies (here we use $w,w'$ to denote the coordinates)
\be
0=\delta  \big\langle   \Psi(w ) \, \phi(w' )   \big\rangle 
=\big\langle {\Psi(w)\,  \bar \Psi(w')  }\big\rangle   Z(w') \Lambda+ \big\langle{  (\slashed\p_w+1) \phi(w)\,  \phi(w') }\big\rangle  Z(w) \Lambda\ , \la{5.18}
\ee
where  we used the supersymmetry transformation  \rf{311}  with  the parameter $\cal E$  in \rf{4.6} and 
$F= \hW'= \phi$. 
Since this equation holds for an arbitrary constant Majorana  spinor $\Lambda$, it should be also true  as  a  matrix equation.
 Multiplying  by $Z^{-1}(w')$ and using the explicit form of $Z$ in \rf{4.6}  we arrive at 
\be\la{5.112}
\cS(w,w') =\EV{\Psi(w) \bar \Psi(w')  } = - \Big[(\slashed{\partial}_w+1) \EV{\phi(w)\, \phi(w')} \Big] R(w,w') \ , 
\ee\be  
\la{5.11}
R(w,w') \equiv  Z(w)\, Z^{-1}(w')=
\frac{1}{2\,\sqrt{\bz\bz'}}
\begin{pmatrix}
i\,(\bt-\bt')+\bz+\bz' & -(\bt-\bt')+i\,(\bz-\bz') \\
-(\bt-\bt')-i\,(\bz-\bz') & -i\,(\bt-\bt')+\bz+\bz'
\end{pmatrix}.
\ee
Thus we get  the following  representation for  the fermion bulk-to-bulk propagator in terms of the bosonic  one in \rf{5.3} 
\be
\la{5.13}
\cS(w,w') = -\Big[(\slashed{\partial}_w+1)\, g(w,w')\Big] R(w,w') =R(w,w')\,\Big[(\slashed{\partial}_{w'}-1)\, g(w,w')\Big]. 
\ee

Let  us  now  comment on the issue of UV regularization. 
In the bosonic 2d Liouville theory  the exponential potential  is special being 
 the eigen-function of the anomalous dimension operator (the second derivative term in 
\rf{3.13})  and also the interaction term in the classical   equation. As a result (using, e.g.,  the background field method) 
  the UV  divergences can be absorbed
into a field redefinition, i.e. there is no need for the   coupling $b$ renormalization. 
 This is equivalent to  replacing the exponential interaction term in (\ref{1.2})
by its normal ordered expression $e^{2b\phi}\to :e^{2b\phi}:$, i.e. to assuming that the  subtracted  value of the 
propagator at the coinciding points $g(x,x)$  is zero (or equal to a scheme-dependent  constant). 
The condition $g(x,x)=0$ is natural  in the present \adst context   where 
it can be achieved  by  using the covariant regularization $u\to u+\eps$ or $\eta\to \eta+\eps$ in \rf{5.3},\rf{56} 
and assuming  minimal subtraction of the resulting logarithmic divergence.\foot{A  detailed  comparison with other 
 approaches to regularization of the Liouville theory in \adst can be found in 
\cite{Beccaria:2019stp}.}

The discussion of UV   divergences  in the supersymmetric Liouville  theory in flat space 
  \cite{Goldschmidt:1981rx,DHoker:1983xdu}  implies that the picture should  be  essentially the same. 
  Assuming a covariant regularization and then minimal subtraction of the log divergence 
     $g(x,x')$ and $\cS(x,x')$ appear as  components of a superpropagator that is vanishing (or constant)
at $x=x'$.\foot{This is equivalent to normal ordering of the exponential interaction in the linear in $\hW$ terms in (\ref{3.11}), before
integrating out the auxiliary field $F$.}
We shall thus assume that  all the divergences  come only  from the  tadpole diagrams 
 and thus  can be eliminated by assuming that 
\be
\la{5.15}
g(x,x) = \cS(x,x) = 0  \ .
\ee
Indeed, in all instances that are considered in the following sections, we shall confirm by explicit calculations that the vanishing
tadpole condition (\ref{5.15}) is enough to make finite all radiative corrections.


\section{Two-point functions}
\la{sec6}

We shall start   with
the computation of  the  one-loop corrections  to the  two-point   boundary correlators  of 
 the elementary scalar $\zeta$ and fermion $\psi$  with the aim of
 checking the order $b^2 $ terms in the duality prediction in \rf{4.40},\rf{441}. 
 Note that the tree-level terms in \rf{441} are correctly normalized as follows from \rf{57}  and \rf{5.9},\rf{555} or \rf{Sb1}:
 \iffa Our aim it to check the one-loop predictions following from  (\ref{4.60}), i.e.
\be
\la{6.1}
\langle\sfPhi(\bt_{1})\sfPhi(\bt_{2})\rangle = \frac{4}{3}\,\left(1-\frac{1}{4}\,b^{2}+\cdots\right)\,\frac{1}{\bt_{12}^{4}},\qquad
\langle\sfPsi(\bt_{1})\sfPsi(\bt_{2})\rangle = 2\,\left(1-\frac{1}{4}\,b^{2}+\cdots\right)\,\frac{1}{\bt_{12}^{3}}.
\ee
Notice that the tree level values are in agreement with the supersymmetry relation (\ref{5.12}). Indeed 
the following  asymptotic bulk propagators solve that relation for $\bz, \bz'\to 0$\
\fi
\begin{align}\la{61} 
&\langle \sfPhi(\bt_{1})\sfPhi(\bt_{2}) \rangle_0\equiv   (\bz\bz')^{-2} \EV{\zeta(x)  \zeta(x')}_0 \Big|_{\bz, \bz'\to 0}=  (\bz\bz')^{-2} g(x,x')\Big|_{\bz, \bz'\to 0}           \to \ \   \frac{4}{3}\frac{1}{(\bt-\bt')^{4}}\ , \\
\la{6.1} 
&\langle \sfPsi(\bt_{1})\sfPsi(\bt_{2}) \rangle_0\equiv   (\bz\bz')^{-3/2} \EV{\psi(x)  \psi(x')}_0 \Big|_{\bz, \bz'\to 0} = (\bz\bz')^{-3/2} \cS_{12}(x,x')\Big|_{\bz, \bz'\to 0} \to  \ \    2\,\frac{1}{(\bt-\bt')^{3}}\ . 
\end{align}

\subsection{Scalar  propagator   correction}

The one-loop correction to the  scalar 2-point function 
in \rf{4.40}  
   is given by the boundary limit of the 
sum of the contributions of the  three bulk diagrams.
For the calculation of the  Witten diagrams  below  it will be often useful to use
the \adst disk parametrization, cf.  (\ref{A17}) and (\ref{54}), and the Fourier representation of  
$g(x,x')$ derived in \cite{Menotti:2003km}.
\begin{align}
\la{6.3}
\langle \zeta(z_{1})\zeta(z_2)\rangle &= g(z_{1},z_{2}) + D_{\rm 1}(z_{1},z_{2})+\mc O(b^{4})\ , \notag \\
D_{\rm 1}(z_{1},z_{2}) &= D_{\rm sc}(z_{1},z_{2}) + 
D_{\rm ins}(z_{1},z_{2}) + D_{\rm f}(z_{1},z_{2}).
\end{align}

\paragraph{Scalar loop:}
The first  contribution  in (\ref{6.3}) is that of  the boson bubble diagram 
(see  (A.15) in \cite{Beccaria:2019stp})
\begin{align}
D_{\rm sc}(z_{1},z_{2}) &= 
\begin{tikzpicture}[line width=1 pt, scale=0.8, baseline=-0.1cm,decoration={
    markings,
    mark=at position 0.53 with {\arrow{>}}}]
\draw (-1.8,0)--(-0.8,0); 
\draw (0.8,0) arc(0:180:0.8);
\draw (-0.8,0) arc(-180:0:0.8);
\draw (1.8,0)--(0.8,0); 
\end{tikzpicture}
= \frac{1}{2}\,(-8b)^{2}\,\int \mathsf{d}^{2}z'\mathsf{d}^{2}z''\,g(z_{1},z')\,g(z',z'')^{2}\,g(z'',z_{2}) = 32\,b^{2}\,\DD(z_{1},z_{2}),
\la{scalarbubble}\no \\
\DD(z_{1},z_{2}) &= \frac{\eta ^2 \log ^2\eta }{64 (\eta -1)^2}-\frac{4 \eta +9}{96 \
(\eta -1)}-\frac{\eta  \log \eta }{48 (\eta -1)}+\log (1-\eta ) \
\Big[\frac{1}{48} -\frac{(\eta +1) \log \eta }{48 (\eta \
-1)}\Big]\notag \\
&\ \ \ -\frac{(\eta +1) \text{Li}_2(\eta )}{96 (\eta -1)}-\frac{39+\pi^{2}}{576}\frac{1+\eta}{1-\eta}\ .
\end{align}

\paragraph{Insertion diagram:}
The second contribution  is  due to the  insertion of the 
$3\,b^{2}\,\frac{\zeta^{2}}{2!}$ vertex  in (\ref{xxx}) (see  (A.16) in \cite{Beccaria:2019stp})
\begin{align}
D_{\rm ins}(z_{1},z_{2}) &= 
\begin{tikzpicture}[line width=1 pt, scale=0.8, baseline=-0.1cm,decoration={
    markings,
    mark=at position 0.53 with {\arrow{>}}}]
\draw (-1.8,0)--(1.8,0); 
\draw[fill=white] (0,0) circle (0.15); 
\draw (0.0-0.15/1.4142,0-0.15/1.4142)--(0.0+0.15/1.4142,0+0.15/1.4142);
\draw (0.0-0.15/1.4142,0+0.15/1.4142)--(0.0+0.15/1.4142,0-0.15/1.4142);
\end{tikzpicture}
= -3b^{2}\,\int \mathsf{d}^{2}z'\,g(z_{1},z')\,g(z',z_{2}) = -3\,b^{2}\,\widehat{B}(z_{1},z_{2}),
\la{scalarInsertion}\no\\
\widehat B(z_{1},z_{2}) &= -\frac{\eta  \log \eta }{6 (\eta -1)}+\log (1-\eta ) \
\Big[\frac{1}{6}-\frac{(\eta +1) \log \eta }{12 (\eta -1)}\Big]-\frac{(\eta 
+1) \text{Li}_2(1-\eta )}{6 (\eta -1)}-\frac{1}{6}\ .
\end{align}

\paragraph{Fermionic loop:}
The third  contribution is that of the 
 fermion loop diagram
 \footnote{For a detailed discussion of Feynman  rules for  Majorana fermion see for instance  
  \cite{Denner:1992me,Denner:1992vza}. }
\begin{align}
D_{\rm f}(z_{1},z_{2}) &= 
\begin{tikzpicture}[line width=1 pt, scale=0.8, baseline=-0.1cm,decoration={
    markings,
    mark=at position 0.53 with {\arrow{>}}}]
\draw (-1.8,0)--(-0.8,0); 
\draw[postaction={decorate}] (0.8,0) arc(0:180:0.8);
\draw[postaction={decorate}] (-0.8,0) arc(-180:0:0.8);
\draw (1.8,0)--(0.8,0); 
\end{tikzpicture}
= (2b)^{2}\frac{1}{2}(-1)\,\int \mathsf{d}^{2}z\mathsf{d}^{2}z'\,\text{Tr}[\cS(z,z')\cS(z',z)]\,g(z_{1},z)\,g(z',z_{2}).
\la{Dfscalar}
\end{align}
To compute this integral it is  convenient to first  go back to the  half-plane coordinates and use (\ref{5.13}).
Then
\begin{align}
\la{6.10}
 \text{Tr}[\cS(w,w')\cS(w',w)] &= 
-\text{Tr}\Big[(\desl+1)g(w,w')R(w,w')
R(w',w)(\desl-1)g(w,w')\Big],\notag \\
&= -\text{Tr}\Big[(\desl+1)g(w,w') (\desl-1)g(w,w')\Big]
= -2\,\Big(\p^{\mu}g \p_{\mu}g -g^{2}\Big),
\end{align}
where all derivatives are over  $w$. 
 As a result,  
\begin{align}
\frac{1}{4b^{2}}\,D_{\rm f}(z_{1},z_{2}) &= \int \disk{z}\,\disk{z'}\,
g_{2}(z_{1},z)
\Big(\p^{\mu}g(z,z')\p_{\mu}g(z,z')-[g (z,z')]^{2}\Big)\,g(z',z_{2})\notag \\
&= -\DD(z_{1},z_{2})+\int\disk{z}\, g(z_{1}, z)B_{\p\p}(z, z_{2}),
\la{Dfres}
\end{align}
where the function $B_{\p\p}$ reads  (see Appendix  D of \cite{Beccaria:2019mev})
\be
B_{\p\p}(z_{1}, z_{2}) = -\frac{3}{8}+\frac{1+\eta}{4\,(\eta-1)}
\log\eta-\frac{1+\eta^{2}}{16(\eta-1)^{2}}\log^{2}\eta.
\la{Bpp}
\ee
This function was found in \cite{Beccaria:2019mev} by 
integrating by parts and assuming  $\nabla^2 g \to  2g$, i.e.
removing the $\delta$-function consistently with the renormalization condition in  (\ref{5.15}).
Using the method in Appendix~A of \cite{Beccaria:2019stp}, a straightforward
calculation then gives the (finite) expression for the disk integral
\begin{align}
\int\disk{z}\,g(z_{1},z)\,B_{\p\p}(z,z_{2}) &= -\frac{(\eta -2) \eta  \log ^2\eta}
{64 (\eta-1)^2}+\frac{\eta  \log \eta }{48 (\eta -1)}+\log (1-\eta ) \Big[\frac{(\eta +1) \log\eta }
{96 (\eta -1)}-\frac{1}{48}\Big]\notag \\ \la{610}
&\ \ \ \  -\frac{(\eta +1) \text{Li}_2(1-\eta )}{96 (\eta -1)}-\frac{11}{192}.
\end{align}

\paragraph{Total one-loop correction:}

Combining the  contributions \eqref{scalarbubble}, \eqref{scalarInsertion} and \eqref{Dfscalar}, we get for \rf{6.3}
\begin{align}
& D_{ \rm 1} = b^{2}\,\Big\{
\frac{\pi ^2 (\eta +1)}{8 (\eta -1)}+\frac{\eta  (3 \eta +1) \log ^2\eta}{8 (\eta -1)^2}
-\frac{3 (\eta +1) \log (1-\eta ) \log \eta }{4 (\eta -1)}-\frac{3 (\eta +1) \text{Li}_2(\eta )}{4 (\eta -1)}+1
\Big\}. \la{611}
\end{align}
Taking the boundary limit, we find  the  1-loop correction to the  boundary  two-point correlator 
\be\la{6.12}
\langle  \sfPhi(\bt_{1})  \sfPhi(\bt_{2})\rangle_{\rm 1} =  \lim_{\sfz_1 , \sfz_2\to 0} \sfz_1^{- 2}\sfz_2^{-2} \, D_{\rm 1}(x_{1},x_{2}) = -\frac13 b^2 \frac{1}{(\sft_{1}-\sft_2)^4}\ . 
\ee
It   agrees with   the  duality prediction  in (\ref{4.40}),\rf{441}.


\subsection{Fermion propagator  correction}

Let  us now  check the one-loop   $b^2$ term in the expression  for the fermionic correlator in \rf{4.40}. 
This requires computing  the boundary limit of the one-loop corrected  $\langle \psi(x_{1})\psi(x_{2})\rangle$ 
(or  $\cS_{12} $) component of the fermion propagator (cf. \rf{555},\rf{6.1})
given by the   sum of the  contributions  of the   scalar exchange  with two $b \zeta \bar \Psi \Psi$  vertices from \rf{xxx} 
 and the insertion of the $b^2 \bar \Psi \Psi$  vertex 
in \rf{xxx}\foot{Here we will use the half-plane parametrization $w= \sft + i \sfz$.}
\begin{align}
\la{6.17}
\langle \Psi(w_{1})\bar\Psi (w_{2})\rangle &= \cS(w_{1},w_{2}) + \cS_{\rm 1}(w_{1},w_{2})+\mc O(b^{4}), \notag \\
\cS_{\rm 1}(w_{1},w_{2}) &= \cS_{\rm sc}(w_{1},w_{2}) + 
\cS_{\rm ins}(w_{1},w_{2}) \ .
\end{align}

\paragraph{Scalar  exchange:}
The first  contribution  in (\ref{6.17})  is given by  
 \begin{align}
\cS_{\rm sc}(w_{1},w_{2}) &= 
\begin{tikzpicture}[line width=1 pt, scale=0.4, rotate=0,baseline=-0.1cm,decoration={
    markings,
    mark=at position 0.53 with {\arrow{>}}}]
\coordinate (A1) at (-3,0);  \coordinate (A2) at (-1.6,0);  \coordinate (A3) at (1.6,0);
\coordinate (A4) at (3,0);
\draw[postaction={decorate}] (A1)--(A2);
\draw[postaction={decorate}] (A2)--(A3);
\draw[postaction={decorate}] (A3)--(A4);
\draw (A3) arc(0:180:1.6);
\end{tikzpicture}
= (2b)^{2}\,I(w_{1},w_{2})  \ , 
\\
I(w_{1},w_{2}) &=\int\disk{w}\disk{w'}\, \cS(w_{1}, w)\,\cS(w, w')\,\cS(w',w_{2})\,g(w,w') \ . \la{6.15}
\end{align}
Using the representation \eqref{5.13}, the integrand in $I$ can be rewritten as
  \begin{align}
I(w_{1},w_{2})
&= - \int\sfd^2 w\,\sfd^2 w'\, \cS(w_{1}, w)\,[(\slashed{\partial}+1) g(w,w') ]\,R(w,w') \cS(w',w_{2})\,g(w,w')
\notag\\
&= - \int\sfd^2 w\,\sfd^2 w'\, \cS(w_{1}, w)\,\big( 1+\tfrac14 \Gamma^\sfz \big)\,R(w,w') \cS(w',w_{2})\,[ g(w,w') ]^2
\notag\\
&\quad 
- \tfrac12 \int\sfd^2 w\,\sfd^2 w'\, \cS(w_{1}, w)\, \slashed D [ g(w,w') ]^2\,R(w,w') \cS(w',w_{2}) \ . 
\end{align}
The second term   can be simplified using  integration by parts: 
then the derivative  can either act  on $\cS(w_1,w)$ and the measure $\sfd^2 w$ on  the left hand side, or  on  $R(w,w') $ on the  right on 
side.\foot{Explicitly, using  \eqref{DiracIBP}, we get: \ \ \ \   $
 I(w_1,w_2)= - \int\sfd^2 w\,\sfd^2 w'\, \cS(w_{1}, w)\,\big( 1+\tfrac14 \Gamma^\sfz \big)\,R(w,w') \cS(w',w_{2})\,[ g(w,w') ]^2
$
$$
- \tfrac12 \int\sfd^2 w\,\sfd^2 w'\, \cS(w_{1}, w)\,(-\overleftarrow{  \slashed D}) [ g(w,w') ]^2\,R(w,w') \cS(w',w_{2}) 
- \tfrac12 \int\sfd^2 w\,\sfd^2 w'\, \cS(w_{1}, w)\,  \big( -\slashed \p R(w,w') \big) \cS(w',w_{2})  [ g(w,w') ]^2\ . \no 
$$
}
In view of   \eqref{5.5}  we thus find  
\begin{align}
I(w_1,w_2)&= - \int\sfd^2 w\,\sfd^2 w'\, \cS(w_{1}, w)\,\big( 1+\tfrac14 \Gamma^\sfz \big)\,R(w,w') \cS(w',w_{2})\,[ g(w,w') ]^2
\notag\\
&\quad 
- \tfrac12 \int\sfd^2 w\,\sfd^2 w'\,  \Big( \cS(w_{1}, w)+\tfrac{2\pi}{\sqrt{g(w)}}\delta^{(2)}(w-w_1)  \Big)  [ g(w,w') ]^2\,R(w,w') \cS(w',w_{2}) 
\notag\\
&\quad 
- \tfrac12 \int\sfd^2 w\,\sfd^2 w'\, \cS(w_{1}, w)\,  (1-\tfrac{1}{2}\Gamma^\sfz)R(w,w') \cS(w',w_{2})  [ g(w,w') ]^2 \no 
\\&
=- \tfrac12 \int\sfd^2 w\,\sfd^2 w'\,  \Big( 4\cS(w_{1}, w)+\tfrac{2\pi}{\sqrt{g(w)}}\delta^{(2)}(w-w_1)  \Big)  [ g(w,w') ]^2\,R(w,w') \cS(w',w_{2}) \no\\
& \equiv  I^{(0)} +I^{(1)} \ ,    \  \la{618}
\end{align}
where we also  used that $\slashed \p R(w,w') = (\slashed D+\frac12\Gamma^\sfz) R(w,w') =(-1+\frac12\Gamma^\sfz) R(w,w') $.
Thus the integral $I(w_1,w_2)$  reduces to the sum of two pieces    where  the first 
 in \rf{618}   is  the one   without $\delta$-function 
\be
I^{(0)}(w_1,w_2)
=-2 \int\sfd^2 w\,\sfd^2 w'\,  \cS(w_{1}, w)  \,R(w,w') \cS(w',w_{2})  \,  [ g(w,w') ]^2 \ . \la{619}
\ee
Taking  both  end points   to the boundary, we  get 
\begin{align}
\mathcal I^{(0)}(\sft_1,\sft_2)&=\lim_{\sfz_1,\sfz_2\to 0 }   {\sfz_1^{ -3/2}\sfz_2^{- 3/2} } { I^{(0)}(w_1,w_2)}
=-2 \int\sfd^2 w\,\sfd^2 w'\,  \sb (\sft_{1}; w)  \,R(w,w') \sb (w';\sft_{2})  \,  [ g(w,w') ]^2
\no  \\&
=18\sft_{12}\Gamma^\sft \cP_- \; \int\sfd^2 w\,\sfd^2 w'\,   \gb (\sft_1; w) \gb (\sft_2; w') [ g(w,w') ]^2\ , 
\la{I0}
\end{align}
where  $    \sb $  is   the bulk-to-boundary propagator in \rf{Sb1},\rf{Sb2}.  It is easy to see 
 that the integral here corresponds to the boundary limit of the scalar bubble diagram contribution in  \eqref{scalarbubble}. 
The term   in \rf{618} that contains the   $\delta$-function  is 
\begin{align}
I^{(1)} (w_1,w_2)&=- \tfrac12 \int\sfd^2 w\,\sfd^2 w'\,  \tfrac{2\pi}{\sqrt{g(w)}}\delta^{(2)}(w-w_1)  [ g(w,w') ]^2\,R(w,w') \cS(w',w_{2}) 
\notag  \\ &=- \tfrac14 \int \sfd^2 w'\,     [ g(w_1,w') ]^2\,R(w_1,w') \cS(w',w_{2}) \ .
\end{align}
Using again (\ref{5.13})  and  computing the integral as in    (A.14) in  \cite{Beccaria:2019stp}   we get 
\begin{align}
I^{(1)}  (w_1,w_2) &=- \tfrac14 \int \sfd^2 w'\,     [ g(w_1,w') ]^2\,R(w_1,w') R (w',w_{2}) (\slashed \p _{w_2} -1 ) g(w',w_2)
\no \\&=- \tfrac14 R(w_1 ,w_{2}) (\slashed \p _{w_2} -1 )  \int \sfd^2 w'\,     [ g(w_1,w') ]^2\, g(w',w_2)
\no \\
&= - \tfrac{1}{32} R(w_1 ,w_{2}) (\slashed \p _2 -1 )  \Big[1
-\frac{\eta\log^{2}\eta}{\,(1-\eta)^{2}}\Big],\qquad\qquad  \eta\equiv \eta (w_1,w_2) \ . \la{625}
\end{align}
Sending the  legs to the boundary  gives finally 
\be
\mathcal I^{(1)}(\sft_1,\sft_2)=\lim_{\sfz_1\to 0, \sfz_2\to 0} \sfz_1^{-3/2}\sfz_2^{-3/2}  I^{(1)}(w_1,w_2) 
=
- \frac18\,\frac{1}{\bt_{12}^{3}}\,\Gamma^{\bt}\cP_{-} \ .
\la{I1}
\ee


\paragraph{Insertion diagram:}
The second  contribution in (\ref{6.17})
is  given by 
\begin{align}
\la{6.36}
S_{\rm ins}(z_{1},z_{2}) &= 
\begin{tikzpicture}[line width=1 pt, scale=0.4, rotate=0,baseline=-0.1cm,decoration={
    markings,
    mark=at position 0.53 with {\arrow{>}}}]
\coordinate (A1) at (-3,0);  \coordinate (A2) at (-1.6,0);  \coordinate (A3) at (1.6,0);
\coordinate (A4) at (3,0);
\draw[postaction={decorate}] (A1)--(0,0);
\draw[postaction={decorate}] (0,0)--(A4);
\draw[fill=white] (0,0) circle (0.25); 
\draw (0.0-0.25/1.4142,0-0.25/1.4142)--(0.0+0.25/1.4142,0+0.25/1.4142);
\draw (0.0-0.25/1.4142,0+0.25/1.4142)--(0.0+0.25/1.4142,0-0.25/1.4142);
\end{tikzpicture}
= b^{2}\,J(w_{1},w_{2}), \qquad\quad 
J(w_{1},w_{2})  = \int\disk{w}\,\cS(w_{1},w)\, \cS(w ,w_{2})  \ . 
\end{align}
Taking the boundary limit and using the relation \eqref{SbSbuseful}, we get
\begin{align}
\no 
\mathcal  J(\sft_1,\sft_2)=\lim_{\sfz_1 , \sfz_2\to 0} \sfz_1^{-3/2}\sfz_2^{-3/2}  J(w_1,w_2) &= \int\disk{w}\,\sb (\sft_{1};w)\, \sb (w;\sft_{2})  
\\&=-9\,  \sft_{12}\,\Gamma^{\bt}\,\cP_{-} \ \int\disk{w} \,\gb (\bt_{1}; x)\,\gb (\bt_{2}; x)\ .
\la{JJ}
\end{align}
\iffa
From (\ref{6.36}) we thus obtain 
\be
\la{6.38}
\langle \sfPsi(\bt_{1})\sfPsi(\bt_{2})\rangle_{\rm ins} = -\frac{9}{2}\,b^{2}\,\bt_{12}\,
\int\disk{z}\disk{z'}\,\gb (\bt_{1}; z)\, \gb (z; \bt_{2}) \ .
\ee
\fi 

\paragraph{Total one-loop correction:}
Summing up  the  contributions in (\ref{I0}),(\ref{I1})  and (\ref{JJ}) with appropriate coupling factors, 
we get the full one-loop correction to the boundary two-fermion correlator. 
Projecting  to the   relevant correlator $\langle     \psi (x_{1})  \psi (x_{2})\rangle$   (or 12 matrix  element of the spinor 
matrix propagator) using $(\Gamma^{\bt}\cP_{-})_{12}=\frac{1}{2}$  we find 
\begin{align}
& \langle     \sfPsi(\bt_{1})  \sfPsi(\bt_{2})\rangle_{\rm 1} 
 = \Big[ (2b)^2\big( \mathcal I^{(0)}+\mathcal I^{(1)} \big) +b^2 \mathcal J \Big]_{12} = 36\,b^{2}\, \bt_{12}\, Y (\bt_1, \bt_2)  -\,\frac{b^2}{4 \bt_{12}^{3}}\ , 
\la{627} \\
& Y = \int\disk{w}\disk{w'}\,\gb (\bt_{1}; w)\,g(w,w')\,\gb (w'; \bt_{2})
-\tfrac{1}{8}\,\int\disk{w}\disk{w'}\,\gb (\bt_{1}; w)\, \gb (w; \bt_{2})
\ . \la{628}
\end{align}
The expression for the combination of  the bosonic  integrals in  \rf{628}   can be  found from   
 the boundary limit of eq. (A17)
in  \cite{Beccaria:2019stp}:
\begin{align}
 Y =   \frac{1}{32}\,\lim_{\bz_{1}, \bz_{2}\to 0}\frac{1}{\bz_{1}^{2}\bz_{2}^{2}}\Big[
\frac{3}{2}+\frac{\eta_{12}^{2}\,\log^{2}\eta_{12}}{2\,(1-\eta_{12})^{2}}-\frac{1+\eta_{12}}{1-\eta_{12}}\,\text{Li}_{2}
(1-\eta_{12})\Big] = -\frac{1}{144}\,\frac{1}{\bt_{12}^{4}},\quad
\end{align}
Thus finally we get 
\be
\langle  \sfPsi(\bt_{1})  \sfPsi(\bt_{2})\rangle_{\rm 1} = b^{2}\,\Big(-\frac{36}{144}-\frac{1}{4}\Big)\,\frac{1}{\bt_{12}^{3}} = 
-\frac{1}{2} b^{2}\,\frac{1}{\bt_{12}^{3}} \ , 
\ee
which is again in  agreement with  the one-loop  prediction for  the fermionic correlator in (\ref{4.40}),\rf{441}.

\section{Three-point functions}
\la{sec7}
In this section we shall  first discuss    the  tree level values of the non-vanishing three-point function 
of  three  scalars and   three-point function  of one scalar and two fermions, then   compute 
 the one-loop correction to the  three-scalar correlator,   
 verifying the duality predictions in \rf{444},\rf{462},\rf{4.62}.

\subsection{Tree level}

At the  tree level, the three-point function $\langle \sfPhi\sfPhi\sfPhi\rangle$ is the same 
 as in the bosonic Liouville theory, 
i.e.  we have~\cite{Beccaria:2019stp}  
\be C_{\sfPhi\sfPhi\sfPhi} = -\tfrac{16}{9}\,b+\mc O(b^{3}) \ ,\la{70}
 \ee  which agrees with \rf{462}. 
This follows from the fact  that the $b \zeta^3$  term in the action \rf{3.22},\rf{xxx} is the same as in the Liouville  theory. 
The  tree-level  correlator  $\langle \sfPsi\sfPsi\sfPhi\rangle$ is given  by  the following Witten diagram (see \rf{57},\rf{5.12}--\rf{Sb2})
\be\la{71}
\begin{tikzpicture}[line width=1 pt, scale=0.4, rotate=0,baseline=-0.1cm,decoration={
    markings,
    mark=at position \midpoint with {\arrow{>}}}]
\coordinate (A1) at (90:2);  \coordinate (A2) at (210:2);  \coordinate (A3) at (-30:2);
\draw[thin, densely dashed] (0,0) circle (2);
\def\midpoint{0.53} \draw[postaction={decorate}] (A2)--(0,0);
\def\midpoint{0.63}\draw[postaction={decorate}] (0,0)--(A3);
\draw (0,0)--(A1);
\draw[fill=black] (A1) circle (0.1); \draw[fill=black] (A2) circle (0.1); \draw[fill=black] (A3) circle (0.1); 
\node[above] at (A1) {$\bt_{3}$};
\node[right] at (A3) {$\bt_{2}$};
\node[left] at (A2) {$\bt_{1}$};
\end{tikzpicture}  
=2b \int \sfd^2 w \,  \sb (\sft_1; w) \sb  (w; \sft_2) \gb  (\sft_3; w)\ .
\ee
Using   \eqref{SbSbuseful}  and  picking up  the  $\psi \psi$  (i.e. 12) component  gives (see \eqref{PsiPhiDiagram})  
\be
\Big[ \int \sfd^2 w \,  \sb (\sft_1; w) \sb  (w; \sft_2) \gb  (\sft_3; w)  \Big]_{12}
=-\tfrac 92\sft_{12} \int \sfd^2 w \,  \gb (\sft_1; w) \gb  (w; \sft_2) \gb  (\sft_3; w) \ . \la{72}
\ee 
The integral  here is  the same as in  the  scalar three-point function ~\cite{Beccaria:2019stp}  so that we get
\be
\EV{\sfPsi(\sft_1)\sfPsi(\sft_2)\sfPhi(\sft_3)}=2b \cdot(-\tfrac 92\sft_{12}) \cdot \tfrac{ 2 }{9}\frac{1}{\sft_{12}^2\sft_{13}^2\sft_{23}^2}=-2b \frac{1}{\sft_{12} \sft_{13}^2\sft_{23}^2}\ , \la{73}
\ee
in agreement with \rf{4.62}.

\subsection{One-loop correction to   three-scalar   correlator}   

According to the  duality relation  (\ref{462}) the   boundary correlator of     three $\zeta$ fields   should  have the 
coefficient 
\begin{align}
\la{7.4}
\Cppp = C_0\,b+ C_1\,b^{3}+\cdots, \qquad\qquad  
C_0 =-\tfrac{16}{9}, \ \ \ \ \ \ \ C_1 =\tfrac{16}{9}.
\end{align}
The one-loop  $b^3$    contribution  comes from several  diagrams. 
In addition to the bosonic (scalar)  loop diagrams   which are the same  as in the Liouville theory computed in  \cite{Beccaria:2019stp}
and give 
\be
C_1^{(\rm sc)}=\tfrac{ 64}{27} \ , \la{74}
\ee
there are also  diagrams  with fermion loop 
and also the  insertion   diagram   with the higher-order  $ b^{3}\zeta^{3}$  vertex  in \rf{xxx}.\foot{It
comes from the $Q$-dependent terms in the action  \rf{3.22}  with subleading term in  $Q$   
in super Liouvile theory \rf{3.16} having extra 1/2  factor  compared to the bosonic Liouville case.}
There  will be four different contributions  to this additional   part of $C_1$   that we will denote with tilde, i.e. 
\be \la{75}
 \tC_1 \equiv  C_1 - C_1^{(\rm sc)}  =
    \tC_1^{(a)}  + \tC_1^{(b)}  +    \tC_1^{(c)} +  \tC_1^{(d)}     = - \tfrac{16}{27}= \tfrac13 C_0 \ .
    \ee 
Here we  gave  the expected value of $ \tC_1$  following from \rf{7.4},\rf{74} 
that we  confirm   below  by computing the four  contributions to it  in turn.

\paragraph{a) Self-energy diagrams:}

The first  type  of one-loop diagrams contributing to $\tC_1$  is 
the  fermionic  loop correction to the  scalar  $\zeta$ field  propagator (or self-energy correction)  
\be \la{777}
\begin{tikzpicture}[line width=1 pt, scale=0.5, rotate=0,baseline=-0.1cm,decoration={
    markings,
    mark=at position \midpoint with {\arrow{>}}}]
\coordinate (A1) at (90:2);  \coordinate (A2) at (210:2);  \coordinate (A3) at (-30:2);
\coordinate (A4) at (0,-0.4);
\draw[thin,dashed] (0,0) circle (2);
\draw (A1)--(A4); \draw (A2)--(A4); \draw (A3)--(A4);
 \draw[fill=white] (0,0.8) circle (0.7);
\def\midpoint{0.555}\draw[postaction={decorate}] (0,0.1) arc(-90:90:0.7);
\def\midpoint{0.555}\draw[postaction={decorate}] (0,1.5) arc(90:270:0.7);
\draw[fill=black] (A1) circle (0.1); \draw[fill=black] (A2) circle (0.1); \draw[fill=black] (A3) circle (0.1); 
\node[above] at (A1) {$\bt_{1}$};
\node[below] at (A2) {$\bt_{2}$};
\node[below] at (A3) {$\bt_{3}$};
\end{tikzpicture} \ 
\ee

It  amounts to the following correction to the bulk to boundary  scalar propagator in \rf{57} 
\be
\la{7.7}
\gb (\bt;  z) \to \,\left(1-\tfrac{1}{12}\,b^{2}+\cdots\right) \gb (\bt;  z) \ . 
\ee
It can be found from  the  value of    $C_{\sfPhi\sfPhi}$  in \rf{441}    (confirmed in \rf{6.12} in  the previous section) 
after subtraction of the bosonic Liouville  theory value:
 from  (\ref{110})  and (\ref{441})
 we get  $(\frac{4}{3}-\frac{1}{3}b^{2}+\cdots)/(\frac{4}{3}-\frac{2}{9}b^{2}+\cdots)
= 1-\frac{1}{12}\,b^{2}+\cdots$. 
Then  from  (\ref{7.7}) it follows that 
\be
\la{7.8}   \tC_1^{(a)}  
= 3 \cdot (  -\tfrac{1}{12})\cdot  C_0 = -\tfrac{1}{4} C_0 \ ,  
\ee
where the factor 3  accounts for the fact that  each of the  three scalar propagators 
may get a fermion loop  correction.

\paragraph{b) Quasi-self-energy diagrams:}
There are also fermion loop  diagrams   with one    $\zeta \bar \Psi \Psi  $  and one  $\zeta^2 \bar \Psi \Psi$ vertices in 
\rf{xxx}, i.e.  
\be
\begin{tikzpicture}[line width=1 pt, scale=0.5, rotate=0,baseline=-0.1cm,decoration={
    markings,
    mark=at position \midpoint with {\arrow{>}}}]
\coordinate (A1) at (90:2);  \coordinate (A2) at (230:2);  \coordinate (A3) at (-50:2);
\draw[thin, densely dashed] (0,0) circle (2);
\draw (A1)--(0,0.7); 
\draw (A2)--(0,-0.7)--(A3);
\def\midpoint{0.53}\draw[postaction={decorate}] (0,-0.7) arc(-90:90:0.7);
\def\midpoint{0.6}\draw[postaction={decorate}] (0,0.7) arc(90:270:0.7);
\draw[fill=black] (A1) circle (0.1); \draw[fill=black] (A2) circle (0.1); \draw[fill=black] (A3) circle (0.1); 
\node[above] at (A1) {$\bt_{1}$};
\node[below] at (A2) {$\bt_{2}$};
\node[below] at (A3) {$\bt_{3}$};
\node[below] at (0,0.7) {$z$};
\node[below] at (0,-0.7) {$z'$};
\end{tikzpicture}   
\la{3ptbubbleA}
 \ee
 The  contribution of this  diagram  contains the integral 
 \be\la{78}
 H(z_{1},z') = \int \mathsf{d}^{2}z\,g(z_{1},z)\,\text{Tr}[\cS(z,z')\,  \cS(z', z)]\ .
 \ee
Using  (\ref{6.10}) we get 
\begin{align}\no
H(z_{1},z') &=      -2 \int \mathsf{d}^{2}z\,g(z_{1},z)\, \p^{\mu}g(z,z') \p_{\mu}g(z,z') +2\int \mathsf{d}^{2}z\,g(z_{1},z)\,  [g(z,z')]^{2} 
\\&= -2\,B_{\p\p}(z_{1},z')+2\,B(z_{1},z'),
\end{align}
where $B(z_{1},z_{2}) = \frac{1}{8}-\frac{\eta\log^{2}\eta}{8(1-\eta)^{2}}$ 
is found  from  (A.14) of \cite{Beccaria:2019stp} and $B_{\p\p}$ (from  Appendix~D of \cite{Beccaria:2019mev})  was given  already in  \eqref{Bpp}. 
Taking the boundary limit one finds 
\be
\lim_{\sfz_1\to 0} \sfz_1^{-2}H(z_{1},z_{2}) = \tfrac{1}{2} \gb (\bt_{1}; z_{2})\ .
\ee
As a consequence, the upper  leg in diagram \eqref{3ptbubbleA}  may be   effectively replaced by a free  propagator 
so that  \eqref{3ptbubbleA} reduces to the tree-level diagram  and  we get 
\be
\la{7.12}
\tC_1^{(b)} = 
( (- 8 b)^{-1}\, b C_0 )\cdot\tfrac{1}{2}\cdot(-1)\cdot(2b)\cdot b^2 \cdot\tfrac{1}{2!}\cdot 3 b^{-3} = \tfrac{3}{8}\, C_0 \ , 
\ee
where $(-1)$ is the fermion loop sign, $2b,\ b^{2}$ are the vertex  couplings, $\frac{1}{2!}$ is a  symmetry factor (fermions are Majorana),  
and we added extra  factor of 3  because there are  three  diagrams like \rf{777}.

\paragraph{c)        Insertion diagram:} 

The cubic $\zeta^{3}$ vertex in the \adst  Liouville action \rf{1.2} expanded near the constant vacuum 
is multiplied by the factor $Qb=1+b^{2}$ \cite{Beccaria:2019stp}.
In the super Liouville theory, the analogous factor is $1+\frac{7}{8}b^{2}+\cdots$, see (\ref{xxx}). 
Hence, the extra contribution
 is simply 
\be
\la{7.13}
\tC_1^{(c)}   = \big(\tfrac{7}{8}-1\big) C_0 = -\tfrac{1}{8}\, C_0 \ . 
\ee
Remarkably,    the total contribution from  the  above three types of diagrams 
 given by the sum  of   (\ref{7.8}), (\ref{7.12}) and (\ref{7.13})  vanishes 
 \be
\la{750}
\tC_1^{(a)}   + \tC_1^{(b)}   + \tC_1^{(c)}   =\big(
-\tfrac{1}{4}+\tfrac{3}{8}-\tfrac{1}{8}
\big)\,  C_0 =0\ .
\ee

\paragraph{d) Fermion loop   correction to 3-vertex:  }
Thus the non-zero value of $\tC_1$ in  \rf{75}  should   be solely  due to the contribution 
of the remaining  most   non-trivial  diagram:
\be
\la{7.15}
\,b^{3} \tC_1^{(d)} = \ \ 
\begin{tikzpicture}[line width=1 pt, scale=0.5, rotate=0,baseline=-0.1cm,decoration={
    markings,
    mark=at position 0.53 with {\arrow{>}}}]
\coordinate (A1) at (90:2);  \coordinate (A2) at (210:2);  \coordinate (A3) at (-30:2);
\coordinate (B1) at (90:1);  \coordinate (B2) at (210:1);  \coordinate (B3) at (-30:1);
\draw[thin, densely dashed] (0,0) circle (2);

\draw[postaction={decorate}] (B1)--(B2);
\draw[postaction={decorate}] (B2)--(B3);
\draw[postaction={decorate}] (B3)--(B1);

\draw (A1)--(B1); \draw (A2)--(B2); \draw (A3)--(B3);
\draw[fill=black] (A1) circle (0.1); \draw[fill=black] (A2) circle (0.1); \draw[fill=black] (A3) circle (0.1); 
\end{tikzpicture}\ \ 
= \frac{(-1)(2b)^{3}}{\mc K_{3}(\bm\theta)}\int \mathsf{d}^{2}z_{1}\mathsf{d}^{2}z_{2}\mathsf{d}^{2}z_{3}\,
\text{Tr}[\cS(z_{1},z_{2})\cS(z_{2},z_{3})\cS(z_{3},z_{1})]\ 
\prod_{i=1}^{3}\gb\big(\sft(\theta_{i}); z_{i}\big)\ .
\ee
Here  we divided by the  kinematical prefactor $\mc K_{3} ={\bt_{12}^{-2}\,\bt_{13}^{-2}\,\bt_{23}^{-2}} $ in the three-point function in \rf{444};
expressed in terms of  the angles on the 
circular (compactified) boundary of \adst  it is  given by 
\be
\mc K_{3}(\bm\theta) \equiv  \mc K(\theta_{1}, \theta_{2}, \theta_{3}) = \prod_{i<j}^{3}
\big| \bt(\theta_{i})-\bt(\theta_{j})\big|^{-2},\qquad\qquad  \sft(\theta)= -\cot \tfrac{\theta}{2} \ . 
\ee
Computing 
the trace in (\ref{7.15})  we get  (using \rf{5.9},\rf{5.10} and $u_{ij}\equiv u(z_i,z_j)$)
\begin{align}
& \text{Tr}[\cS(z_{1},z_{2})\cS(z_{2},z_{3})\cS(z_{3},z_{1})] = 
\frac{1}{2\,(u_{12}+2)^{2}}\frac{1}{2\,(u_{13}+2)^{2}}\frac{1}{2\,(u_{23}+2)^{2}}\notag \\
&\times \Big\{
\F_2(u_{12}) \Big[(u_{12}+u_{13}-u_{23}) \F_1(u_{23}) 
\F_2(u_{13})+(u_{12}-u_{13}+u_{23}) \F_1(u_{13}) 
\F_2(u_{23})\Big]\notag \\
& \;\; -\F_1(u_{12}) \Big[(u_{12}+u_{13}+u_{23}+4) \F_1(u_{13}) 
\F_1(u_{23})+(u_{12}-u_{13}-u_{23}) \F_2(u_{13}) \F_2(u_{23})\Big]
\Big\}\ .
\end{align}
To   compute the  remaining integral in \rf{7.15}  we use numerical integration that 
can be performed by the {\tt Suave} routine of the {\tt Cuba} library \cite{Hahn:2004fe}
as discussed also in \cite{Beccaria:2019stp,Beccaria:2019mev}. Our best estimate is 
\be\la{765}  \tC^{(d)}_1 = -0.590\pm 0.002  \ . \ee
At the same time,  
according \rf{75},\rf{750}  the  contribution of this diagram should be 
\be
\la{11.13}
\tC^{(d)}_1 
= -\tfrac{16}{27} = -0.59259\dots\ .
\ee
The Monte Carlo numerical integration  \rf{765} is compatible with the analytical prediction 
within the statistical uncertainty and with  a deviation below 0.5 \% that we consider satisfactory.


\section{Four-point functions}
\la{sec8}

Let us now consider the tree-level  four-point correlators \rf{4143}--\rf{4.45}. 
The tree-level expression of  the 4-scalar  correlator     is given in  \rf{4143} and it  is the same as in the 
bosonic Liouville theory \ci{Ouyang:2019xdd,Beccaria:2019stp}. 
As discussed   in section \ref{sec4.3},  the  other two 
 non-vanishing four-point  correlators  $   \EV{\sfPsi(\bt_1)\sfPhi(\bt_2)\sfPsi(\bt_3) \sfPhi(\bt_4 )}$ and 
$\EV{\sfPsi(\bt_1)\sfPsi(\bt_2)\sfPsi(\bt_3) \sfPsi(\bt_4 )}$
have their kinematical structure fixed by the conformal  symmetry and their coefficients 
related as in \rf{466}  to 
$C_{\sfPhi\sfPhi\sfPhi\sfPhi}$ in  $
\langle \sfPhi(\bt_{1})\sfPhi(\bt_{2})\sfPhi(\bt_{3}) \sfPhi(\bt_{4})\rangle$   by the  boundary  supersymmetry (\ref{4.22}). 
Thus  their direct computation 
(with  some
details  discussed   in  
Appendix  \ref{4pt2B2F})
is  simply a check of consistency  of our perturbation theory rules for the fermions.  

For example, let consider  the tree-level  computation of   the (connected part of)  
4-fermion correlator $ \EV{\sfPsi(\bt_1)\sfPsi(\bt_2)\sfPsi(\bt_3) \sfPsi(\bt_4 )}$.
As follows from the structure of the action \rf{xxx}  here the only cointribution comes 
from  the  three scalar exchange diagrams  with  the s-channel one being 
 \be
 \begin{tikzpicture}[line width=1 pt, scale=0.5, rotate=0,baseline=-0.1cm,decoration={
    markings,
    mark=at position \midpoint with {\arrow{>}}}]
\coordinate (A1) at (135:2);  \coordinate (A2) at (45:2);  
\coordinate (A3) at (-45:2);   \coordinate (A4) at (-135:2);
\coordinate (B1) at (-1,0); \coordinate (B2) at (1,0);
\draw[thin,dashed] (0,0) circle (2);
\draw[dashed,red] (A1)--(B1)--(A4); \draw[dashed,red] (A3)--(B2)--(A2); \draw (B1)--(B2);
\draw[fill=black] (A1) circle (0.1); \draw[fill=black] (A2) circle (0.1); 
\draw[fill=black] (A3) circle (0.1); \draw[fill=black] (A4) circle (0.1); 
\def\midpoint{0.53} \draw[postaction={decorate}] (B1)--(A4);
\def\midpoint{0.63}\draw[postaction={decorate}] (A1)--(B1);
\def\midpoint{0.53} \draw[postaction={decorate}] (B2)--(A2);
\def\midpoint{0.63}\draw[postaction={decorate}] (A3)--(B2);
\node[left] at (A1) {$\bt_{1}$};
\node[right] at (A2) {$\bt_{4}$};
\node[right] at (A3) {$\bt_{3}$};
\node[left] at (A4) {$\bt_{2}$};
\end{tikzpicture} 
\la{82}
 \ee  
Using the relations \eqref{SbSbuseful},\eqref{PsiPhiDiagram}, the  contribution of this diagram 
 reduces to that of the four-scalar diagram  with the scalar exchange:   
\begin{align}
  \EV{\sfPsi(\bt_1)\sfPsi(\bt_2)\sfPsi(\bt_3) \sfPsi(\bt_4 )}_s&= 
 (2b)^2  \cdot  \frac {-9 \sft_{12}}{2}  \cdot \frac {-9 \sft_{34}}{2}\!\int\!{\sf d}^2w \,{\sf d}^2w'\,    \gb  (\sft_1,\! w ) 
            \gb  (\sft_2, \! w ) g  (w,\! w' )  \gb  (\sft_3, \!w ')            \gb  (\sft_4, \! w' ) 
\notag \\&= 81 \sft_{12} \sft_{34}b^2 \cdot  2\pi\cdot (\frac43)^4 \cdot \frac{1}{(4\pi)^2} \cdot\sft_{34}^{-2} D_{ 2211}(\sft_1,\sft_2,\sft_3,\sft_4) 
 \\& =
  3b^2\frac{ \sft_{12} \sft_{34} }{ \sft_{13}^4\sft_{24}^4 } \bar D_{2211}(\chi)\ , 
 \end{align}
 where  $\chi=\frac{\sft_{12}\sft_{34}}{\sft_{13}\sft_{24}}$  and the   $D$ and  $\bar D$ functions  are defined as, e.g., in \cite{Beccaria:2019dws,Ouyang:2019xdd}. 
 The  contributions of the t and u channels   can be obtained by permuting the legs. Taking into account the fermi statistics, they are 
$
  \EV{\sfPsi(\bt_1)\sfPsi(\bt_2)\sfPsi(\bt_3) \sfPsi(\bt_4 )}_t    =- \EV{\sfPsi(\bt_1)\sfPsi(\bt_3)\sfPsi(\bt_2) \sfPsi(\bt_4 )}_s    ,\ \ 
  \EV{\sfPsi(\bt_1)\sfPsi(\bt_2)\sfPsi(\bt_3) \sfPsi(\bt_4 )}_u    =- \EV{\sfPsi(\bt_1)\sfPsi(\bt_4)\sfPsi(\bt_3) \sfPsi(\bt_2 )}_s 
$.
As a result,  summing the three  contributions we get   the connected part of the  tree-level correlator 
\be\la{83}
 \EV{\sfPsi(\sft_1)\sfPsi(\sft_2)\sfPsi(\sft_3)\Psi(\sft_4)}_{0,\text{conn}} 
 =3b^2\frac{1}{\sft_{13}^3\sft_{24}^3} \, \frac{1}{\chi (1- \chi )} \ . 
\ee
This is  in agreement with the tree-level of the  duality prediction in \rf{4.45},\rf{4.47}.

We discuss the tree-level  computation of the mixed $\EV{\sfPsi\sfPhi\sfPsi\sfPhi}$ correlator in Appendix~\ref{4pt2B2F}.
The result is again in agreement with the leading-order value in \rf{443},\rf{447}, i.e. 
\be\la{84}
 \EV{\sfPsi(\sft_1)\sfPhi(\sft_2)\sfPsi(\sft_3)\Phi(\sft_4)}_{0,\text{conn}} 
 ={4\ov 3}b^2  \frac{1}{\sft_{13}^3\sft_{24}^4} \, \frac{3- 4 \chi + 4 \chi^2 }{2\chi^2 (1- \chi )^2} \ . 
\ee

\def \bu {$\bullet$}\def \ads {\adst}

\iffa

\section{Concluding remarks}
\la{sec9}

about  other models 

{WZW model in \adst }

 special role of Liouville scalar with non-trivial 

 conformal  transformation?   conformal   sigma-models? 

\bu WZW in flat  space:   chiral decomposition,  trivial  S-matrix 

   $g= e^{ t_a \vp_a}$,     $\vp_a $  massless 

$L= \del_+ \vp_a \del_- \vp_a + {1\ov 3\sqrt k} f_{abc} \vp_a \del_+ \vp_b \del_- \vp_c + ...$

\bu   \ads: \ \ \ $m^2=0$ $\to \Delta=1$:  $\vp_a $   dual to chiral currents $J_a$ 

\bu correlators  of $\vp_a$ in \ads   $\to $   correlators of  chiral  currents  

$ \langle J_a (z_a)  J_b (z_2)  \rangle = {\delta_{ab} \ov z^2_{12}}, \ \ \ \ \ \ \ \ 
 \langle J_a (z_a)  J_b (z_2)  J_c (z_3)  \rangle = {1 \ov \sqrt k}  {f_{abc}  \ov z_{12}   z_{23}   z_{31}    }
$

 match  $\langle J ... J \rangle $  restricted to the boundary  of   half-plane 

\bu relation  is more  direct:  $J_a = \tr ( t_a g^{-1} \del g) = \del \vp_a + ... $

\ads: $\vp_a({\rm z}, t) \big|_{\rm z\to 0} \to  {\rm z}\,  \Phi_a (t) + ...$, \ \ \   $J_a \to  \Phi_a $

(cf. Liouville field $m^2=2$ dual to stress tensor $T$) 

 \iffa 
 \bu checked, e.g.,  on example of $SL(2)$ WZW   model 
 \be \no  L=k \big( \pa\varphi \bar\pa\varphi   + e^{ 2\phi}    \pa \eta \bar\pa \bar\eta\big) \ee
\fi


\section*{Acknowledgments}

We would like to thank  S. Giombi, S. Kuzenko and R. Metsaev   for 
useful  discussions of related questions. 
MB was supported by  the INFN grant GSS (Gauge Theories, Strings and Supergravity).
HJ was supported by Swiss National Science Foundation.
AAT was supported by the STFC grant ST/P000762/1.

 \appendix
 \section{Majorana fermion action  in Euclidean  \adst}
 \la{app:conventions}
 
Let us list our conventions for the fermionic fields. 
We start with  the flat Euclidean space  with metric $ds^2 = dx^a dx^a \equiv d\sft^2 + d \sfz^2=  dw\,d\bar w$.
We shall use the  following representation for the  Clifford  algebra of Dirac matrices  
$\{ \Gamma^a, \Gamma^b \} =2\delta^{ab}$
\be
\Gamma^\sft=\Gamma^1=\PBK{0&1 \\ 1 & 0},\qquad  \qquad \Gamma^\sfz=\Gamma^2=\PBK{0&-i \\  i & 0},\la{a1}
\ee
with  the chirality matrix $\Gamma_*$  and  the charge  conjugation matrix $\mathcal C$
(satisfying $ \mathcal C\,\Gamma^a\, \mathcal C^{-1}=-(\Gamma^a) ^T$)
defined as 
\be\la{a2}
\Gamma_*=i \Gamma^1 \Gamma^2=  \PBK{1&0 \\0  & -1}\ , \qquad 
\mathcal C=\Gamma^2,\qquad \qquad \mathcal C=\mathcal C^\dagger =\mathcal C^{-1}, \qquad 
\ee
The Dirac spinor  and its charge-conjugate then are 
\be
\Psi=\PBK{\psi_1  \\ \psi_2},\qquad \qquad  \Psi^c = \mc C\,\Psi^*=  \PBK{ -i  \bar \psi_2    \\  i \bar \psi_1  },
\ee
where $\psi_1,\psi_2$ are complex  and  $\bar \psi $  denotes complex   conjugation of $\psi$.


Since $(\Psi^c)^c=-\Psi$, a Majorana fermion may be defined  as satisfying 
  $\Psi^c=\Gamma_* \Psi$; then 
 $\psi_2=-i \bar \psi_1  $ (cf., e.g.,  Appendix A in  \cite{Benini:2012ui}).  Hence, the explicit form of  a Majorana  fermion $\Psi$  and  its Majorana conjugate $\bar \Psi$ are
     \be
  \Psi=    \PBK{   \psi   \\ - i\bar\psi  }, \qquad \qquad 
  \bar \Psi \equiv i \Psi^T \mathcal C=  \PBK{ i \bar \psi     &  \psi  }\ .
  \ee
For any two arbitrary Majorana spinors $\Psi_{1}$ and $\Psi_{2}$,
the product  $\bar \Psi_{1} \Psi_{2}$ is invariant under $SO(2)$ rotation and the following
identities hold
 \be
 \bar \Psi_{1} \Psi_{2}=\bar \Psi_{2} \Psi_{1}, \qquad\qquad     \bar \Psi_{1} \Gamma^a\Psi_{2}= - \bar \Psi_{2}  \Gamma^a\Psi_{1} .
 \ee
 The    action for a  Majorana $\Psi$  in flat 2d space  is 
  \be\la{a6}
   \mc S=\int d^2 x   \bar \Psi (\Gamma^a \p_a -m ) \Psi
   =2\int d^2 x    \Big( \psi \bar\p \psi + \bar\psi   \p \bar\psi  - im \bar\psi  \psi\Big),
  \ee
  where 
    \be
w=\sft+i\sfz,\qquad  \p \equiv \p_w =\tfrac12 (\p_\sft -i \p_\sfz), \qquad  \bar\p \equiv \p_{\bar w} =\tfrac12 (\p_\sft +i \p_\sfz)\ .
  \ee
Let us now consider the  \adst  space with metric  
      \be
  ds^2 =\frac{d\sft^2 +d\sfz^2 }{\sfz^2}\ . 
  \ee
The zweibein is $e_\mu^a=\frac{1}{\sfz}\, \delta_\mu^a$, $e^\mu_a=\sfz\, \delta^\mu_a$\,,\footnote{We will use Greek letters to denote a curved index and Latin letters to denote  a flat index.}
so that the  spin-connection is $\omega_\mu^{ab}=\frac{1}{\sfz} (\delta_\sfz^a \delta_\mu^b- \delta_\sfz^b \delta_\mu^a   )$. The Dirac operator is then 
  \be
  D_\mu= \p_\mu +\tfrac12 \omega_\mu ^{bc}\Omega_{bc}, \qquad\qquad 
  \slashed D\equiv \gamma^\mu D_\mu=e^\mu_a \Gamma^a \Big( \p_\mu +\tfrac12 \omega_\mu ^{bc}\Omega_{bc}\Big)   =\sfz \Gamma^a \p_a -\tfrac12 \Gamma^\sfz,
  \ee
where $\Omega_{ab}=\frac14[\Gamma^a, \Gamma^b]$ and the ``curved'' $\gamma_\mu$ matrices are  defined as
\be
\gamma^\mu=e^\mu_a \Gamma^a, \qquad 
\frac{1}{\sfz} \gamma^\sft= \sfz\, \gamma_\sft= \Gamma^\sft,
\qquad
\frac{1}{\sfz}  \gamma^\sfz= \sfz\, \gamma_\sfz=  \Gamma^\sfz .
\ee
The action for a Majorana spinor  in  \adst is then (cf. \rf{a6})
   \beqn
 \mc S&=&\int d^2 x\sqrt{g}\,   \bar \Psi \,(  \slashed D-m )\, \Psi
=2\int \frac{d\sft d\sfz}{\sfz^2}  \,  \big( \sfz\,  \psi \bar\p \psi +\sfz\,  \bar\psi   \p \bar\psi  - im \bar\psi  \psi\big).
\label{A.16}
 \eeqn
 The spin connection term  here dropped  out because of the anti-commuting property of the fermions.  
  
%

\section{Propagators in AdS}\label{app:propagators}
\la{secB}

\subsection{Scalar and fermion propagators in AdS$_{d+1}$}
Consider a Euclidean AdS$_{d+1}$ with the metric
\be
ds^2=\frac{1}{z_0^2} \Big(dz_0^2 +d\vec{z}^{2} \Big) = 
\frac{1}{z_0^2} \Big(dz_0^2 +\sum_{i=1}^d dz_i^2 \Big)\ .
\ee
Here we  use   subscript  ``0''   to label the radial direction  $z_0$ 
and $\vec z=(z_1, z_2, ...,z_d)$ (with $z_0\equiv \sfz$ and $z_1\equiv \sft$ in \adst case). The 
tangent-space 
$\Gamma$-matrices satisfy $\{ \Gamma^a, \Gamma^b\}=2\delta^{ab}$ with $a=0,1, ...,d$. 

The bulk-to-bulk propagator $G_\Delta$  for  a scalar field with mass $m$  defined via 
\be
\big( -\nabla^2   +m^2\big) G_\Delta( z,z')=\frac{1}{\sqrt {g(z)}}\delta^{(d+1)}(z-z')\ , 
\ee
is given by (see,  e.g.,  \cite{DHoker:1999mqo})
\be
G_{\Delta}(z,z') =G_{\Delta}(u(z,z')) =   C_\Delta \Big(\frac{2}{u} \Big)^\Delta {}_2F_1\Big(\Delta, \Delta-\frac{d}{ 2}+\frac{1}{2}; 2\Delta-d+1; -\frac{2}{u}\Big),  
\ee
\be
m^2=\Delta(\Delta-d), \qquad
u(z,z')=\frac{(z_0-z'_0)^2+(\vec z-\vec z')^2}{2z_0 z'_0},
\qquad 
 C_\Delta= \frac{1}{2^{2\Delta+1}\pi^{d/2}  }  \frac{\Gamma(\Delta)}{\Gamma(\Delta-\frac{d}{2} +1)}\ .
\ee
Its equivalent form is 
\be
G_{\Delta}(u)=(2\xi)^\Delta C_\Delta \,  {}_2F_1\Big( \frac{\Delta}{2}, \frac{\Delta+1}{2},\Delta-\frac{d}{2}+1, \xi^2 \Big), 
\qquad \xi \equiv \frac{1}{u+1}\ .
\ee

The bulk-to-boundary propagator is 
\be
\label{Kfunction}
K_\Delta(z, \vec x) = c_\Delta \tilde K_\Delta(z, \vec x) , \qquad
\tilde K_\Delta(z, \vec x) = \Bigg( \frac{z_0}{z_0^2+(\vec z-\vec x)^2}\Bigg)^\Delta, \qquad
c_\Delta=\frac{\Gamma(\Delta)}{ \pi^{d/2}\Gamma(\Delta-\frac{d}{2}) }.
\ee
Up to a normalization factor, it is   the boundary limit of the bulk-to-bulk propagator 
\be
K_\Delta(z, \vec x) =\mathcal N\,   \lim_{x_0\rightarrow 0}x_0^{-\Delta} G_\Delta(z,x) , \qquad
\mathcal N=\frac{c_\Delta}{4^\Delta C_\Delta}=2\Delta -d\ .
 \ee


For a Dirac field (with  mass parameter $m >0$) with the action 
$\int d^{d+1} z \sqrt{g} \,\bar \Psi (\slashed D-m )\Psi$ 
the propagator  satisfying\foot{Here $\slashed D f=z_0 \Gamma^a \p_a f -\frac{d}{2} \Gamma^0 f,\;  f \overleftarrow{\slashed  D} =z_0 \p_a f\Gamma^a   -\frac{d}{2} f\Gamma^0$.}
\be
(\slashed{D}_{z}-m)\,S(z,z') = S(z,z')(-\overleftarrow{\slashed{D}}_{z'}-m) 
= \frac{1}{\sqrt g}\,\delta^{(d+1)}(z-z') ,
\ee
was  presented in various equivalent forms in ~\cite{Camporesi:1992tm,Muck:1999mh,Kawano:1999au,Basu:2006ti}, e.g., 
\begin{align}
S(z,z') &=-\frac{1}{\sqrt{z_0 z'_0}} \Big[ (\slashed z \cP_- -  \cP_+ \slashed z'  )G_{\Delta_-}'(z,z')
+(\slashed z \cP_+ -  \cP_- \slashed z' )G_{\Delta_+}'(z,z')
\Big],\la{b8}
\\
 S(z,z') &= -\Big(\slashed D+\Gamma^0+m  \Big)  \Big[z_0^{-1/2}\Big(G_{\Delta_-}(z,z') \cP_- +G_{\Delta_+}(z,z') \cP_+   \Big)  z'^{ 1/2}_0   \Big],
  \la{b9}
\\
 S(z,z')&=  -\frac{1}{ \pi^{d/2}\,2^{m+ (d+3)/2}}\,\frac{\Gamma(m+\frac{d+1}{2})}{\Gamma(m+\frac{1}{2})}\,
  \frac{1}{(u+2)^{m+ (d+1)/2}} \frac{1}{\sqrt{z_0 z'_0}} 
\no
\\ 
&\ \ \qquad  \times  \Big[ 
 \Big(z_a \Gamma^a \Gamma^0+ \Gamma^0 \Gamma^a z' _a\Big) {\sf F}_1(u)
 -   (z-z')_a \Gamma^a  {\sf F}_2(u)
 \Big]\ .
 \la{b10}
 \end{align}
 Here      $\slashed z=\Gamma_a z^a$, $ \Delta_\pm=m+\tfrac{d}{2} \pm \tfrac12 $ ,  and
\begin{align} \la{b11}
  \cP_\pm   &=\tfrac12 (1\pm \Gamma^{  0}),
  &G'_\Delta(z,z')  &=\frac{dG_\Delta(u(z,z'))}{du} , 
\\
\label{solF2}
  {\sf F}_{1}  (u)  &=  {}_2F_1(m+ \tfrac{d+1}{2},m;2m+1;\tfrac{2}{u+2}),
& {\sf F}_{2} (u)  & =   {}_2F_1(m+\tfrac{d+1}{2},m+1;2m+1;\tfrac{2}{u+2})\ .
\end{align}
The bulk-to-boundary fermionic propagator   is given by  \cite{Kawano:1999au}
\begin{align}
\Sigma_\Delta(z, \vec x) &=U(z, \vec x) K_\Delta (z, \vec x) \cP_-\, ,
\la{Sigma1}
\\
\bar \Sigma_\Delta(z, \vec x)& =   \cP_+  K_\Delta (z, \vec x)  U(z, \vec x)  \ ,  \qquad   \qquad U(z, \vec x) =\frac{1}{\sqrt {z_0}} \Gamma^a (z_a-x_a)\Big|_{x_0=0}\ , 
\la{Sigma2}
\end{align}
where $\Delta=\Delta_+=m+\frac{d}{2}+\frac12$.
 Its relation to the bulk-to-bulk propagator is
\be 
\Sigma(z, \vec x) =\lim_{x_0 \rightarrow 0} x_0^{-\Delta_f} S(z,x), \qquad
\bar \Sigma(z, \vec x) =\lim_{x_0 \rightarrow 0} - x_0^{-\Delta_f} S(x,z )\ , 
\ee
where  $\Delta_f=\Delta_+-\tfrac12 =m+\tfrac{d}{2}$  is the dimension of a CFT$_d$  operator dual to the fermion field.

%
 
 \def \mathcalN { } 
 
\subsection{Component form of fermion propagator in AdS$_2$} \label{app:componentPropgators}

To check consistency of the above general expressions when applied to the Majorana fermion in \adst case,   here 
 we shall explicitly determine  the  fermion propagator in \adst  by starting with the 
 component action \rf{a6} for a Majorana fermion, i.e. 
     \begin{align}
 \mc S&=\mathcalN   \int d^2 x\sqrt{g}  \bar \Psi \big(  \slashed D-m \big) \Psi
=2\mathcalN \int \frac{d\sft d\sfz}{\sfz^2}   \Big( \sfz \psi \bar\p \psi +\sfz \bar\psi   \p \bar\psi  - im \bar\psi  \psi\Big)
\notag\\ &=2\mathcalN \int  d\sft d\sfz   
\Big(   \eta \bar\p \eta+   \bar\eta   \p  \bar \eta - \frac{i  m}{\sfz} \bar\eta \eta  \Big),
\qquad \qquad  \psi=\sfz^{1/2} \eta, \qquad \bar\psi=\sfz^{1/2} \bar\eta\ .
\la{b16} \end{align}
The equations of motion for the $\eta$ fields are 
 \be\label{EoMeta}
\bar\p \eta - \frac{   m}{w-\bar w} \bar\eta=0, \qquad\qquad 
  \p \bar\eta + \frac{   m}{w-\bar w} \eta=0,
 \ee 
  implying, in particular, 
 \be\label{eometa}
\Big[ \p\bar\p + \frac{1}{w-\bar w} \bar\p + \frac{ m^2}{(w-\bar w)^2} \Big]\eta=0\ .
 \ee
Let us make the following   Ansatz   for the propagator, i.e.
the free  two-point correlator  ($w=\sft + i \sfz$),   
   \begin{align}\label{eomF}
 & \EV{\eta(w,\bar w)  \eta(w',\bar w')}=  \frac{\bar w- \bar w'}{ {(w-\bar w)(w'-\bar w')}}   F(u)\ , \\
 & u=  \frac{ (\sfz-\sfz')^2+(\sft-\sft')^2}{2\sfz \sfz'} =-2\frac{(w-w')(\bar w-\bar w')}{(w-\bar w)(w'-\bar w')}\ .
 \end{align}
 Then using  \eqref {eometa} 
 we get  the following differential equation
 for $F(u)$
 \be
 \la{B.6}
\Big[ u(u+2) \frac{d^2}{du^2}+(4+3u) \frac{d}{du} +(1-m^2) \Big] F(u)=0\ .
 \ee
 The most general solution to (\ref{B.6}) is  ($U\equiv  { 2\ov 2 + u }$)
%
 \be
 \la{solF}
 F(u)=  {\sf c}_1 U^{m+1}   \, _2F_1(m+1,m+1;2 m+1;U) + {\sf c} _2   U^{1-  m} \, _2F_1(1-m,1-m;1-2 m;U)\ . 
 \ee
 To compute the mixed propagator $\EV{\eta\bar\eta}$   we may use 
 \eqref{EoMeta} in order to write
 \be
 \EV{\bar\eta(w,\bar w) \eta (w',\bar w')}=\frac{w-\bar w}{ m} \bar\p\, \big\langle   \eta(w,\bar w) \eta (w',\bar w')\big\rangle \ . 
 \ee
 From  \eqref{eomF} and \eqref{solF}, we then obtain  
 \begin{align}
 \EV{\bar\eta(w,\bar w)    \eta (w',\bar w')}= \frac{  w-\bar w'   }{(w-\bar w) ( w'-\bar  w')} 
 \Big[& - {\sf c}_1 U^{1+ m} \, _2F_1(m,m+1;2 m+1;U)
 \notag \\
& +{\sf c}_2  U^{1-m} \, _2F_1(1-m,-m;1-2 m;U)\Big]\ .
 \end{align}
The correlators  $\EV{\bar\eta \bar\eta}$ and $\EV{\eta\bar\eta}$ can 
be obtained by the complex conjugation. 
%
The coefficients ${\sf c}_1,{\sf c}_2$ in (\ref{solF}) may be determined by inspection of the 
short distance limit. 
Assuming   $m>0$,  we find  
 \be
 F(u)=   \frac { 2 {\sf c}_1     }{u}\frac{    \Gamma (2 m+1)}{\Gamma (m+1)^2}+\cdots\ , \quad
    \EV{\eta(w,\bar w)  \eta(w',\bar w')} 
 = -{\sf c}_1 \frac{    \Gamma (2 m+1)}{\Gamma (m+1)^2} \frac{1}{w-w'}+\cdots\ , \quad 
 w\to w' \ . 
 \ee
On the other hand, taking  $u\rightarrow \infty$ 
in  \rf{solF}   we  will get a  divergence in $F(u)\sim {\sf c}_2 u^{ m -1}$ 
 (which  also yields a  divergence  when setting one leg to the boundary); this implies we should  set ${\sf c}_2=0$.
 The short-distance behavior of two-point function in \adst\ should be  the same as in  flat space 
 (in particular, it should be independent of the mass). 
 Comparing to the flat space two-point function, we  find $ {\sf c}_1$  
 \be
 {\sf c}_1=  - \frac{1}{4\pi \mathcalN}\frac {\Gamma (m+1)^2}   {    \Gamma (2 m+1)}
 =- \frac{1}{ \mathcalN \sqrt \pi  2^{2+2m}}\frac {    \Gamma (  m+1)} {    \Gamma (  m+\frac12)}~.
 \ee
 As a result, we find from \rf{eomF},\rf{solF}
   \begin{align}
g_{\psi\psi}(w,w') &=  \EV{ \psi(w,\bar w) \psi (w',\bar w')}
 = C
  \frac{ \bar  w-   \bar w'   }{ \sqrt{-(w-\bar w) ( w'-\bar  w')}}
    \frac{ \sfF_2(u) }{(2+u)^{m+1}} \ ,
       \\
g_{\bar\psi\bar\psi}(w,w') &=  \EV{ \bar\psi(w,\bar w) \bar\psi (w',\bar w')}
 = C 
  \frac{   w-    w'   }{ \sqrt{-(w-\bar w) ( w'-\bar  w')}}
    \frac{ \sfF_2(u) }{(2+u)^{m+1}} \  ,\no 
        \\
g_{\psi\bar\psi}(w,w') &=  \EV{ \psi(w,\bar w) \bar\psi (w',\bar w')}
 =C 
  \frac{  - ( w-   \bar w' )  }{ \sqrt{-(w-\bar w) ( w'-\bar  w')}}
    \frac{ \sfF_1(u) }{(2+u)^{m+1}}\   ,\no 
        \\
g_{\bar\psi\psi}(w,w') &=  \EV{\bar \psi(w,\bar w) \psi (w',\bar w')}
 = C 
  \frac{ - ( \bar w-   w' )  }{ \sqrt{-(w-\bar w) ( w'-\bar  w')}}
    \frac{ \sfF_1(u) }{(2+u)^{m+1}}   \ ,\la{b30}
 \end{align}
where
\be
 C\equiv  \tfrac{1}{ \mathcalN \sqrt \pi  \,2^{2+ m}}\frac {    \Gamma (  m+1)} {    \Gamma (  m+\frac12)} ,
\ee
and
\be  
{\sf F}_{1}(u  ) = {}_{2}F_{1}(m+1, m, 2m+1, \tfrac{2}{u+2}),\qquad 
{\sf F}_{2}(u ) = {}_{2}F_{1}(m+1, m+1, 2m+1, \tfrac{2}{u+2}) \ .
\ee 
 It is easy to verify that the propagators in \rf{b30}
  agree with the $d=1$ limit of the general expression in 
  \eqref{b10}  (up to a factor of 2 due to the Majorana condition
  assumed in \rf{b16}). 
 
In the main text \eqref{555}, we consider the  action \rf{b16}  with $m=1$   and 
 an extra overall normalization  factor  $\frac{1}{4\pi}$, implying that  the  above  propagators  
 are  to be multiplied  by $4 \pi$.

Let us note that in  the massless case $m=0$, we need to choose ${\sf c}_1={\sf c}_2$ to ensure $\EV{\eta\bar\eta}=0$ as here  the two fields decouple in \rf{b16}.  The resulting non-zero  propagators are 
 the same  as on a flat half-plane (reflecting conformal invariance of the massless case)
\be 
 \EV{ \psi(w,\bar w) \psi (w',\bar w')}
 =\frac{1}{8\pi\mathcalN }  \frac { \sqrt{-(w-\bar w) ( w'-\bar  w')}} { w-     w'   }, \qquad
 \EV{ \bar\psi(w,\bar w)\bar \psi (w',\bar w')}
 =\frac{1}{8\pi\mathcalN }  \frac { \sqrt{-(w-\bar w) ( w'-\bar  w')}} { \bar w- \bar    w'   }\ .
\ee

\section{Tree-level calculation of  $\EV{\sfPhi^2\sfPsi^2}$ boundary correlation function }    \la{4pt2B2F}

Here we  describe the calculation of  the tree level four-point correlator 
of two scalars and two fermions  in the super Liouville theory
in \adst\!\!. 
This requires to evaluate  the contribution of 
 the fermion exchange  diagram. 
 We did not find a direct way to 
 exploit the supersymmetry relation (\ref{5.13}) to reduce its  calculation to
 that of  a scalar exchange diagram
 (although in principle this should be possible). Instead, 
 we will compute  it  following  the same strategy as used  in \cite{Kawano:1999au}.

\subsection{Fermion exchange diagram in AdS$_{d+1}$}

Let us first compute the  fermion 
exchange diagram in the general  case of AdS$_{d+1}$ by exploiting inversion transformations
as  in~\cite{Kawano:1999au} and express then  it in terms of 
differential operators acting on the scalar exchange  contribution. 
Our notation in this subsection follows  Appendix~\ref{app:propagators}. 
Stripping out couplings, the  contribution of this diagram to  the mixed scalar-fermion  
boundary correlator  is~\foot{Here $z$ and $w$ denote generic bulk points in  AdS$_{d+1}$.}
  \beqn
 && A(\vec x_1, \vec x_2, \vec x_3, \vec x_4 ) =
 \begin{tikzpicture}[line width=1 pt, scale=0.5, rotate=0,baseline=-0.1cm,decoration={
    markings,
    mark=at position \midpoint with {\arrow{>}}}]
\coordinate (A1) at (135:2);  \coordinate (A2) at (45:2);  
\coordinate (A3) at (-45:2);   \coordinate (A4) at (-135:2);
\coordinate (B1) at (-1,0); \coordinate (B2) at (1,0);
\draw[thin,dashed] (0,0) circle (2);
\draw[dashed,red] (A1)--(B1)--(A4); \draw[dashed,red] (A3)--(B2)--(A2); \draw (B1)--(B2);
\draw[fill=black] (A1) circle (0.1); \draw[fill=black] (A2) circle (0.1); 
\draw[fill=black] (A3) circle (0.1); \draw[fill=black] (A4) circle (0.1); 
\draw (B1)--(A4);
\def\midpoint{0.63}\draw[postaction={decorate}] (A1)--(B1);
\def\midpoint{0.63}\draw[postaction={decorate}] (B1)--(B2);
\def\midpoint{0.53} \draw[postaction={decorate}] (B2)--(A2);
\draw (A3)--(B2);
\node[left] at (A1) {$\vec x_{1}$};
\node[right] at (A2) {$\vec x_{3}$};
\node[right] at (A3) {$\vec x_{4}$};
\node[left] at (A4) {$\vec x_{2}$};
\node[left] at (B1) {$z$};
\node[right] at (B2) {$w$};
\end{tikzpicture}
\notag\\&=&
 -\int d^{d+1} z\sqrt{g(z)}d^{d+1} w\sqrt{g(w)}\; 
K_{\Delta_2} (z,\vec x_2) \ \bar\Sigma_{\Delta_1}(z, \vec x_1) S(z,w) \Sigma_{\Delta_3}(w,\vec x_3) 
\  K_{\Delta_4} (w,\vec x_4)\ . \qquad \la{c1}
 \eeqn
 Here $\Delta_r$ are dimensions of the  corresponding fields  and the exchanged field has mass $m$. 
  Using the boundary  translational invariance, we can set $ {\vec x}_3=0$, i.e.   
\be\la{c2}
A(\vec x_1, \vec x_2, \vec x_3, \vec x_4 ) =A( {\vec x}_{13},  {\vec x}_{23}, 0,  {\vec x}_{43} ), \qquad\qquad   {\vec x}_{i3} = {\vec x} _i - {\vec x}_3 \ .
\ee 
 Let us now  consider the  inversion transformation  of a 
bulk and a boundary point  
 \be
 z^a \rightarrow \hat z^a=\frac{z^a}{|z|^2}=\frac{z^a}{z_0^2+\vec z^2}\ ,\qquad
\vec x \rightarrow \hat {\vec x} =\frac{\vec x}{|\vec x|^2} \ ,
 \ee
under which the AdS measure and the geodesic length are left invariant
 \be
 \int d^{d+1} z \sqrt{g(z)}= \int d^{d+1} \hat z \sqrt{g(\hat z)}, \qquad u(z,w)=u(\hat z, \hat w),
 \ee
while the bulk-to-boundary propagators transform according to 
 \begin{align}  \label{Kinversion}
 K_{\Delta}(\hat z,\hat {\vec x}  ) &=  |\vec x|^{2\Delta}  K_{\Delta}(z,\vec x), \qquad
 & K_\Delta (\hat w, \hat{\vec x}=0)&= c_\Delta w_0^\Delta  ,
 \\
 \bar \Sigma(\hat z, \hat{\vec x})&= -\frac{\slashed {\vec x}}{|\vec x|^{2-2\Delta}}\cP_- K_\Delta(z,\vec x) U(z,\vec x)  \frac{\slashed z}{|z|}, \qquad
 & \Sigma(\hat w, \hat{\vec x}=0) &= c_\Delta  \frac{\slashed w}{|w|} w_0^{\Delta-1/2}\cP_-\ .
 \end{align}
The fermion bulk-to-bulk propagator in \eqref{b9}  changes as 
 \be
 S(\hat z,\hat w) =\frac{\slashed z}{|z|} \Big(\slashed D+\Gamma^0-m  \Big)  \Big[z_0^{-1/2}
 \Big(G_{\Delta_-}(z,w) \cP_+ +G_{\Delta_+}(z,w) \cP_-   \Big)  w_0^{ 1/2}  
 \frac{\slashed w}{|w|} \Big] \ .
 \la{Shatinversion}
 \ee
Using these  inversion transformation rules, we can express  the  arguments of propgators in \rf{c1},\rf{c2}  in terms of the inverted coordinates $\hat x_{i3}$
 \beqn
A(\vec x_1, \vec x_2, 0, \vec x_4 ) &=& -\int d^{d+1} z\sqrt{g(z)}d^{d+1} w\sqrt{g(w)}\; 
K_{\Delta_2} (z,\vec x_{23})   \bar\Sigma_{\Delta_1}(z, \vec x_{13}) S(z,w) \Sigma_{\Delta_3}(w,0)   K_{\Delta_4} (w,\vec x_{43})
\notag\quad \\&=&
c_{\Delta_3}   \frac{ \hat{\slashed {\vec x}}{}_{\;13}}{|\hat{\vec x}_{13}|^{2-2\Delta_1 } |\hat{\vec x}_{23}|^{-2\Delta_2 } |\hat{\vec x}_{43}|^{-2\Delta_4 } } 
 \cP_-\int d^{d+1}  z\sqrt{g(  z)}d^{d+1}  w\sqrt{g( w)}\; 
\notag\\& &
\  K_{\Delta_2} (\hat z, \hat{\vec x}_{23})   K_{\Delta_1}(\hat z, \hat{\vec x}_{13}) \ 
U(\hat z, \hat{\vec x}_{13})  \frac{\slashed {\hat z}}{|\hat z|}   S( z,  w) \frac{\slashed {\hat w}}{|\hat w|} 
 \hat w_0^{\Delta_3-1/2} \  K_{\Delta_4} (\hat w,\hat{\vec x}_{43}) \cP_-
\notag\\&=&
c_{\Delta_3}   \frac{ {\slashed {\vec x}}_{13}}{| {\vec x}_{13}|^{2\Delta_1 } | {\vec x}_{23}|^{ 2\Delta_2 } | {\vec x}_{43}|^{2\Delta_4 } } 
 \cP_-\int d^{d+1} \hat z\sqrt{g(\hat z)}d^{d+1}\hat  w\sqrt{g(\hat w)}\; 
\notag\\& &
\   K_{\Delta_2} (\hat z, \hat{\vec x}_{23})   K_{\Delta_1}(\hat z, \hat{\vec x}_{13}) \ 
U(\hat z, \hat{\vec x}_{13})  \frac{\slashed {\hat z}}{|{\hat z}|}   S(\hat{\hat z},\hat{\hat w}) \frac{\slashed {\hat w}}{|{\hat w}|} 
 \hat w_0^{\Delta_3-1/2} \  K_{\Delta_4} (\hat w,\hat{\vec x}_{43}) \cP_- ,
\eeqn
where we used that $\hat{\hat z}=z$, i.e. that the inversion squares to the identity.
To simplify  notation, we will use $\vec y_i=\hat{\vec x}_{i3}$ and make the
replacements  $\hat z\rightarrow z, \hat w\rightarrow w$. Then 
\begin{align}
A(\vec x_1, \vec x_2, \vec x_3, \vec x_4 )&=c_{\Delta_3}   \frac{ {\slashed {\vec x}}_{13}}{| {\vec x}_{13}|^{2\Delta_1 } | {\vec x}_{23}|^{ 2\Delta_2 } | {\vec x}_{43}|^{2\Delta_4 } } 
B(\vec y_1, \vec  y_2, \vec  y_4) \ , 
\\
 B(\vec y_1, \vec  y_2, \vec  y_4)&= \cP_-\int d^{d+1}   z\sqrt{g(  z)}d^{d+1}   w\sqrt{g(  w)}\; 
K_{\Delta_2} (  z, \vec y_2)   K_{\Delta_1}(  z, \vec y_1) \no\\ 
&\qquad \qquad \qquad \qquad \times U(  z,  {\vec y}_{1 })  \frac{\slashed {  z}}{|{  z}|}   S(  \hat z, \hat  w) \frac{\slashed {  w}}{|{  w}|} 
   w_0^{\Delta_3-1/2} \  K_{\Delta_4} (  w,\vec y_4) \cP_-
  \notag   \\&= 
 \cP_-\int d^{d+1}   z\sqrt{g(  z)}d^{d+1}   w\sqrt{g(  w)}\; 
K_{\Delta_2} (  z, \vec y_2)   K_{\Delta_1}(  z, \vec y_1) \ 
U(  z,  \vec y_1)  
\notag\\& \qquad\qquad\qquad \times 
 \Big(\slashed D+\Gamma^0-m  \Big)  \Big[z_0^{-1/2}   G_{\Delta_+}(z,w)     \Big] 
   w_0^{\Delta_3 } \  K_{\Delta_4} (  w,\vec y_4) \cP_-  \ , \la{c10}
\end{align}
where \rf{c10} we used  the inversion relation  \eqref{Shatinversion} 
for $S(\hat z, \hat w)$ as well as the projection properties 
 $\cP_-^2 =\cP_-, \,  \cP_+^2=\cP_+, \, \cP_+\cP_-=\cP_-\cP_+=0$.
 
%
Next, we may use integration by parts relations like\foot{Here $f$ and $g$ are arbitrary matrix functions that decay rapidly enough at infinity, so no surface terms are needed.}
 \begin{align}
& \int \frac{d^{d+1} z }{z_0^ {d+1}} f(z) z_0 \Gamma^a \p_a g(z)=
 \int \frac{d^{d+1} z }{z_0^{d+1}} \Big( - \p_a  f(z) z_0 \Gamma^a +d\,  f\Gamma^0\Big) g(z)\ ,
 \no \\
& \int d^{d+1}    z\sqrt{g(  z)}\; f\slashed \p g = \int d^{d+1}    z\sqrt{g(  z)}\; f \big(-\overleftarrow{  \slashed \p} +d \Gamma^0\big)g\ ,\no \\
&  \int d^{d+1}    z\sqrt{g(  z)}\; f\slashed D g= \int d^{d+1}    z\sqrt{g(  z)}\; f \big(-\overleftarrow{  \slashed D}  \big)g \ . 
\la{DiracIBP}
 \end{align} 
 Here  $\overleftarrow{\slashed  D} ,\ \overleftarrow{\slashed  \p} $ do  not act on the measure.
 One can also show that 
  \begin{align}
& \frac{1}{\sqrt{z_0}} \cP_- \Big(K_{\Delta'}(z,\vec y)   K_\Delta(z,\vec x)  U(z,\vec x) \Big) \overleftarrow{\slashed  D}  \cP_-=  -( \Delta-\Delta'-\frac{d+1}{2})      \cP_-  K_{\Delta'}(z,\vec y) K_\Delta(z,\vec x)\no 
\\
  &\qquad\qquad \qquad \qquad\qquad \qquad 
  -2  (\Delta '-\frac{d}{2})     \cP_- U(z,\vec x)  U(z,\vec y)  \cP_-   K_{\Delta'+1} (z,\vec y) K_\Delta(z,\vec x) \cP_-\ \ .
 \end{align} 
Using  (\ref{DiracIBP}) and the following projections, 
\be
 \frac{1}{\sqrt{z_0}} \cP_-U(z, \vec x) \cP_- = - \cP_-, \quad \cP_-\Gamma^0=-\cP_-, \quad
\cP_- U(  z,  \vec y )\Big(- \Gamma^0+m \Big) \cP_-= - (1+m) \cP_- ,
\ee
one can write $B$ in \rf{c10}  as
%
%
 \be
 B(\vec y_1,\vec y_2,\vec y_4)=(2\Delta_2-d) J (\vec y_1,\vec y_2,\vec y_3)
 +\Big( m+\Delta_1-\Delta_2-\frac{d}{2} +\frac12 \Big) I  (\vec y_1,\vec y_2,\vec y_3) ,
 \qquad \qquad
 \ee
 where 
 \beqn
I(\vec y_1,\vec y_2,\vec y_4) &= &
\int d^{d+1}   z\sqrt{g(  z)}d^{d+1}   w\sqrt{g(  w)}\; 
  K_{\Delta_2}(z,\vec y_2) K_{\Delta_1}(z,\vec y_1) 
  G_{\Delta_+}(z,w)     w_0^{\Delta_3 }   K_{\Delta_4} (  w,\vec y_4)  \cP_-, 
\qquad\quad\\
J(\vec y_1,\vec y_2,\vec y_4) &= &
\int d^{d+1}   z\sqrt{g(  z)}d^{d+1}   w\sqrt{g(  w)}\; 
  K_{\Delta_2+1}(z,\vec y_2) K_{\Delta_1}(z,\vec y_1) 
  G_{\Delta_+}(z,w)     w_0^{\Delta_3 }   K_{\Delta_4} (  w,\vec y_4)  
  \notag\\&&\qquad\qquad\qquad\qquad\qquad\qquad
   \times\ \cP_- U(z,\vec y_1)  U(z,\vec y_2)  \cP_- \ .\la{c16}
   \eeqn
%
%
%
Since 
 $
 U(z, \vec y)  
 =U(z, \vec x) - {\slashed {\vec y} -\slashed {\vec x}\ov   \sqrt{z_0}}
 \ $, 
 one can further show that 
 \beqn
 \cP_- U(z, \vec y) U(z, \vec x) \cP_-
=\Big[     \frac{(z-x)^2}{z_0}  
 -  \frac{ (\slashed {\vec y} -\slashed {\vec x}) (\slashed {\vec z} -\slashed {\vec x})    }{z_0}    \Big]  \cP_-  \ .
 \eeqn
We also have
 \be
  \frac{ (\slashed {\vec z} -\slashed {\vec x})    }{z_0}  K_{\Delta+1}(z, \vec x) 
  = \frac{1}{2\Delta-d } \slashed \p_{\vec x} K_\Delta(z, \vec x), \quad
   \frac{(z-x)^2}{z_0}    K_{\Delta+1}(z, \vec x)  = \frac{2\Delta}{ 2\Delta-d }K_\Delta(z, \vec x)
\ .  \ee
These formulae enable one to express $J$ in \rf{c16} as 
 \be
 J(\vec y_1,\vec y_2,\vec y_4) =  \frac{1}{2\Delta_2-d} \Big( 2 \Delta_2 -\slashed {\vec y}_{12} \slashed \p_{\vec y_2}\Big)  I( \vec y_1,\vec y_2,\vec y_4) \ ,
 \ee
so that  we can finally write $B$ as  
  \begin{align}
B(\vec y_1,\vec y_2,\vec y_4)& =    \Big(  -\slashed {\vec y}_{12}  \slashed\p_{\vec y_2} 
+\Delta_1+\Delta_2 +m+\frac12-\frac{d}{2}   \Big)  I( \vec y_1,\vec y_2,\vec y_4) 
\notag\\&=    \Big(  -\slashed {\vec y}_{12}  \slashed\p_{\vec y_2} 
+\Delta_1+\Delta_2 +m+\frac12-\frac{d}{2}   \Big)  \mathcal I( \vec y_1,\vec y_2,\vec y_4)  \cP_-,
 \end{align}
 where
  \be
\mathcal I(\vec y_1,\vec y_2,\vec y_4) = 
\int d^{d+1}   z\sqrt{g(  z)}d^{d+1}   w\sqrt{g(  w)}\; 
  K_{\Delta_2}(z,\vec y_2) K_{\Delta_1}(z,\vec y_1) 
  G_{\Delta_+}(z,w)     w_0^{\Delta_3 }   K_{\Delta_4} (  w,\vec y_4) \ .
 \ee
This expression can be simplified further by using again  the inversion transformation.
From the relations~\eqref{Kinversion} for the  scalar propagators, we can write
  \begin{align}
&\mathcal I(\vec y_1,\vec y_2,\vec y_4) = \frac{1}{c_{\Delta_3}}
 |\hat{\vec y}_1|^{2\Delta_1}|\hat{\vec y}_2|^{2\Delta_2} |\hat{\vec y}_4|^{2\Delta_4}\;
  W^s_{\Delta_1\Delta_2\Delta_3\Delta_4,\Delta_+}(\hat{\vec y}_1, \hat{\vec y}_2, \hat{\vec y}_3=0, \hat{\vec y}_4)\ ,
 \\
  &W^s_{\Delta_1\Delta_2\Delta_3\Delta_4,\Delta_+}   (\vec y_1,\vec y_2,\vec y_3,\vec y_4) \notag \\
 \equiv &\int d^{d+1}    z\sqrt{g(   z)}d^{d+1}    w\sqrt{g(   w)}\; 
  K_{\Delta_1}(  z, {\vec y}_1)   K_{\Delta_2}(  z, {\vec y}_2)
  G_{\Delta_+}(  z,  w)       K_{\Delta_3}(  w ,0)    K_{\Delta_4} (    w, {\vec y}_4) \ ,
\la{c23}  \end{align}
 where \rf{c23} is the contribution of   the exchange diagram involving only scalar fields.  
 Let us also note that  the derivative of $\mathcal I$  may be written as 
 \begin{align}
& \frac{\p }{\p y_2^j} \mathcal I(\vec y_1,\vec y_2,\vec y_4) =- |\hat{\vec y}_2 |^2 
  \frac{\p }{\p \hat y_2^j} \mathcal I(\vec y_1,\hat{\vec y}_2,\vec y_4) 
  \notag\\&=
  -  \frac{1}{c_{\Delta_3}}
 |\hat{\vec y}_1|^{2\Delta_1}|\hat{\vec y}_2|^{2\Delta_2} |\hat{\vec y}_4|^{2\Delta_4}\;
  \Big(   |\hat{\vec y}_2|^{  2}  \frac{\p }{\p \hat y_2^j}  +2\Delta_2 \hat y_2^j    \Big) 
  W^s_{\Delta_1\Delta_2\Delta_3\Delta_4,\Delta_+}(\hat{\vec y}_1, \hat{\vec y}_2, \hat{\vec y}_3, \hat{\vec y}_4) 
 \end{align}
 Thus finally we obtain  for \rf{c1}
 (note that  $ \vec y_i =\hat {\vec x}_{i3}=  {{\vec x}_{i3}\ov |{\vec x}_{i3}|^2}$, 
 and thus $ \hat {\vec y}_i ={\vec x}_{i3} $):
 \begin{align}
A(\vec x_1, \vec x_2, \vec x_3, \vec x_4 )&=
  c_{\Delta_3}   \frac{ {\slashed {\vec x}}_{13}}{| {\vec x}_{13}|^{2\Delta_1 } | {\vec x}_{23}|^{ 2\Delta_2 } | {\vec x}_{43}|^{2\Delta_4 } } 
 \Big(  -\slashed {\vec y}_{12}  \slashed\p_{\vec y_2}+\Delta_1+\Delta_2 +m+\frac12-\frac{d}{2}   \Big)  \mathcal I( \vec y_1,\vec y_2,\vec y_4)  \cP_-
\notag\\&=
    {\slashed {\vec x}}_{13} \cP_-  \Big[    \Big( \hat{\slashed {\vec x}}_{\;\; 13} - \hat{\slashed {\vec x}}{}_{\;\; 23} \Big)
  \Big(   |  {{ \slashed{\vec  x}}{}_{23}} |^{  2}  \slashed\p {}_ {{ {\vec  x}}{}_{23}}  +2\Delta_2 { \slashed{\vec  x}}{}_{23}   \Big) \no \\
 & \qquad \qquad \qquad 
+\Delta_1+\Delta_2 +m+\frac12-\frac{d}{2}   \Big]    W^s_{\Delta_1\Delta_2\Delta_3\Delta_4,\Delta_+}
( {\vec x}_{13},  {\vec x}_{23}, 0,  {\vec x}_{43})
\notag \\&=  
  {\slashed {\vec x}}_{13} \cP_-  \Bigg[   \Big( \frac{ {\slashed {\vec x}}_{ 13}  } {|{\vec x}_{ 13}|^2}   - 
\frac{ {\slashed {\vec x}}_{ 23}  } {|{\vec x}_{ 23}|^2}   \Big)
  \Big(   |  {{ \slashed{\vec  x}}{}_{23}} |^{  2}  \slashed\p {}_ {{  {\vec  x}}{}_{2 }}  +2\Delta_2 { \slashed{\vec  x}}{}_{23}   \Big)  \no \\
&  \qquad \qquad \qquad 
+\Delta_1+\Delta_2 +m+\frac12-\frac{d}{2}   \Bigg]  
 W^s_{\Delta_1\Delta_2\Delta_3\Delta_4,\Delta_+}
( {\vec x}_{1  },  {\vec x}_{2 },  {\vec x}_{3 },  {\vec x}_{4 })\ . 
\la{c25}
 \end{align}
 As   anticipated, this allows us  to rewrite the fermion exchange diagram in terms of 
 a suitable differential operator acting on the purely scalar exchange diagram
 which  is much easier   to compute 
 using  known general results  (see,  e.g.,  \cite{DHoker:1999mqo}).

\subsection{$\EV{\sfPhi^2\sfPsi^2}$ correlator in super Liouville theory  in AdS$_2$}

The  correlator $\EV{\sfPhi^2\sfPsi^2}$ involving two fermions and two scalars
receives   contributions from the following  four diagrams
  \be
  \la{c26}
 \begin{tikzpicture}[line width=1 pt, scale=0.5, rotate=0,baseline=-0.1cm,decoration={
    markings,
    mark=at position \midpoint with {\arrow{>}}}]
\coordinate (A1) at (135:2);  \coordinate (A3) at (45:2);  
\coordinate (A2) at (-45:2);   \coordinate (A4) at (-135:2);
\coordinate (B1) at (-1,0); \coordinate (B2) at (1,0);
\draw[thin,dashed] (0,0) circle (2);
\draw[dashed,red] (A1)--(B1)--(A4); \draw[dashed,red] (A3)--(B2)--(A2); \draw (B1)--(B2);
\draw[fill=black] (A1) circle (0.1); \draw[fill=black] (A2) circle (0.1); 
\draw[fill=black] (A3) circle (0.1); \draw[fill=black] (A4) circle (0.1); 
\draw (B1)--(A4);
\def\midpoint{0.63}\draw[postaction={decorate}] (A1)--(B1);
\def\midpoint{0.63}\draw[postaction={decorate}] (B1)--(B2);
\def\midpoint{0.53} \draw[postaction={decorate}] (B2)--(A2);
\draw (A3)--(B2);
\node[left] at (A1) {$\bt_{1}$};
\node[right] at (A2) {$\bt_{3}$};
\node[right] at (A3) {$\bt_{4}$};
\node[left] at (A4) {$\bt_{2}$};
\node[below] at (0,-2.5) {$\mathcal A_s$};
\end{tikzpicture} 
+
 \begin{tikzpicture}[line width=1 pt, scale=0.5, rotate=0,baseline=-0.1cm,decoration={
    markings,
    mark=at position \midpoint with {\arrow{>}}}]
\coordinate (A1) at (135:2);  \coordinate (A3) at (45:2);  
\coordinate (A2) at (-45:2);   \coordinate (A4) at (-135:2);
\coordinate (B1) at (0,1); \coordinate (B2) at (0,-1);
\draw[thin,dashed] (0,0) circle (2);
\draw[fill=black] (A1) circle (0.1); \draw[fill=black] (A2) circle (0.1); 
\draw[fill=black] (A3) circle (0.1); \draw[fill=black] (A4) circle (0.1); 
\draw (B2)--(A4);
\def\midpoint{0.63}\draw[postaction={decorate}] (B2)--(A2);
\def\midpoint{0.63}\draw[postaction={decorate}]  (B1)--(B2);
\def\midpoint{0.53} \draw[postaction={decorate}] (A1)--(B1);
\draw (A3)--(B1);
\node[left] at (A1) {$\bt_{1}$};
\node[right] at (A2) {$\bt_{3}$};
\node[right] at (A3) {$\bt_{4}$};
\node[left] at (A4) {$\bt_{2}$};
\node[below] at (0,-2.5) {$\mathcal A_t$};
\end{tikzpicture} 
+
 \begin{tikzpicture}[line width=1 pt, scale=0.5, rotate=0,baseline=-0.1cm,decoration={
    markings,
    mark=at position \midpoint with {\arrow{>}}}]
\coordinate (A1) at (135:2);  \coordinate (A3) at (45:2);  
\coordinate (A2) at (-45:2);   \coordinate (A4) at (-135:2);
\coordinate (B1) at (0,1); \coordinate (B2) at (0,-1);
\draw[thin,dashed] (0,0) circle (2);
\draw[fill=black] (A1) circle (0.1); \draw[fill=black] (A2) circle (0.1); 
\draw[fill=black] (A3) circle (0.1); \draw[fill=black] (A4) circle (0.1); 
\draw[fill=black] (A4) circle (0.1); 
\draw (B2)--(A4);
\def\midpoint{0.63}\draw[postaction={decorate}] (B1)--(A2);
\draw (B1)--(B2);
\def\midpoint{0.53} \draw[postaction={decorate}] (A1)--(B1);
\draw (A3)--(B2);
\node[left] at (A1) {$\bt_{1}$};
\node[right] at (A2) {$\bt_{3}$};
\node[right] at (A3) {$\bt_{4}$};
\node[left] at (A4) {$\bt_{2}$};
\node[below] at (0,-2.5) {$\mathcal A_u$};
\end{tikzpicture}
+
 \begin{tikzpicture}[line width=1 pt, scale=0.5, rotate=0,baseline=-0.1cm,decoration={
    markings,
    mark=at position \midpoint with {\arrow{>}}}]
\coordinate (A1) at (135:2);  \coordinate (A2) at (45:2);  
\coordinate (A3) at (-45:2);   \coordinate (A4) at (-135:2);
\coordinate (B1) at (0,0); \coordinate (B2) at (0,0);
\draw[thin,dashed] (0,0) circle (2);
\draw[dashed,red] (A1)--(B1)--(A4); \draw[dashed,red] (A3)--(B2)--(A2); \draw (B1)--(B2);
\draw[fill=black] (A1) circle (0.1); \draw[fill=black] (A2) circle (0.1); 
\draw[fill=black] (A3) circle (0.1); \draw[fill=black] (A4) circle (0.1); 
\draw[fill=black] (B1) circle (0.081); 
 \draw (B1)--(A2);
\def\midpoint{0.63}\draw[postaction={decorate}] (A1)--(B1);
\def\midpoint{0.53} \draw[postaction={decorate}] (B2)--(A3);
 \draw (A4)--(B2);
\node[left] at (A1) {$\bt_{1}$};
\node[right] at (A2) {$\bt_{4}$};
\node[right] at (A3) {$\bt_{3}$};
\node[left] at (A4) {$\bt_{2}$};
\node[below] at (0,-2.5) {$\mathcal A_\text{contact}$};
\end{tikzpicture} \ .
 \ee 
A non-trivial calculation is required for the first diagram $\mc A_{s}$ only. Indeed, the second diagram $\mc A_{t}$
can be obtained by crossing, while the other two diagrams $\mc A_{u}$ and $\mc A_{\rm contact}$
do not involve a bulk fermion propagator and can  be computed by using the relation
 (\ref{PsiPhiDiagram}). 
 
 The contribution of the  first diagram $\mathcal A_s$  is obtained from $A$ defined in (\ref{c1}) by specializing the general result (\ref{c25})
 to the particular $d=1$ case with  $\Delta_1=\Delta_2=\Delta_3=\Delta_4=\Delta_+=2$, and taking the $(12)$ component of the spinor matrix  to project to the
 $\sfPsi\sfPsi$ channel. 
 Finally, we have to insert the coupling  and  normalization factors to match our conventions
 for the fermionic propagators used  in the main text as  compared to
 Appendix  \ref{app:propagators}. 
 All steps are straightforward: 
 the specialisation of (\ref{c25}) gives 
 \be
 \la{cc27}
 A (\sft_1, \sft_2, \sft_3,\sft_4 ) 
=  {\sft} _{13}\, \Gamma^\sft  \cP_-
  \Big[  \Big(\frac{1}{\sft_{13}} - \frac{1}{\sft_{23}}   \Big)   \Big(  2\Delta_2  \sft_{23} + \sft_{23} ^2   \p_{ \sft_{ 2}}  \Big) 
+\Delta_1+\Delta_2 +m+\frac12-\frac{d}{2}   \Big]    W^s_{2222,2}(\sft_{1 } , \sft_{2 }, \sft_3,\sft_{4 })  , 
 \ee
 where  
  \beqn
 W^s_{2222,2}(\sft_1,\sft_2,\sft_3,\sft_4)    
  &=&  \int  \frac{d ^2 z}{z_0^2}   \frac{d ^2 w}{w_0^2}  \; 
  K_{2}( z, {\sft}_1)  K_{ 2}(  z, {\sft}_2) 
  G_{2}( z,  w)       K_{2}( w , {\sft}_3)    K_{2} (  w, {\sft}_4) 
 \notag \\&=&
  c_2^4  \  \frac14 |\sft_3-\sft_4|^{-2} D_{2211 }( {\sft}_1, {\sft}_2, {\sft}_3, {\sft}_4)
\notag  \\&=&
    -\frac{\left( {\chi} ^3-3  {\chi} +2\right) \log \left| 1- {\chi} \right| + {\chi}  \left( 
    -{\chi} ^2\, 
    \log \left|  {\chi} \right| + {\chi} -1\right)}{2 \pi ^3 ( {\chi} -1)^2  {\chi} ^3  \sft_{13}^4 \sft_{24}^4}\ .
 \eeqn
 Here    $\chi=\frac{\sft_{12}\sft_{34}}{\sft_{13}\sft_{24}}$ is the  1d cross-ratio, $c_2$ is given in \eqref{Kfunction},
 and the (standard) AdS integral is computed  as in  \cite{DHoker:1999mqo} ($D$-functions are discussed
 in \cite{DHoker:1999kzh,Dolan:2003hv}). Evaluating (\ref{cc27}) gives
 \be
A (\sft_1, \sft_2, \sft_3,\sft_4 ) =  {\sft} _{13}\, \Gamma^\sft  \cP_- \frac{\left(2 \chi ^4-5 \chi ^3+7 \chi -4\right) \log \left| 1-\chi \right| -\chi  (2 \chi -5) \left(\chi ^2 \log \left| \chi \right| -\chi +1\right)}{2 \pi ^3 (\chi -1)^2 \chi ^3  \sft_{13} ^4  \sft_{24} ^4}.
 \ee
Adding the coupling and normalization factors,\footnote{Explicitly, 
$ ({2\pi\ov 3})^2  $ is the rescaling of two scalar bulk-to-boundary propagator 
$g_{_{\del}}(\sft;w')=\frac{4}{3c_2} K(\sft,w')=\frac{2\pi}{ 3} K(\sft,w') $; \ $ ( 2\pi)^3$ 
comes from  different normalizations of the  3 fermionic propagators, 
  $\sb (w,\sft')=2\pi \Sigma(w,\sft') $ and $\sb(\sft,w')=-2\pi \bar\Sigma(w',\sft)$; 
$\frac{1}{(4\pi)^2}$ arises from the rescaling of the integration measure;    
$(2b)^2$ is the coupling factor  of the two vertices; 
the projection  to $\sfPsi$ gives $(\Gamma^\sft  \cP_-)_{12}=\ha $. }
we get the final expression for the contributions of the s-channel fermion exchange 
   \beqn
\mathcal   A_s (\sft_1, \sft_2, \sft_3,\sft_4 ) &=&  (2b)^2  \  ( 2\pi)^3 \ \Big(\frac{2\pi}{3 }\Big)^2 \  \frac{1}{(4\pi)^2}  
 \Big[A (\sft_1, \sft_2, \sft_3,\sft_4 ) \Big]_{12}
\notag \\&=&  {\sft} _{13}\frac{4 \pi ^3 b^2}{9}\frac{\left(2 \chi ^4-5 \chi ^3+7 \chi -4\right) 
 \log \left| 1-\chi \right| -\chi  (2 \chi -5) \left(\chi ^2 \log \left| \chi \right| -\chi +1\right)}{2 \pi ^3 (\chi -1)^2 \chi ^3  \sft_{13} ^4  \sft_{24} ^4}\ .  \la{c30}
 \qquad
 \eeqn
 The t-channel  contribution  in (\ref{c26}) is  obtained by exchanging 
 the 2,4 legs 
 \be\la{c31}
   \mathcal   A_t(\sft_1, \sft_2, \sft_3,\sft_4 ) =   \mathcal  A_s(\sft_1, \sft_4, \sft_3,\sft_2 )  \ .
 \ee
 The  third (u-channel scalar exchange) diagram   in (\ref{c26}) as well as the last contact diagram 
 can be computed  using  (\ref{SbSbuseful},\ref{PsiPhiDiagram}):  
 \begin{align}
 \mathcal  A_u&=2b \  (-8b) \Big[ \int {\sf d}^2 w \,{\sf d}^2 w'\;  \sb (\sft_1, w) \sb(w, \sft_3) g(w,w')  \gb(\sft_2, w') \gb(\sft_4,w') \Big]_{12}
\notag  \\&=
2  b \  (-8b) \  \Big(-\frac92 \sft_{13}\Big)\int {\sf d}^2 w \,{\sf d}^2 w'\; \gb(\sft_1, w) \gb(w, \sft_3) g(w,w')  \gb(\sft_2, w') \gb(\sft_4,w')
\notag   \\ &= 
72b^2 \sft_{13 } \   2\pi\  \Big(\frac43\Big)^4 \  \frac{1}{(4\pi)^2} \ 
 \frac14 |\sft_2-\sft_4|^{-2} D_{2121 }( {\sft}_1, {\sft}_2, {\sft}_3, {\sft}_4)
 \notag  \\ &= 
 \frac{64 b^2}{9\pi  }      \frac{\sft_{13}}{\sft_{24}^2}D_{2121 }( {\sft}_1, {\sft}_2, {\sft}_3, {\sft}_4)\ , \la{c32} \\
 \mathcal  A_\text{contact}&=
2b^2 \  \Big(-\frac92 \sft_{13}\Big)\int {\sf d}^2 w \; \gb(\sft_1, w) \gb(w, \sft_3) \gb(\sft_2, w ) \gb(\sft_4,w )
\notag  \\ &=-9 b^2 \sft_{13 } \    \Big( \frac43\Big)^4 \  \frac{1}{(4\pi) } \ 
 D_{2222 }( {\sft}_1, {\sft}_2, {\sft}_3, {\sft}_4)
= 
 -  \frac{64 b^2}{9\pi  }        \frac{\sft_{13}}{\sft_{24}^2}D_{2222 }( {\sft}_1, {\sft}_2, {\sft}_3, {\sft}_4)\ .\la{c33}
 \end{align}
 Summing up all the contributions \rf{c30},\rf{c31},\rf{c32},\rf{c33}  to  (\ref{c26}), we arrive at the final result
 \be
\EV{\sfPsi(\sft_1)  \sfPhi(\sft_2)  \sfPsi(\sft_3)  \sfPhi(\sft_4)   } _{0,\text{conn}} 
=\mathcal A_s  +\mathcal A_t    +\mathcal A_u+\mathcal A_\text{contact}
=\frac{2b^2}{3} \frac{1}{ \sft_{13}^3 \sft_{24}^4  } 
 \frac{4\chi^2-4\chi+3}{\chi^2(\chi-1)^2} \ ,\la{c34}
 \ee
which is in full agreement with the prediction from  \eqref{443} and \eqref{447}.

\ed